\documentstyle[eqsecnum,array,multicol,prl,aps,epsf]{revtex}
\draft
\tighten
\begin{document}

\title{Paramagnetic Meissner effect and related dynamical phenomena}

\author{Mai Suan Li}

\address{Institute of Physics, Polish Academy of Sciences,
Al. Lotnikow 32/46, 02-668 Warsaw, Poland}

\address{
\centering{
\medskip\em
{}~\\
\begin{minipage}{14cm}
The hallmark of superconductivity is the diamagnetic response to external 
magnetic field. In striking contrast to this behavior, a paramagnetic response
or paramagnetic Meissner effect was observed in ceramic high-$T_c$
and in conventional superconductors.
The present review is given on this interesting effect and related phenomena.
We begin with a detailed discussion of experimental results on
the paramagnetic Meissner effect in both granular and
conventional superconductors. There are two main mechanisms leading to
the paramagnetic response: the so called $d$-wave and the flux compression.
In the first scenario, the Josephson critical current between
two $d$-wave superconductors becomes negative or equivalently one has
a $\pi$ junction. The paramagnetic signal occurs due to
the nonzero spontaneous supercurrent circulating in a loop consisting
of odd number of $\pi$ junctions.  
In addition to the $d$-wave mechanism we present the flux compression mechanism
for the paramagnetic Meissner effect. The compression may be due to either
an inhomogeneous superconducting transition or flux trap inside the giant
vortex state. The flux trapping which acts like a total nonzero spontaneous
magnetic moment causes the paramagnetic signal. The anisotropic pairing
scenario is believed to be valid for granular materials while the flux trap
one can be applied to both conventional and high-$T_c$ superconductors. 
The study of different phenomena 
by a three-dimensional lattice model of randomly distributed
$\pi$ Josephson junctions with finite self-inductance occupies the main
part of our review. By simulations one can show that
the chiral glass phase in which chiralities are frozen in time and in
space may occur in granular superconductors possessing $d$-wave pairing
symmetry.
Experimental attempts on the search for the chiral glass phase are analysed.
Experiments on dynamical
phenomena such
as AC susceptibility, compensation effect, anomalous microwave absorption, 
aging effect,  AC resistivity and enhancement of critical current
by external electric fields are considered.
These phenomena were studied by Monte Carlo and Langevin dynamics 
simulations which show
satisfactory agreement  with experiments.
We present a resistively shunted junction model describing rich dynamics of
Josephson junction networks.
{}~\\
{}~\\
%{\noindent PACS numbers: 75.40.Gb, 74.72.-h}
\end{minipage}
}}

\maketitle

%\begin{flushright}
%e-mail: masli@ifpan.edu.pl
%\end{flushright}

%\noindent
%KEYWORDS: chiral glass, high-$T_c$ superconductors,
%Josephson junction, screening, frustration,
%second harmonics, Monte Carlo simulation.

\begin{multicols}{2}

\tableofcontents

\section{Introduction}

The phenomenon of superconductivity was first discovered 
in 1911 by  H. Kamerlingh Onnes. He observed that when mercury was put 
into a liquid
helium bath and cooled down to very low temperatures, its resistivity dropped
to zero %\cite{Kamerlingh}
 once the temperature becomes lower than a critical
value $T_c$. It was soon observed that superconductors, except from 
zero resistivity, also exhibit unusual magnetic properties.
In 1933 W. Meissner and R. Ochsenfeld 
% \cite{Meissner}
found that when a superconducting 
specimen is placed in a low enough magnetic field and is subsequently cooled
through its transition temperature, the magnetic flux is totally expelled
from its interior.
This phenomenon was called Meissner effect. In the conventional
Meissner phase 
the magnetic susceptibility is negative or the
response to the external magnetic field is diamagnetic.

The revolutionary discovery of high temperature superconductivity by Bednorz and
Muller \cite{Bednorz} ignited an explosion of efforts in study of
ceramic materials. One of such efforts led to the discovery of
 the so called {\em paramagnetic
Meissner effect} (PME).
Contrary
to the standard diamagnetic response to the external field, a paramagnetic 
signal was observed in certain ceramic superconductors
upon cooling in low enough fields 
\cite{Svelindh89,Braunisch92,Braunisch93}. 
In other words,
the field cooled (FC) susceptibility becomes positive whereas the zero field
cool (ZFC) magnetization remains negative.
This effect is now referred to as the PME
or as the Wohlleben effect \cite{Khomskii94}.
The existence of the PME is quite surprising because according to
the Ginzburg-Landau
phenomenology for continuous media \cite{Tinkham}, the lowest free energy states
should be diamagnetic.

The anomalous PME
in ceramic superconductors has been interpreted in the framework of different
models, such as spontaneous supercurrents due to vortex fluctuations
combined with pinning \cite{Svelindh89}, orbital glass \cite{Kusmartsev92},
the presence of so called
$\pi$ contacts 
\cite{Braunisch93,Sigrist92,Sigrist95,Chen94,Khomskii94,Dominguez},
and Josephson junctions \cite{Shrivastava94a,Shrivastava94b}.
In this review we focus on the most prominent
$d$-wave
mechanism proposed by  Sigrist and Rice \cite{Sigrist92,Sigrist95}.
According to their theory , the nature of the unusual paramagnetic behavior may
be related to the existence of $\pi$-junctions characterized by negative
Josephson couplings (conventional 0-junctions have positive couplings). 
It was argued that such $\pi$-junctions are a consequence
of $d_{x^2-y^2}$ type pairing symmetry \cite{Sigrist92}.
If a loop consists of odd number of $\pi$ junctions then one can show that
the spontaneous current (or magnetic moment) circulating around it becomes
nonzero. 
Such a frustrated $\pi$ loop would,
therefore, behave like a paramagnet and have the paramagnetic response 
to the external magnetic field. The existence of
the $d$-wave pairing seems to be supported by phase-sensitive and 
phase-insensitive experiments \cite{Tsuei00}.

The situation becomes ambiguous when the PME was observed even in 
conventional superconductors such as Nb \cite{Thompson,Kostic}
and Al \cite{Geim}. This has prompted the appearance of a less-exotic 
mechanism based on flux capture inside a superconducting sample and its
consequent compression with lowering temperature 
\cite{Larkin,Khalil97,Moshchalkov}. The flux trap can be caused by 
inhomogeneities \cite{Larkin,Khalil97} but could also be an intrinsic
property of any finite-size superconductor due to the sample boundary
\cite{Moshchalkov}. The trapped fluxes play the same role as 
spontaneous supercurrents in the $d$-wave mechanism and give rise to
the PME. 

At present, it is not clear if one needs the exotic $d$-wave mechanism to
explain the PME in ceramic materials or the flux trapping is adequate
for both conventional low-$T_c$ and high-$T_c$ superconductors 
\cite{Geim,Sigrist98}. The study of dynamical phenomena such as
the microwave absorption (MWA) \cite{Braunisch92}, the compensation effect
\cite{Heinzel},
the aging \cite{aging} and
enhancement of critical current \cite{Orlova99}, may shed light on this 
important issue. The anomalous MWA, e.g., was observed
\cite{Braunisch92} only in the ceramic samples showing the paramagnetic signal.
Simulations by different groups show that
all of dynamical effects may be captured by the multi-loop model of
the Josephson network \cite{Dominguez} with $\pi$-junctions. So  
the study of dynamics effects in ceramic superconductors supports the 
$d$-wave mechanism.

Recently we have proposed that a novel thermodynamic phase
 might occur in zero external 
field in unconventional superconductors \cite{Kawamura95,KawLi97}. 
This phase is characterized by a broken time-reversal symmetry and called a
chiral glass phase. The key idea is that the random distribution of 0- and
$\pi$-junctions between adjacent $d_{x^2-y^2}$ superconducting particles 
causes the frustration effect and induces the chiral glass transition. In
the chiral glass phase, similar to spin glasses, chiralities \cite{Villain}
are frozen in space and in time.  Using the multi-loop model \cite{Dominguez}
one can show that this phase is stable even under the screening
\cite{KawLi97}. Since
the existence of the chiral
glass would favour the $d$-wave mechanism for the PME, its study is very 
important for classification of the symmetry of the order parameter.
The experimental search for the chiral glass phase in
ceramic samples of different groups gives controversial results 
\cite{Matsuura1,Matsuura2,Papadopoulou,Ishida}. In this review
we also briefly discuss the difference between the chiral glass and
other glassy phases in impure type-II superconductors 
such as the gauge glass, vortex and Bragg glass.

It should be noted that results obtained before 1995  were summarized in the
review article of Sigrist and Rice \cite{Sigrist95}. However,
these authors discussed only the $d$-wave mechanism of the paramagnetic 
Meissner effect and related experiments in ceramic materials and
interesting dynamical phenomena correlated with the PME
were not addressed.
The present review focuses
on the progress achieved after 1995. The main developments of this period
are:

a) the discovery of the PME in conventional superconductors \cite{Geim};

b) more detailed study of the PME and related dynamical phenomena in ceramic
materials;

c) flux compression mechanism \cite{Larkin,Moshchalkov} which is 
not based on the $d$-wave
pairing symmetry;

d) theoretical and experimental search for a chiral glass phase.

The article is organized as follows. In section 2 we present a brief summary
of experimental results on the PME in both  ceramic and conventional 
superconductors. In section 3 we review the mechanisms for occurrence of
$\pi$ junction. One of them is relied on the scattering on magnetic impurities
and the another, more intrinsic, mechanism is based on the anisotropic pairing
of superconducting electrons. 
Section 4 concerns theory and simulations of the PME 
in ceramic superconductors employing single-loop and multi-loop models of
the Josephson junction network with $\pi$ junctions.
In section 5 we discuss the flux compression mechanism for the PME
based on the Bean critical model and the Ginzburg-Landau equation approach.
Section 6 is focused on the simulation and experimental search for the
chiral glass phase in granular materials.
Experiments and simulations on dynamical phenomena
related to the PME are presented in section 7. The multi-loop model is shown
to provide an unique tool to describe all of these effects.

\section{Experiments on paramagnetic Meissner effect}

\begin{figure}
\epsfxsize=3.2in
\centerline{\epsffile{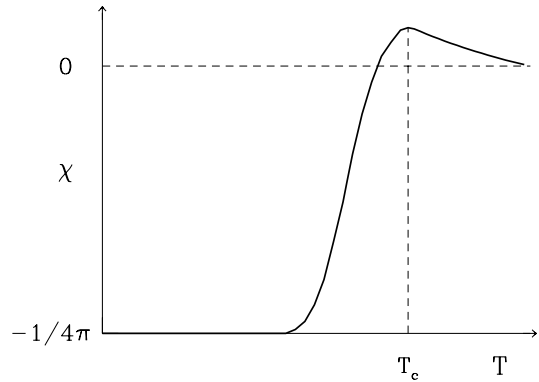}}
%\vspace{0.2in}
\caption{Schematic temperature dependence of the susceptibility of
superconductors without the PME}
\label{st_sus}
\end{figure}

In the
Meissner phase the total magnetic field $B$ inside conventional superconductors
vanishes and we have
\begin{eqnarray}
\vec{B} \; \; = \; \; \vec{H}  + \; 4\pi \chi \vec{M} \; \; = \; \; 0 \; ,
\nonumber\\
\chi \; \; = \; \; \frac{\partial{M}}{\partial{H}} \; \; = \; \;
-\frac{1}{4\pi} \; ,
\end{eqnarray}
where $H, M$ and $\chi$ are the external field,
the induced magnetization and the susceptibility.
Due to the finite London penetration depth (of order of 500 $\AA$)
 the magnetic fluxes do not, however,
entirely pull out from the interior and $\chi$ approaches $-1/4\pi$ as
$T \rightarrow 0$. It is shown schematically in Fig. \ref{st_sus}.
Above $T_c$ one has the paramagnetic behavior $\chi = c/T$ \cite{Tinkham}.
The susceptibility is slightly positive even in the narrow temperature
region below $T_c$ but it has nothing to do with the PME where the FC 
susceptibility remains positive up to $T=0$.

\vspace{-1cm}

% FIGURE 1

\subsection{High-temperature superconductors}

Approximately 100 different cuprate materials, many of which are
superconducting, have been discovered since 1986.
Cuprates have
layered structure, with quite similar $a$- and $b$-axes and considerably
longer $c$-axis. They consist of CuO$_2$ planes, normal to $c$-axis, separated
by layers containing other types of atoms. The superconductivity occurs in
the CuO$_2$ layers and it is sensitive to the oxygen content.
The layered structure gives rise to strong anisotropy of many physical
quantities.

\end{multicols}

\begin{center}
TABLE 1. Ceramic superconductors showing the PME

\vspace{0.2cm}

\begin{tabular}{llllr}\hline
          &  & $T_c$ & $J_c $ & grain size \\
compound & source & (K) & (A/cm$^2$) & ($\mu$ m)\\ \hline
Bi$_2$Sr$_2$CaCu$_2$O$_{8.182}$ & \cite{Braunisch93}  & 84 & 10$^5$ - 10$^6$
\cite{Freitag99} & 2 - 3 \\
Bi$_{1.73}$Pb$_{0.27}$Sr$_2$Ca$_2$Cu$_3$O$_y$ & \cite{Braunisch93} & 110 &
10$^5$ - 10$^6$ &  4 \\
YBa$_2$Cu$_3$O$_{7-\delta}$ & \cite{Riedling94} &  89 & --- &  --- \\
Nd$_{2-x}$Ce$_x$CuO$_y$ & \cite{Okram97} &  21 & -- & --  \\
La$_2$CuO$_{4+\delta}$ & \cite{Chou93} & 32 & 10$^4$ - 10$^5$ & ---\\
Ba-K-Bi-O & \cite{Golovashkin} & 32 & -- & -- \\ \hline

\end{tabular}
\end{center}
\begin{multicols}{2}

\begin{figure}
\epsfxsize=2.8in
\centerline{\epsffile{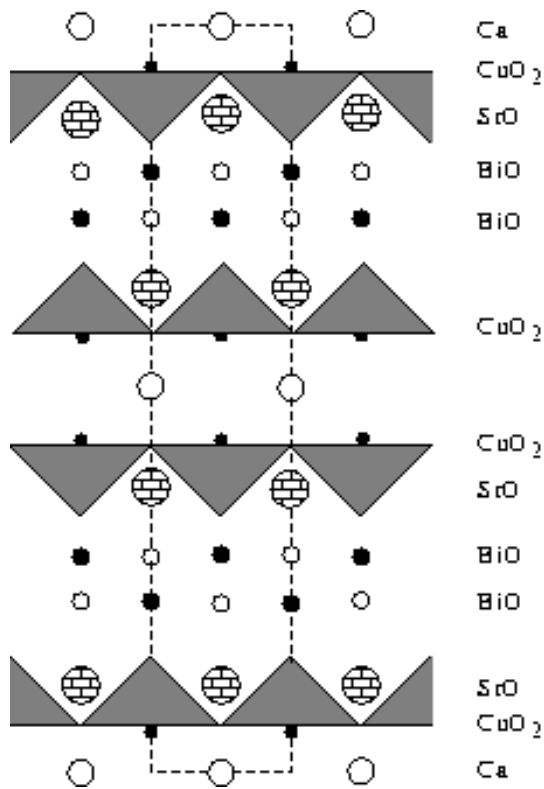}}
\vspace{0.2in}
\caption{Schematic structure of the crystal lattice 
Ba$_2$Sr$_2$CaCu$_2$O$_{8+x}$ where triangles represent pyramids with an 
oxygen  atom in each corner. The lattice constants
$a = 5.141 \AA, b = 5.418 \AA$ and $c = 30.66 \AA$. Adapted from
Ref. \protect\cite{Evie_thes}.}
\label{struc_fig}
\end{figure}

The PME was  first observed in the
Bi$_2$Sr$_2$CaCu$_2$O$_{8+\delta}$ (Bi2212)
the structure of which is schematically shown in Fig. \ref{struc_fig}. The lattice constants
are $a=5.141 \AA, b=5.148 \AA$ and $c=30.66 \AA$. The onset of 
superconductivity occurs at $T_c \approx 87$K and the transition width is about 5 K. In general,
 $T_c$ depends
on the oxygen content \cite{Braunisch93}.

% FIGURE 2

The Bi2212 samples are prepared by two techniques:
 the reaction sintering \cite{Svelindh89,Braunisch93,Niskanen93}
 and the melt-cast
\cite{Braunisch93}. As shown by Scanning Electron Microscopy,
sintered Bi2212 have a porous structure with a typical grain size of
1 - 4 $\mu$m \cite{Braunisch93} (see also Tab. 1) and typical thickness of
0.2 - 0.4 $\mu$m \cite{Evie_thes}. High resolution transmission electron
microscopy studies \cite{Freitag99} have shown that the samples obtained
by the melt-cast technique have an extremely polydomain microstructure on
a $\mu$m length scales. The grains are significantly larger and
 more densely packed compared to
the sintered samples. The single crystalline domains are preferably
$c$-axis oriented  with extremely sharp interfaces (width $\le 1$ nm)
which provide good contacts between the domain in $ab$-planes.
Such kind
of interfaces result in Josephson junctions with quite large critical
currents ($10^5 - 10^6$A/cm$^2$ \cite{Freitag99,Knauf96}). Depending on
the morphology of the samples the Bi-2212 may exhibit the PME or not.

\begin{figure}
\epsfxsize=3.6in
\centerline{\epsffile{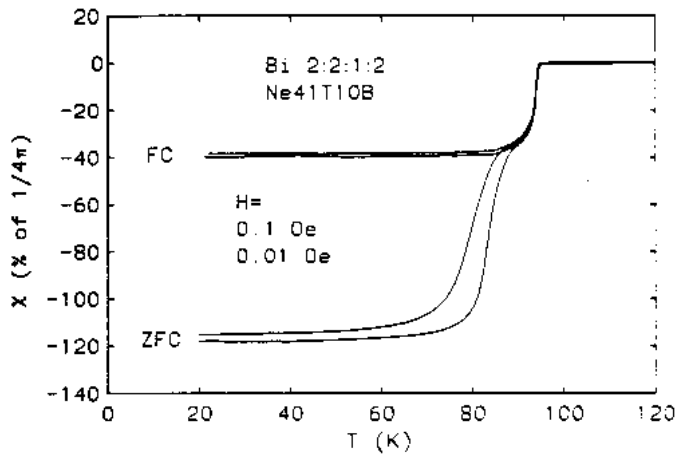}}
%\vspace{0.1in}
\caption{ZFC and FC susceptibilities as a function of temperature of
a high-quality ceramic Bi-2212
in very low fields. The reduced FC flux expulsion resembles the behavior
commonly observed at intermediate field.
After Braunish {\em et al.} \protect\cite{Braunisch93}.}
\label{nonpme_fig}
\end{figure}

Fig. \ref{nonpme_fig} shows the standard Meissner behavior of the dc 
susceptibility in 
granular
superconductor Bi$_2$Sr$_2$CaCu$_2$O$_8$ under FC and ZFC conditions. The 
results were obtained on a high-quality sintered sample 
\cite{Braunisch92,Braunisch93}. In the FC regime one has a fractional Meissner
effect due to a certain amount of magnetic flux trapped in the voids of 
the multiply connected sample. The screening in the ZFC mode is, however, 
larger than 100$\%$ of $-1/4\pi$. It is because the screening currents, which
are induced by applying the field below $T_c$ at the start of the measurement,
screen also voids present in those granular materials \cite{Braunisch93}.
Therefore the screened volume is larger than the effective volume with respect 
to the mass of the sample as calculated by the x-ray density.

% FIGURE 3

% FIGURE 4
\begin{figure}
\epsfxsize=3.6in
\centerline{\epsffile{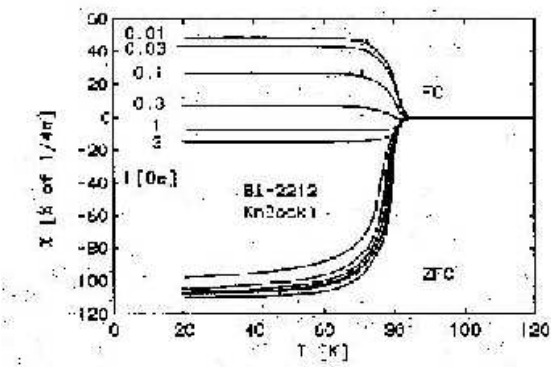}}
%\vspace{0.1in}
\caption{ZFC and FC susceptibilities as a function of temperature of
a granular superconductor Bi$_2$Sr$_2$CaCu$_2$O$_8$ exhibiting the PME
in very low fields. Reprinted from \protect\cite{Braunisch92}.}
\label{pme_fig}
\end{figure}

\begin{figure}
\epsfxsize=3.6in
\centerline{\epsffile{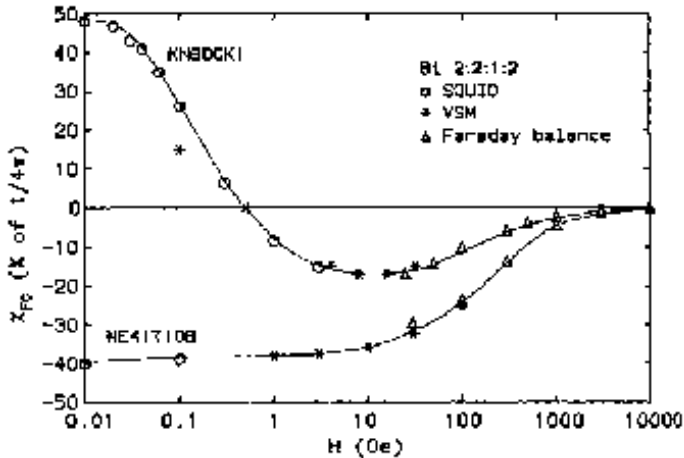}}
%\vspace{0.1in}
\caption{ \label{pme_H}
Field dependence of the FC susceptibility of the melt-process sample KnBock1 and
of the sintered sample Ne41T10B at 20K. The applied field was varied between
10 mOe and 10 kOe. The measurements were performed with the SQUID
magnetometer, the vibrating sample magnetometer (VSM), and a Faraday balance
(see \protect\cite{Braunisch93} for details). The PME of KnBock1 sample emerges in fields
$H < 2$ Oe. Reprinted from \protect\cite{Braunisch93} .}
\end{figure}

The unusual behavior of the dc susceptibility in Bi-2212 compound obtained by
the melt-cast technique is shown in Fig. \ref{pme_fig}. 
For very small external fields
($H \sim 0.01 - 1 $ Oe) a paramagnetic response in the FC conditions apperears
below the superconducting transition temperature. The smaller the $H$, the
stronger the paramagnetic signal, as seen from Fig. \ref{pme_H}.

% FIGURE 5

For fields $H \geq 1$ Oe, $\chi$ becomes
slightly diamagnetic. The fact that even small fields are sufficient to enforce
nearly complete polarization reveals that orbital currents spontaneously
created in the superconducting state should interact weakly. 
It was shown that the magnetization consists of two parts: the conventional
diamagnetic part and the paramagnetic one \cite{Braunisch93,Kusmartsev92a}
\begin{eqnarray}
M(H) \; \; & =& \; \; M_0(H) \; + \; \chi_{dia} H  \; , \\ \nonumber
M_0(H) \; \; &=& \; \;  \frac{M_s H}{(H_0 +H)^{\alpha}} \; , \\ \nonumber
\chi \; \; &=& \; \; \chi_{dia} \; + \; \frac{M_s}{(H_0 +H)^{\alpha}} \; . 
\end{eqnarray}
Here $\chi_{dia}$, $M_s, \alpha$ and $H_0$ depend on the samples ($H_0$ may be
interpreted as the field below which the thermal and interaction effects
suppress complete polarization). For Bi-2212, e.g.,  
$\chi_{dia}$ is about $17\%$ of $-1/4\pi, 
\alpha \approx 1, H_0=0.16$ Oe and $M_s$=0.095 G/4$\pi$.

As to the ZFC experiments, at a first glance one may think that the Meissner
effect is similar to that in ordinary ceramic superconductors. Namely,
$\chi_{ZFC}$ depends on $T$ but not on the external field. However, a more 
careful
inspection \cite{Svelindh89,Niskanen93} shows its rather weak 
field dependence.   
This nonlinear effect gets enhanced near the phase transition to the 
superconducting state.

\begin{figure}
\epsfxsize=3.6in
\centerline{\epsffile{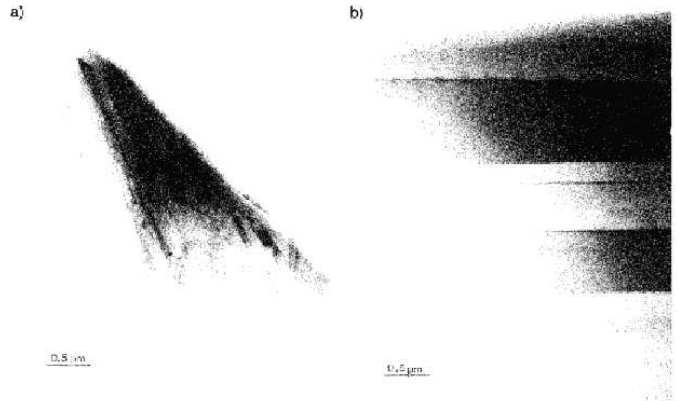}}
%\vspace{0.1in}
\caption{ \label{morpho}
High-Resolution Transmission Electron
Microscopy photos of samples with PME (a) and without PME (b) on the scale
of several microns. After Freitag {\em et al.} \protect\cite{Freitag99}.}
\end{figure}

Recently Freitag {\em et al.} \cite{Freitag99} have found 
a clear qualitative distinction between the morphology of the samples with and
without the PME when approaching a length scale of several microns. It is
shown in Fig. \ref{morpho} where High-Resolution Transmission Electron 
Microscopy photos of two respective samples are shown.
The samples with PME are characterized by an extremely polycrystalline
"chaotic" structure with typical sizes of single crystalline domains 
$\le \mu$m (see Fig. \ref{morpho}a). In contrast, as seen in
Fig. \ref{morpho}b, samples without PME show a
pattern of large well-ordered single crystalline domains. 
The density of the domain boundaries is much larger for the former samples
than for the latter.

% FIGURE 6

The another class of materials where
the PME has been also observed is high-quality, twined single crystals
YBa$_2$Cu$_3$O$_{7-\delta}$ \cite{Riedling94,Lucht95}. 
The study of this material  
would help to discriminate the $d$-wave and the grain boundary origin
of $\pi$-junctions.
Since crystals are single the later mechanism 
can be ruled out. The PME has found to appear only with the 
magnetic field parallel
to the $c$ axis, i.e., the spontaneous currents are confined to the $ab$ plane
\cite{Riedling94}. It is clear from Fig. \ref{pme_YBa},
the magnitude of the observed signal here is considerably smaller than that
for the ceramic samples (less than 3$\%$ of the full shielding value as 
compared to up to 50$\%$) \cite{Braunisch92,Braunisch93,Heinzel}.
The field below which a PME is observed is, however, much higher than that in 
the ceramic samples (0.7mT compared to about 0.05 mT).
Since the sample is a single crystal one has a complete flux expulsion in
the ZFC regime.

A positive FC magnetization has also been
 observed in Nd$_{2-x}$Ce$_x$CuO$_y$ \cite{Okram97}, La$_2$CuO$_{4+\delta}$
\cite{Chou93,Caparroz01}, and
Ba-K-Bi-O compounds \cite{Golovashkin}. In the later case
the magnitude of of $\chi_{FC}$ is much smaller than the one observed 
in BSCCO samples at weak fields but it remains positive for much larger
applied fields. 

Table 1 collects ceramic high-$T_c$ superconductors showing the PME.

\subsection{Conventional superconductors}

One of the scenarios for the PME in granular hole-doped cuprates discussed
above is that the FC paramagnetic response appears from current loops with
$\pi$ phase shifts of the superconducting order parameter at some 
grain-boundary junctions. It was argued that such behavior would be expected to
occur in a $d$-wave superconductor, but not in a conventional $s$-wave
superconductor. The test of this hypothesis led to the discovery of
the PME in conventional superconductors like Nb 
\cite{Minhaj94,Thompson,Kostic,Pust}, Al \cite{Geim}
and the Nb-AlO$_x$-Nb tunnel junctions
\cite{Moreira97,Barbara99}. The occurrence of the PME in these systems 
is probably
due to the confined geometry and flux trapping. 
In this section we discuss the main experimental findings for 
the conventional superconductors.

% FIGURE 7
\begin{figure}
\epsfxsize=2.8in
\centerline{\epsffile{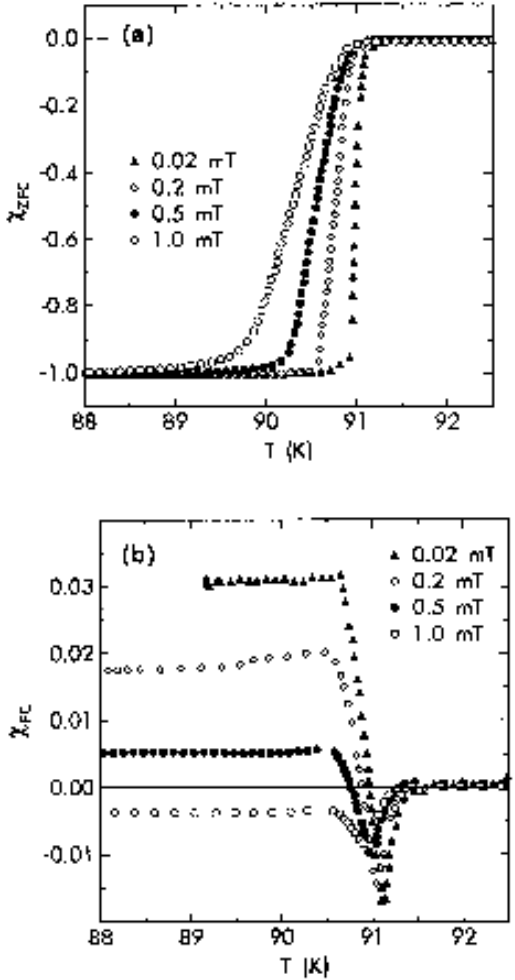}}
%\vspace{0.1in}
\caption{ \label{pme_YBa}
Susceptibility of single crystal YBa$_2$Cu$_3$O$_{7-\delta}$ in various
applied fields vs temperature $T$. From Riedling {\em et al.}
\protect\cite{Riedling94}.}
\end{figure}

\begin{figure}
\epsfxsize=3.6in
\centerline{\epsffile{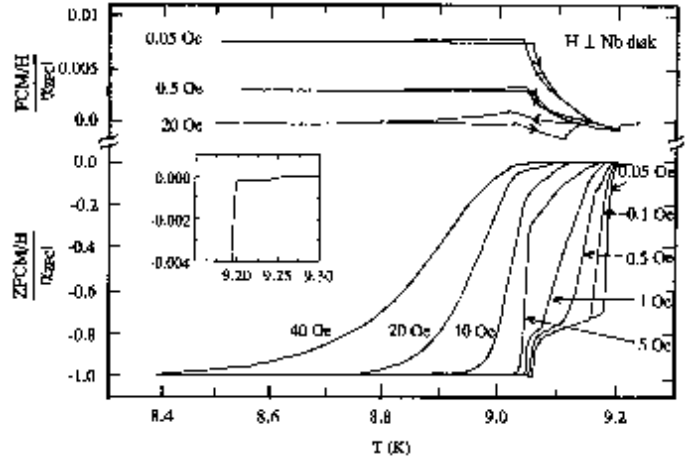}}
%\vspace{0.1in}
\caption{ \label{pme_Nb}
The FC (upper) and ZFC (lower) susceptibility for a macroscopic Nb disk
with magnetic fields applied normal to the disk surface. The data are
scaled to the complete shielding value of $\chi_{ZFC}$. The inset shows an
enlargement of $\chi_{ZFC}$ around $T=9,26$ K. From Thompson {\em et al.}
\protect\cite{Thompson}}
\end{figure}

In the experiments of Minhaj {\em et al.} \cite{Minhaj94,Thompson}
Niobium disks of diameter 6.6 mm and
thickness 0.127 mm were used to study the PME. 
The dc FC magnetization shows the paramagnetic response below the 
superconducting transition temperature $T_c \approx 9,2$ K when the applied
magnetic  field is normal to the disk. It is demonstrated in Fig. \ref{pme_Nb}.
The paramagnetic signal is much weaker compared to its counterpart in
BiSrCaCuO compounds and it becomes diamagnetic for fields larger than
20 Oe.

% FIGURE 8

Comparing Figs. \ref{pme_YBa} and \ref{pme_Nb} one can see the remarkable
similarity in the temperature dependence of the susceptibility of Nb disks
and single crystal YBa$_2$Cu$_3$O$_{7-\delta}$.
With increasing temperatures the ZFC data indicate the presence of two 
different superconducting transitions at $T_1 \approx 9.2 $K and 
$T_p \approx 9.06 $K.
The strong field dependence of $\chi_{ZFC}$
indicates that the local field is larger than
the lower critical field $H_{c1}(T)$ in the interval $T_p < T <T_1$.
The appearance of the two transition temperatures is also seen in the
FC measurements: 
at $T_1$
the paramagnetic moment first appears (vanishes) and a lower temperature 
$T_p$ defines the temperature where the positive
 moment no longer increases
\cite{Thompson,Wenger00}. 
The sharp increase of positive FC magnetization
upon cooling below $T_1$ is fairly spontaneous similar to the onset of
global diamagnetic screening currents at $T_c$ rather than the viscous 
nature exemplified by flux flow. 
Below $T_p$ this moment apparently does not change with temperature  and
can be regarded as an additive constant to the FC magnetic moment of a non-PME
superconductor.

\begin{figure}
\epsfxsize=3.6in
\centerline{\epsffile{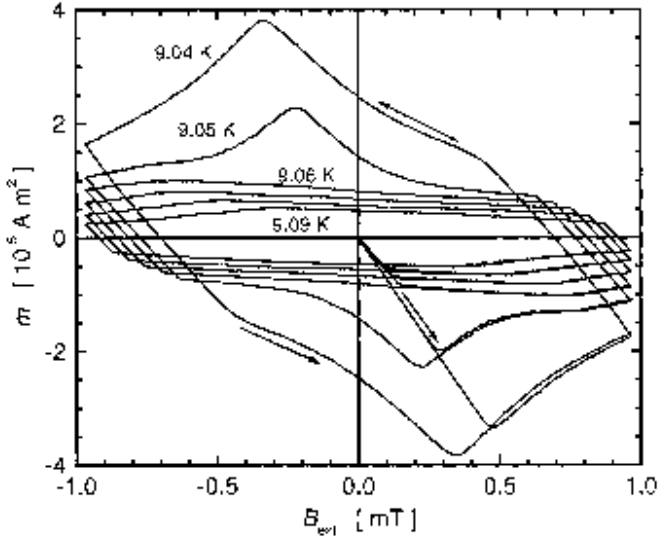}}
\vspace{0.2in}
\caption{ \label{hyster_Nb}
Magnetic hysteresis loops of the Nb disks in the temperature range
$9.04 \le T \le 9,09$ K for 0.01 K increments. The remarkable change of
their shapes between 9.05 and 9.06 K, i.e., at $T=T_p$.
From Pust {\em et al.} \protect\cite{Pust}.}\end{figure}

To gain more insight on the nature of the FC paramagnetic moment
for the Nb disks, magnetic hysteresis loops were measured 
\cite{Thompson,Pust}. A set of such loops recordered in the temperature range 
from 9.04 to 9.09 K, i.e., around $T_p$ are shown in Fig. \ref{hyster_Nb}.
Between $T_c$ and $T_1$ the system is superconducting but only a small
diamagnetic screening current can be induced by ramping the external field.
Below $T_1$ the critical current is expected to increases significantly and
the loops exhibit a strange, nearly parallelogramlike shape the size of
which increases with lowering temperature \cite{Pust}. Just below $T_p$ the
hysteresis loops change their shapes drastically and they are more reminiscent
of a type-II superconductor with the magnetization becoming less diamagnetic as
the flux penetrates into the bulk for $H \ge H_{c1}$.

% FIGURE 9

Interesting similarity to the YBa$_2$Cu$_3$O$_{7-x}$ single crystals 
\cite{Riedling94} results in the hysteresis behavior of the FC magnetization
between cooling and warming cycles. As one can see from Fig. \ref{pme_Nb}
the FC magnetizations are more positive than its warming counter part.
On cooling, the flux should overcome activation-type process in order to 
be expelled and consequently the FC magnetization is more positive
than the equilibrium one. In the warming cycle, the flux cannot, however, 
easily penetrate the sample and the magnetization is more diamagnetic.
This behavior is common in type-II superconductors and is consistent with
a theoretical interpretation based on the critical-state model \cite{Clem93}.

\begin{figure}
\epsfxsize=3.6in
\centerline{\epsffile{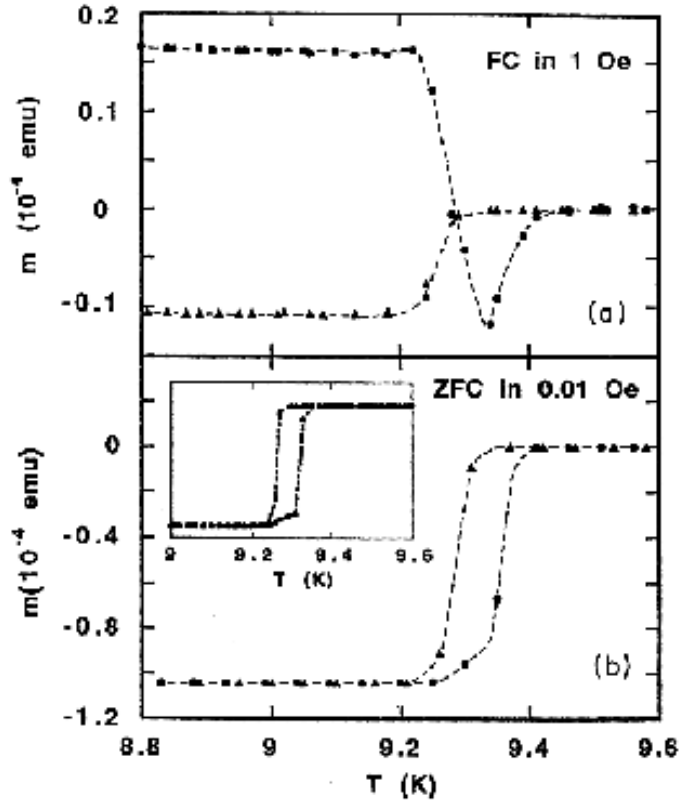}}
\vspace{0.2in}
\caption{ \label{surface_Nb}
FC (a) and ZFC (b) magnetizations of a Nb disk measured before
(solid squares)  and after (solid triangles)
polishing the sample surface. The values of the external magnetic field are
shown next to the curves.
Inset: a high point density ZFC measurement at 0.05 Oe from another sample
cut. The PME disappears after both faces of the sample were polished.
From Kostic {\em et al.} \protect\cite{Kostic}.}
\end{figure}

It has been found that the PME in Nb samples is very sensitive to their
surface and geometry \cite{Thompson,Kostic,Pust}. In particular,
the paramagnetic response occurs in Nb disks cut from some sheets of
rolled Nb. Disks cut from other source materials do not show the PME
\cite{Kostic}. Fig. \ref{surface_Nb} shows the FC magnetizations obtained
before and after polishing the sample faces. Interestingly, the FC
 paramagnetic signal disappears after both faces of the sample were polished.
This gives strong evidence that the surface pinning of the magnetic flux plays
an important role. One may think that the PME is likely to arise from 
inhomogeneously trapped flux, and is unlikely to have any relationship with
$d$-wave superconductivity.

% FIGURE 10

We now discuss the PME in small (micrometre-size) superconductors. The 
experiments were carried out for Nb and Al discs \cite{Geim} with diameters
from 0.3 to 3 $\mu$m and thicknesses from 0.03 to 0.15 $\mu$m. The thickness
is comparable to the characteristic
penetration depth $\lambda \sim 500 \AA$ \cite{Terentiev99}.
Study of small samples is interesting in two ways. First,
an application of the Koshelev-Larkin model \cite{Larkin} to data for Nb discs
\cite{Kostic} suggested that the paramagnetic magnetization must be small
for samples much thicker than $\lambda$; however, for very thin films one
can expect a much larger effect due to macroscopic penetration of Meissner
currents into the sample interior.
Second, confinement of superconductivity in a small volume leads to pronounced
quantization , so that a mesoscopic superconductor resides in one of 
well-resolved states, depending on temperature and magnetic field
(superconducting states are characterized by a different number and
distribution of vortices \cite{Geim97,Schweigert98,Schweigert98a,Bolech95}).
This allows one to measure magnetization of individual vortex states and
rules out the pinning effect.

% FIGURE 11
\begin{figure}
\epsfxsize=3.6in
\centerline{\epsffile{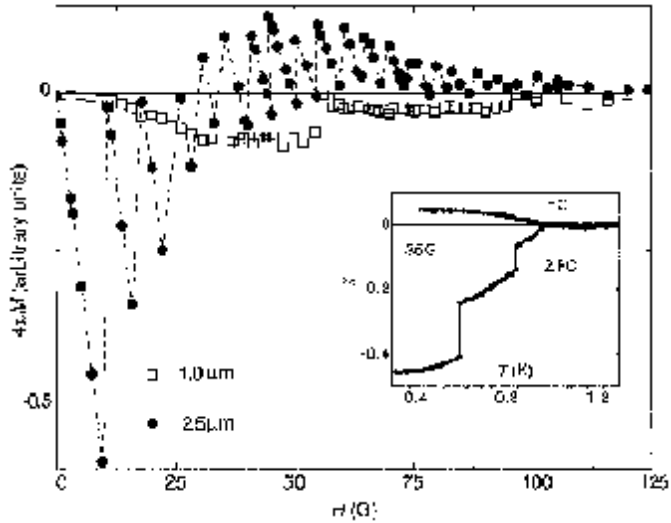}}
\vspace{0.2in}
\caption{ \label{pme_meso}
The field dependence of the Meissner response for Al disks of diameter 1.0
(open squares) and
2.5 $\mu$mi (closed circles) at T=0.4 K. The dashed line is a guide to the eye. The inset
compares FC and ZFC magnetization of the 2.5 $\mu$m disk at the field where
the paramagnetic response is close to its maximum. The jumps in the ZFC
curve correspond to entry of individual vortices into the disk interior.
After Geim {\em et al.} \protect\cite{Geim}.}
\end{figure}

The typical field dependence of the FC paramagnetic moment of mesoscopic 
samples is shown in Fig. \ref{pme_meso} \cite{Geim}. 
The strong oscillation clearly seen
for the larger sample of diameter 2.5 $\mu$m is due to size quantization. 
Each jump corresponds to a change in the number of vortices inside the disk,
which can either form an array of single quantum vortices or assemble 
into a single giant vortex 
\cite{Moshchalkov,Geim97,Schweigert98,Schweigert98a,Bolech95}.
The latter configuration is expected at fields between the second and third
critical field, $ H_{c2} < H < H_{c3}$, where $H_{c3}$ is related to the surface
superconductivity \cite{Moshchalkov}. The smaller sample (diameter 1.0 $\mu$m)
does not show the paramagnetic response over the entire field interval.
This qualitatively different behavior may be explained by the fact that,
in smaller sample the superconductivity is suppressed by  $\sim 3$ flux quanta
entering the whole disk area while $\sim 20 \Phi_0$
($\Phi_0$ is the flux quantum) are needed to
spoil superconductivity of the larger disk 
\cite{Geim,Geim97,Schweigert98,Schweigert98a}.
Thus, the diamagnetic response is always observed in low magnetic fields and 
the PME occurs only in intermediate fields allowing at least several flux
quanta penetrate into the interior. This seems to contradict the previous 
studies on macroscopic samples, where the paramagnetic signal is recorded
in very low fields and gradually disappeared with increasing field. Such
discrepancy is due to the large number of flux quanta (many thousands) may
enter the macroscopic samples even in lowest fields.

% FIGURE 12
\begin{figure}
\epsfxsize=3.6in
\centerline{\epsffile{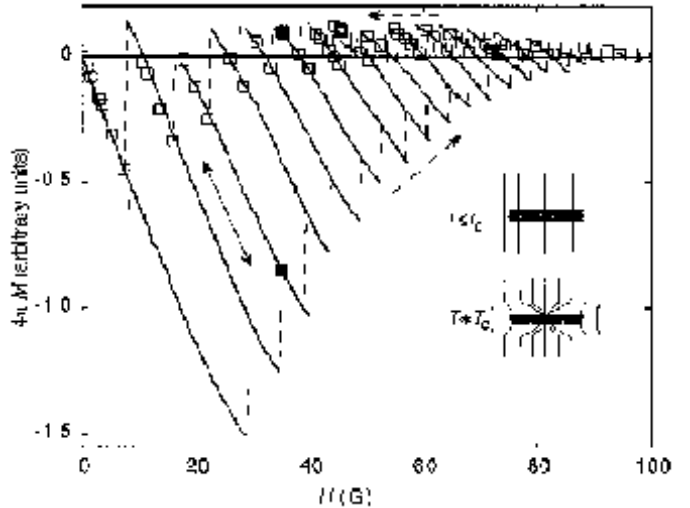}}
%\vspace{0.1in}
\caption{ \label{pme_geim_ct}
The magnetization measured by cooling in a field and by sweeping the field
at a constant temperature (0.4 K). Arrow shows the direction of the sweep.
The FC data shown by open squares are for the 2.5 $\mu$m disk of Fig.
\ref{pme_meso}. The solid curves were measured by pausing at various fields
and then sweeping the field up and down.
When the field swept continuously, the magnetization evolves along one
of the solid curves until it reaches the end of this curve
and jumps to the next one belonging to another vortex state. Then, the process
repeats itself all over again as shown by dashed lines. The closed squares
at 35 G indicate the low-temperature states for the FC and ZFC curves of
Fig. \ref{pme_meso}. The inset illustrates the compression of a giant
vortex into a smaller volume which allows extra flux to enter the sample at the
surface.
From Geim {\em et al.} \protect\cite{Geim}.}
\end{figure}

The nature of the PME becomes more evident from measurements of the 
magnetization by sweeping the magnetic field at a constant temperature
\cite{Geim} (see Fig. \ref{pme_geim_ct}).
Instead of a single magnetization curve characteristic for macroscopic samples
one has a family of curves corresponding to different vortex states.
Since several superconducting states are possible at the same magnetic field,
the states in which the PME occurs are metastable. Only the state with
the most negative magnetization is thermodynamically stable 
\cite{Schweigert98,Schweigert98a}. The metastability results from the sample
surface which favours  a large superconductivity and opposes both vortex escape
and entrance \cite{Palacios00}. As shown below, 
combining the ideas of flux compression \cite{Larkin}
and the trap in giant vortex states \cite{Moshchalkov} one can explain 
the experimental findings for
mesoscopic superconductors \cite{Geim}.

\section{Mechanisms for $\pi$ junction}

\subsection{Scattering on paramagnetic impurities}

The effect of magnetic impurities present in the barrier between 
$s$-wave superconductors on the tunneling current was first studied by
Kulik \cite{Kulik65} (see also \onlinecite{Shiba69}). 
Later Bulaevskii {\em et al.} \cite{Bulaevskii}
have realized  that the presence of such impurities
may lead to a spin-flip 
intermediate state which actually shifts the phase by $\pi$.

Consider a SIS contact with magnetic impurities in the thin insulator layer.
The Hamiltonian of this system is as follows \cite{Kulik65,Bulaevskii}
\begin{eqnarray} 
H_T \; \; = \; \; \sum_{\vec{k},\vec{k}',\vec{n},s,s'} \; ( t_{\vec{k}\vec{k}'}
\delta_{ss'} 
 + \, v_{\vec{k}\vec{k}'\vec{n}}
\vec{\sigma}_{ss'} S_{\vec{n}} ) a^+_{\vec{k}s} b_{\vec{k}'s'}
\; \; \nonumber\\
+ \; \; c.c \; . 
\label{Bulaev1}
\end{eqnarray} 
Here $a_{\vec{k}s} (b_{\vec{k}s})$ are electron annihilation operators in
superconducting layer A (B); $S_{\vec{n}}$ is spin of impurities localized
at point $\vec{n}$ and $\vec{\sigma}$ are Pauli matrices.

Using the Green function technique \cite{Abrikosov}, in the second order
of perturbation theory we obtain the stationary Josephson current $J$ and the
contact energy $E$ in the following form

\begin{eqnarray}
J \; &=& \; J_c\sin \phi \, , \; 
E \; = \; -\frac{\hbar}{2e}J_c\cos \phi , \; J_c = J_0 - J_s \nonumber\\
J_0 \; &=& \; 2\pi^2 e  t^2 N^2(0) \Delta (T) 
\textrm{th}\frac{\Delta (T)}{2T} \, , \nonumber\\
J_s \; &=& \; 2\pi^2 e \sum_{\vec{n}} \, v^2_{\vec{n}}
S(S+1) N^2(0) \Delta (T) \textrm{th}\frac{\Delta (T)}{2T} \; ,
\label{Bulaev2}
\end{eqnarray}
where $t^2$ and $v^2_{\vec{n}}$ are average values of 
$t^2_{\vec{k}\vec{k}'}$ and $v^2_{\vec{k}\vec{k}'\vec{n}}$ on the Fermi surface
respectively, $N(0)$ - the density of states on the Fermi surface,
2$\Delta(T)$ - the energy gap and $\phi$ - the difference between 
superconducting phases. $J_0$ is the standard Ambegaokar-Baratoff
term and $J_s$ describes the electron tunneling
with the spin flip \cite{Kulik65}.

Since $v_{\vec{k}\vec{k}'\vec{n}} \sim [\epsilon_d (\epsilon_d + U)]^{-1}$
\cite{Anderson66}, where $\epsilon_d$ - energy of one electron and
$U$ - the energy of the Coulomb repulsion between two electrons, one can
choose $\epsilon_d$ so that $J_c$ becomes negative ($J_s > J_0$)
\cite{Bulaevskii}.
The another possibility to have $J_c < 0$ is that,
as follows from Eq. (\ref{Bulaev2}),
the concentration of
magnetic impurities should be high enough making the spin flip term
to be dominant.

Rewriting the Josephson current as
\begin{eqnarray}
J \; \; = \; \; \left\{ \begin{array}{ll}
J_c \sin (\phi + 0) & \mbox{if $J_c > 0$}\\
|J_c| \sin (\phi + \pi) & \mbox{if  $J_c <0$}
\end{array}
\right. 
\label{Bulaev3}
\end{eqnarray}
one can understand the meaning of terminology {\em $\pi$-junction}.
Namely, in the case when the critical current $J_c < 0$ the current (or the contact energy)
has the same form as in the conventional case
with $J_c > 0$ \cite{Sigrist95} but the phase is shifted by $\pi$.  
For this reason an junction with negative critical current is called $\pi$ 
junction whereas $J_c > 0$ corresponds to the standard 0 junction. 

An another possibility of occurrence of negative critical currents
was also pointed out by Spivak and Kivelson \cite{Spivak91}.
The main difference from the Bulaevskii approach \cite{Bulaevskii}
is that they described an impurity  by the Anderson model in a regime where 
the localized magnetic moments exist. Spivak and Kivelson argued that
$J_c$ may become negative in a dirty system due to an interplay of disorder and 
electron correlations. Similar results were obtained earlier by
Altshuler {\em et al} \cite{Altshuler83}.

The existence of $\pi$-junctions leads to non-trivial physical phenomena.
In the next section we will show that the spontaneous current in
 the ground state of a frustrated loop 
of odd number of $\pi$ junctions is nonzero. 
This can serve as
a reason for the paramagnetic response to the external magnetic field.

\subsection{$d$-wave picture of a $\pi$-junction}

In all cases discussed above the appearance of the $\pi$ junction depends
not on the intrinsic properties of a superconductor, but rather on external 
conditions like degree of disorder, impurity concentration etc.
Furthermore, the spin flip mechanism, for example, may require the high 
density of
magnetic impurities causing the strong magnetic interaction which would 
invalidate elastic tunneling.

There is, however, another, intrinsic and more exciting scenario 
for $\pi$-junctions: the spontaneous moments
may naturally occur in ceramic superconductors with
the $d$-wave pairing symmetry \cite{Sigrist92}. This idea was inspired by the
the works of Geshkenbein and co-workers \cite{Geshkenbein86,Geshkenbein87}
on heavy-fermion superconductivity showing that special arrangements
of junctions may lead to the $\pi$ shift.
In this chapter,
using the Ginzburg-Landay theory for the Josephson junction
with  two connected $d$-wave superconductors we show that 
the critical current may be positive or negative.

It is well established that the phonon mediated electron-electron interaction
gives rise to spin-singlet pairing with $s$-wave symmetry \cite{Bardeen57}
in conventional low-$T_c$ superconductors. The question of symmetry of 
the order parameter of high-$T_c$ superconductors is under intense debate
\cite{Annett96,Klemm00}.
Although the mechanism of high-$T_c$ superconductivity remains unknown
experiments on Shapiro steps \cite{Esteve87}, the magnetic-flux states of YBCO 
\cite{Gough87}, the Andreev-reflection \cite{Hoevers88} and spin
susceptibilities \cite{Takigawa89,Barrett90} give strong evidence for
spin-singlet Cooper pairing in ceramic materials. We shall focus on this type
of pairing. 

Within the Ginzburg-Landau formalism
the normal metal - superconductor phase transition may be described by the
scalar complex order parameter $\Psi(\vec{r})$ which is the pair wave function. 
In momentum-space representation, $\Psi(\vec{k}) \propto <a_{\vec{k}\uparrow}
a_{-\vec{k}\downarrow}>$ and is related to the gap function $\Delta_{\vec{k}}$
through $\Psi(\vec{k})=\Delta_{\vec{k}}/2E_{\vec{k}}$, where $E_{\vec{k}}$
is the quasiparticle excitation energy.
The energy gap $\Delta_{\vec{k}}$ can serve
as a well defined superconducting order parameter or as a measure of long-range
phase coherence in the pair state. Its symmetry can be experimentally determined,
even without detailed knowledge about the microscopic origin of superconductivity
\cite{Tsuei00}.

Since the transition to the superconducting state is second order it
 should be accompanied with a continuous
symmetry breaking. The symmetry group $\cal H$ describing the pair
state must be a subgroup of the group $\cal G$ describing the normal state,
i.e. $\cal H \subset \cal G$ \cite{Landau79}. Considering pairing in
a crystal one has
 $\cal G$ = G$\times$ R$\times$ I$\times$ T$\times$ U(1), where G is 
the finite crystallographic point 
group, R the symmetry group of spin rotation, I the space inversion operation,
 T the time-reversal symmetry operation
and U(1) the
one-dimensional global gauge symmetry.
For BCS superconductors the gauge symmetry U(1) is broken in the superconducting
state due to the existence of off-diagonal long-range order. 
In an unconventional superconductor, in addition to U(1) one or more symmetries
may be broken at $T_c$.
Since spin-orbit interaction in ceramic superconductors is rather weak the
pair wave function should be either singlet (spin $S=0$, $\Psi(-\vec{k})=
\Psi(\vec{k})$) or triplet  ($S=1$, $\Psi(-\vec{k})=-\Psi(\vec{k})$).
The mixed state like $a \Psi^s + b \Psi ^t$ is forbidden.
Due to the spatial inversion the singlet pair wave function does not change
under the inversion operation ($I \Psi ^s = \Psi ^s$) whereas the triplet
one changes its sign ($I \Psi ^t = -\Psi ^t$).

Point group classification of pair states in cuprate superconductors has been
extensively studied \cite{Annett96,Sigrist87,Annett90,Annett91,Li93,Wenger93,Jha97}.
They are divided into two groups: tetragonal crystal lattice with point-group
symmetry $D_{4h}$ and orthorhombic crystal lattice with point-group symmetry 
$D_{2h}$. La$_{2-x}$Sr$_x$CuO$_4$, Tl$_2$Ba$_2$CaCu$_2$O$_8$, HgBa$_2$CaCuO$_4$ and some YBa$_2$Cu$_3$ compounds belong to the first group whereas
YBa$_2$Cu$_3$O$_{7-\delta}$ and Bi$_2$Sr$_2$CaCu$_2$O$_8$ possess 
$D_{2h}$ symmetry. All possible pair wave functions of $D_{4h}$ and
$D_{2h}$ groups are given in \cite{Annett91}.
It should be noted that superconductivity of cuprates basically originates
from the CuO$_2$ square/rectangular layers. Therefore we focus on the
symmetry of possible pair states on the square lattice which is characterized
by point-group symmetry $C_{4v}$.      
The list of spin-singlet even parity pair states for the
tetragonal point group $C_{4v}$ is given in Tab. 2. All representations
$A_{1g}, A_{2g}, B_{1g}$ and $B_{2g}$ are one-dimensional.

%\end{multicols}
\vspace{0.3cm}

\noindent TABLE 2. Even-parity pair states allowed by $C_{4v}$ symmetry

%\vspace{-0.3cm}

\begin{center}
\begin{tabular}{lllr}\hline
           Name & $\; \; \; $ State & $\; \; \; $ gap function $\Delta(\vec{k})$ & $\; \; \; $ Nodes \\ \hline
$s$-wave & $\; \; \; $ $^1A_{1g}$ & $\; \; \; $  1 & $\; \; \; $ none \\
$g$ & $\; \; \; $ $^1A_{2g}$ & $\; \; \; $ $k_xk_y(k_x^2 - k_y^2)$ & $\; \; \; $ line\\
$d_{x^2-y^2}$ & $\; \; \; $ $^1B_{1g}$ & $\; \; \; $ $k_x^2 - k_y^2$ & $\; \; \; $ line\\
$d_{xy}$ & $\; \; \; $ $^1B_{2g}$ & $\; \; \; $ $k_x k_y$ & $\; \; \; $ line\\ \hline
\end{tabular}
\end{center}

\vspace{0.3cm}

Since there is strong experimental evidence for $d_{x^2-y^2}$
symmetry \cite{Harlingen95,Tsuei00} in the high-temperature superconductors
(see below) we
will concentrate our study on this type of symmetry. Neglecting the possible 
orthorhombic
distortion the pair wave function is described by $B_{1g}$ representation
of $C_{4v}$ or $D_{4h}$ group. In momentum space the pair wave function
is

\begin{equation}
\Psi(\vec{k}) \; = \; <c_{\vec{k}\uparrow} c_{-\vec{k}\downarrow}>
\; = \; \cos k_x - \cos k_y \; .
\label{pairfunction}
\end{equation}

\begin{figure}
\epsfxsize=2.5in
\centerline{\epsffile{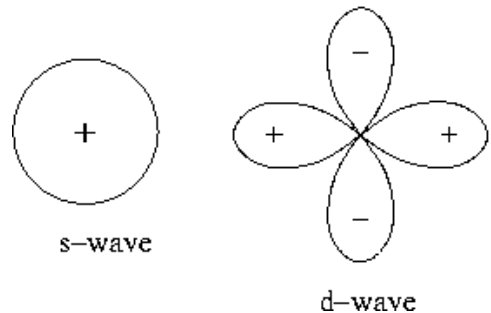}}
\vspace{0.2in}
\caption{Schematic description of $s$- and $d_{x^2-y^2}$-wave functions.
The order parameter may be positive (+) or negative (-).}
\label{sd_function}
\end{figure}

The schematic  description of the pair wave function of $s$ and $d_{x^2-y^2}$
symmetry is shown in Fig. (\ref{sd_function}).
The anisotropic function $\cos k_x - \cos k_y$ has the same symmetry property 
in $C_{4v}$ as $k_x^2 - k_y^2$. For this reason, this state is called the
$d$-wave with the angular momentum $L=2$, although the classification with
respect to angular momentum has no real meaning under the discrete 
crystal field symmetry. Since function (\ref{pairfunction}) is described
by one-dimensional representation, the time-reversal symmetry can not be
broken at $T_c$. Breaking of this symmetry is possible only for 
multi-dimensional representations \cite{Yip93}.

We now consider the phenomenological theory of the Josephson effect in the
case of a junction between two unconventional superconductors based on
symmetry arguments. So far as $d_{x^2-y^2}$-wave corresponds to one-dimensional
representation $B_{1g}$ their complex order parameters 
$\eta_1$ and $\eta_2$ should
be one-component

\begin{equation}
\eta_1 \; = \; |\eta_1| e^{i\phi_1} \, , 
\eta_2 \; = \; |\eta_2| e^{i\phi_2} \; .
\end{equation}
An intensive discussion of the experimental evidence for XY-universality class 
identifying the order
parameter as a complex scalar belonging to a one dimensional irreducible 
representation may be found in Ref. \cite{Schneider00}.

% FIGURE 13

\begin{figure}
\epsfxsize=2.5in
\centerline{\epsffile{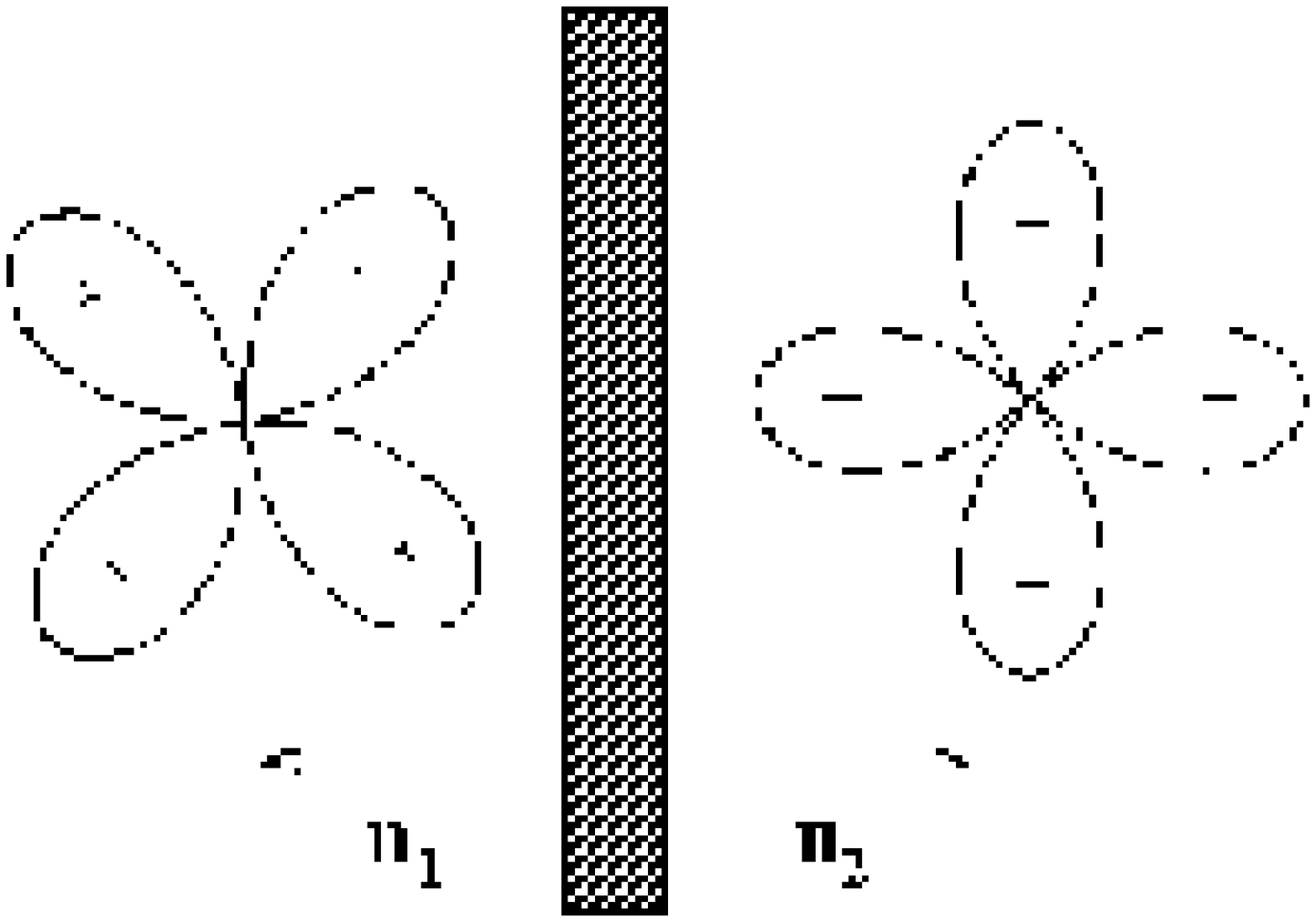}}
\vspace{0.2in}
\caption{Josephson junction between two $d_{x^2-y^2}$-wave superconductors.
Vectors $\vec{n}_1$ and $\vec{n}_2$ defined in the basis of 
local crystal axes are
normal to the interface.}
\label{loop12}
\end{figure}

The free energy of two subsystems (see Fig. \ref{loop12}) joined by a junction 
is determined by the expression \cite{Sigrist95,Mineev99}

\begin{eqnarray}
{\cal F} \; \; = \; \; {\cal F}_1 \, + \, {\cal F}_2 \, + 
\, {\cal F}_{12} \; , \nonumber \\
{\cal F}_i \; \; = \; \; \int_{\vec{r}\vec{n}_i>0} \, d\vec{r} \, F_i
\; \; (i=1,2) \; , \nonumber\\
F_i \; \; = \; \; \alpha_0\left( \frac{T-T_c}{T_c}\right) |\eta_i|^2 + 
\frac{\beta}{2} |\eta_i|^4 \, + \nonumber \\
\, K_{i1} ( |D_x \eta_i|^2 + |D_y \eta_i|^2) \,
+ \, K_{i2}  |D_z \eta_i|^2 + \frac{B^2}{8\pi} \; , \nonumber \\
{\cal F}_{12} \; \; = \; \oint \, dS \, t_0 g_1(\vec{n}_1) g_2(\vec{n}_2)
( \eta_1^* \eta_2 \, + \, \eta_1 \eta_2^* ) \; .
\label{free_ener}
\end{eqnarray}
Here $\alpha_0, \beta, K_{i,j} (i,j=1,2)$ and $t_0$ are phenomenological
parameters; outward normal to the junction surface vectors   
$\vec{n}_i$ ($\vec{n}_1=-\vec{n}_2$) are defined in the basis of
local crystal axes; 
$D_{x,y,z}$ are the components of the gauge-invariant 
gradient $\vec{D}=(\nabla - 2\pi i \vec{A}/\Phi_0)$, 
where $\vec{A}$ is the vector
potential ($\vec{B} = \nabla\times \vec{A}$).
${\cal F}_1$ and ${\cal F}_2$ are scalar under the operations of the complete
symmetry of the system ${\cal G}$ including the crystal symmetry $D_{4h}$
(or $C_{4v}$), time reversal, spin rotation and gauge symmetry
\cite{Sigrist91,Annett90,Annett91}. The interface term ${\cal F}_{12}$
was chosen in the lowest order and so that it generates the correct boundary 
conditions \cite{Geshkenbein86,Annett90,Yip90,Sigrist91}. Functions
$g_1(\vec{n}_1)$ and $g_2(\vec{n}_2)$ should be invariant under all operations
of the symmetry group of the respective half-space. This condition is
satisfied if we choose them to have the same symmetry properties as the 
corresponding order parameters. For the $d_{x^2-y^2}$ we have
\begin{eqnarray}
g_i(\vec{n}_i) \; = \; 
   n_{ix}^2 - n_{iy}^2 \sim 
\cos n_{ix} - \cos n_{iy} \; (i=1,2).
\label{g_orien}
\end{eqnarray}

Varying the gradient term of Eq. (\ref{free_ener}) with respect to $\vec{A}$, 
we obtain the current density 
in the left and right half-spaces

\end{multicols}
\begin{eqnarray}
\vec{j}_1 \; = \; - c \frac{\delta {\cal F}_{1,grad}}{\delta \vec{A}}
\; = \; - \frac{2\pi c}{\Phi_0} Re \eta_1^* [K_{11}(D_x + D_y) + K_{12} D_z]
\eta_1 \, , \nonumber\\
\vec{j}_2 \; = \; - c \frac{\delta {\cal F}_{2,grad}}{\delta \vec{A}}
\; = \; - \frac{2\pi c}{\Phi_0} Re \eta_2^* [K_{21}(D_x + D_y) + K_{22} D_z]
\eta_2 \, .
\label{current12}
\end{eqnarray}
\begin{multicols}{2}
The boundary conditions for the Ginzburg-Landau equations at the junction 
interface are obtained by varying the full free energy (\ref{free_ener})
 with respect to
$\eta_1^*$ and $\eta_2^*$
\end{multicols}
\begin{eqnarray}
i[ K_{11} (n_{1x} D_x + n_{1y} D_y) + K_{12} n_{1z} D_z ] \eta_1|_S \; \; =
\; \; -t_0 g_1(\vec{n}_1)g_2(\vec{n}_2) \eta_2|_S \; , \nonumber\\ 
i[ K_{21} (n_{2x} D_x + n_{2y} D_y) + K_{22} n_{2z} D_z ] \eta_2|_S \; \; =
\; \; -t_0 g_1(\vec{n}_1)g_2(\vec{n}_2) \eta_1|_S \; .
\label{boundary}
\end{eqnarray}
\begin{multicols}{2}

From Eqs. (\ref{current12}) and (\ref{boundary}) we have the current density 
though the Josephson junction

\begin{eqnarray}
\vec{j} \; \; = \; \; (\vec{j}_1 + \vec{j}_2)\vec{n}_1 |_S \; \; = \; \;
J_c \sin (\phi_1 - \phi_2) \, , \nonumber\\
J_c \; = \; \frac{4\pi c t_0}{\Phi_0} g_1(\vec{n}_1) g_2(\vec{n}_2) 
|\eta_1| |\eta_2| \; .
\label{current12a}
\end{eqnarray}
Taking Eqs. (\ref{g_orien}) and (\ref{current12a}) into account one can see
that the critical current $J_c 
\propto (n^2_{1x}-n^2_{2x})(n^2_{1y}-n^2_{2y})$ may be positive or negative depending
on orientation of vectors $\vec{n}_1$ and $\vec{n}_2$. A negative $J_c$
is equivalent to an intrinsic phase shift of $\pi$ in the junction.
Because $g_i(\vec{n}_i)$ has essentially the same symmetry as the order
parameters, the Josephson effect allows one to probe the phase of the pair
wave function. The existence of $\pi$ junctions may serve as an indicator
for the $d$-wave superconductivity. 

% FIGURE 14

\subsection{Experimental verification of $d$-wave pairing}

So far as the $d$-wave pairing may play a key role in understanding the PME
in ceramic superconductors,
in this part we discuss experiments which support its existence.
Those experiments are 
divided into two classes: phase insensitive
and phase sensitive ones \cite{Tsuei00}.

\noindent {\em Phase insensitive experiments.}
There are two kinds of
phase-insensitive experiments. One of them probes the gap features at the
Fermi surface. It includes, in particular, the angle-resolved \cite{Shen95},
tunnel \cite{Lesueur98}, Raman scattering \cite{Devereaux95}
 and point-contact \cite{Yanson91} spectroscopies.
The results obtained by these techniques for a number of specific hole-doped
oxides, are usually interpreted as manifestation of the $d$-wave pairing
although it is impossible in some cases to distinguish it from the anisotropic
$s$-wave symmetry \cite{Tsuei00}.  
Here we focus on the second group of phase-insensitive experiments
based on the fact that the low temperature behavior
of thermodynamic and transport properties 
of $d$-wave superconductors is governed 
not by  exponential as in the BCS theory but by power laws.

At low temperatures the specific heat of of BCS superconductors behaves as
$C_s(T \rightarrow 0) \sim T^{-3/2} \exp(-\Delta_0/T)$, where
$\Delta_0$ is the isotropic $s$-gap at $T=0$ \cite{Tinkham}. Simple calculations
\cite{Monien87} show, however,
that for the hexagonal  $d_{x^2-y^2}$
superconductors $C_s \sim T^2$ due to lines of nodes on the Fermi surface.
There is some controversy about experiments of different groups.
The $C_s \sim T^3$ was found in Bi-compounds \cite{Chakraborty89} whereas
the linear behavior $C_s \sim T$ was reported by other investigators
\cite{Reeves87,Feng88,vonMolnar88}. Recent experiments on single
crystals \cite{Moler94,Moler97} and ceramic samples \cite{Wright99} of YBCO
show, however, that $C_s \sim T^2$.

In the presence of the magnetic field the quadratic temperature dependence
rolls over to a $T$ \cite{Volovik93}, i.e. $C_s =k \gamma_n T \sqrt{H/H_{c2}}$
for $T/T_c \ll \sqrt{H/H_{c2}}$, where $k$ is a constant of order one and
$\gamma_n$ is the coefficient of the linear-$T$ term in normal state
\cite{Tinkham}. This linear-$T$ and $H^{1/2}$ dependence was confirmed by
experiments \cite{Moler97,Wright99,Reeves89}.
Furthermore Revaz {\em et al} \cite{Revaz98} found that the anisotropic
component of the field-dependent specific heat $C_s(T,H \| c)-C_s(T,H \perp c)$
of single crystal YBCO obeyed a scaling relation predicted for a 
superconductor with a line of nodes \cite{Volovik97,Simon97}.

In the low temperature limit the London
penetration depth $\lambda_L(T)$ of the conventional superconductors
approaches its zero value $\lambda_L(0)$ exponentially:
$\delta\lambda_L(T) = (\lambda_L(T)-\lambda_L(0))/\lambda_L(0) 
\sim T^{-1/2} \exp(-\Delta_0/T)$.
The theory
of $d_{x^2-y^2}$ superconductivity gives $\delta\lambda_L(T) \sim T$ for
pure materials and $\delta\lambda_L(T) \sim T^2$ for dirty ones 
\cite{Annett91a,Scalapino95}.
For nominally clean YBCO samples linear dependence was observed \cite{Hardy93}
whereas for Zn- and Ni-doped  as well as nonhomogeneous crystals the
quadratic behavior was reported 
\cite{Hardy93,Ma93,Lee94,Bonn94,Walter98,Kamal98,Karpinska00}. Although
these results agree with the $d$-wave symmetry of the gap function, they cannot
distinguish between the $d$- and $s$-wave because the linear dependence
of $\delta\lambda_L(T)$ could also arise from promixity effects between 
alternating $s$-wave superconducting and normal layers in the cuprates
\cite{Atkinson95,Klemm95}.

The another way to probe the symmetry
of the order parameter is to study the low-$T$ behavior of the thermal
conductivity of electrons $\kappa _e (T)$.
$\kappa _e (T) \sim T^{-1} \exp(-\Delta_0/T)$ for the BCS superconductors
but $\kappa _e (T) \sim T$ for the
$d_{x^2-y^2}$ materials \cite{Graf96,Durst00}. This linear
dependence was shown to be insensitive to the impurity scattering 
\cite{Durst00}.
Unfortunately, it is difficult to extract the thermal conductivity component 
 $\kappa _e (T)$ from experiments due to the complex mutual action of electrons,
phonons and impurities \cite{Graf96,Houssa97}. Nevertheless, the existing
experiments confirm the linear temperature dependence
in Zn-doped YBCO 
\cite{Uher92,Gold94,Taillefer97},
below $T^*= 200$mK in Bi$_2$Sr$_2$Ca(Cu$_{1-x}$Ni$_x$)$_2$O$_8$
\cite{Movshovich98}
and in BiSrCaCuO \cite{Chiao00}.

As to the ultrasonic 
attenuation coefficient $\alpha_s$ in superconducting state with anisotropic
pairing,
the theory \cite{Vekhter99} predicted 
$\alpha_s \sim T$ at low temperatures. Experiments performed for 
YBa$_2$Cu$_3$O$_{7-x}$ \cite{Bhattacharya88,Xu88} and 
La$_{1.8}$Sr$_{0.2}$CuO$_{4-x}$ showed that $\alpha_s \sim T^n$
with a large scatter of exponent $n$ for each substance.

Finally,  as $T \rightarrow 0$, the nuclear spin-lattice relaxation 
rate $T_1^{-1}$ of the $d_{x^2-y^2}$
superconductors was shown to scale with temperature as $T_1^{-1} \sim T^3$
\cite{Scalapino95}. Such a temperature dependence was confirmed by
experiments \cite{Scalapino95,Bulut92}

It should be noted that low temperature asymptotics of various thermodynamic
and transport properties of $s$-wave superconductors can roll over power-laws
\cite{Gabovich99,Gabovich02} provided the gap values are widely distributed
due to structure domains, charge stripes, charge-density waves or other
mesoscopic inhomogeneities.
So the interpretation of the $d$-wave symmetry based only on the low-$T$ 
asymptotics
remains ambiguous.

\noindent {\em Phase sensitive experiments}.
Recently, a number of phase-sensitive experimental techniques has been 
developed to determine the order parameter symmetry of cuprate superconductors
\cite{Harlingen95,Scalapino95,Annett96,Tsuei00} starting from the  
pioneering work of Wollman {\em et al} \cite{Wollman93}.
The key idea is relying on the sign changes in the Josephson critical current
$I_c$ as proposed first by Geshkenbein and Larkin \cite{Geshkenbein86} as a test
for axial $p$-wave pairing symmetry in the heavy fermion superconductors.

\begin{figure}
\epsfxsize=3.2in
\centerline{\epsffile{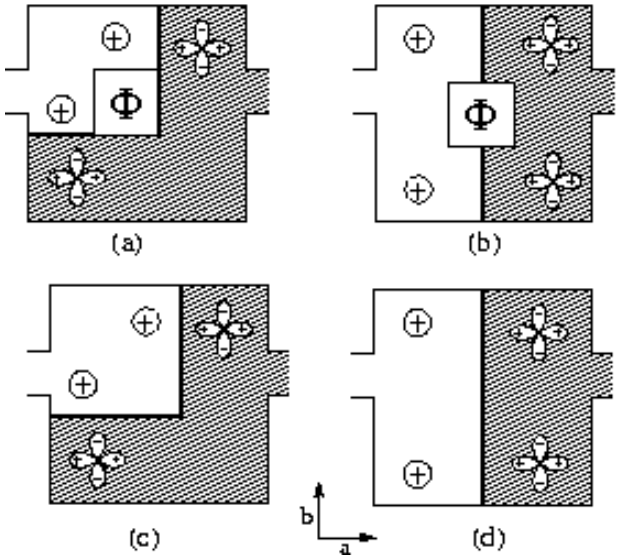}}
\vspace{0.2in}
\caption{Experimental geometry for the SQUID and single junction experiments
 (see Ref. \protect\cite{Wollman93,Wollman95}):
(a) Configuration of the corner SQUID interferometer experiment used to
determine the relative phase between orthogonal crystal directions a and b.
(b) The edge SQUID used as a control sample, in which both junctions are
on the same crystal face.
(c) The corner configuration for a single junction experiment.
(d) The same as in (c) but for the edge geometry.
}
\label{squid}
\end{figure}

In the "corner SQUID" design shown in Fig. (\ref{squid}a) Josephson junctions
are made between $s$-wave Pb thin films and two orthogonally oriented $ac$
and $bc$ plane faces of single crystal YBCO \cite{Wollman93}. If YBCO
is a $d$-wave superconductor, then, according to Eq. (\ref{current12a}),
there should be a $\pi$ phase shift between weak links on adjacent faces
of the crystal. For the "edge SQUID" geometry in Fig. (\ref{squid}b)
no phase shift is expected because both two junctions
are on the same crystal face.

We consider the dc "corner" SQUID with junction critical currents $I_a$ 
and $I_b$. Then a bias current through the SQUID is the sum of the currents
passed by junctions $a$ and $b$
\begin{equation}
I \; \; = \; \; I_{a} \sin \phi_a \, + \, I_{b} \sin \phi_b \; .
\label{squid_cur}
\end{equation}
On the other hand, the quantum phase coherence around the SQUID loop forces 
the constraint on the gauge-invariant phase differences across the junctions
on the $ca$ and $cb$ faces of crystal \cite{Tinkham,Wollman93}, namely,
\begin{equation}
\phi_a - \phi_b + 2\pi \left( \frac{L_a I_a}{\Phi_0} -
\frac{L_b I_b}{\Phi_0} + \frac{\Phi_{ext}}{\Phi_0} \right) + \delta_{ab} \; =
 \; 2\pi n
\; ,
\label{squid_phase}
\end{equation}
where $L_a$ and $L_b$ are the effective self-inductances of the two arms of the
ring, $n$ is integer,
 $\delta_{ab}$ accounts for the intrinsic phase shift inside the YBCO 
crystal between pairs tunneling into the crystal in the $a$ and $c$ directions.
One can expect that $\delta_{ab} = \pi$
and $\delta_{ab} = 0$ for the "corner" and "edge SQUID", respectively.

% FIGURE 15

In order to find the maximum current one can substitute $\phi_b$ from
Eq. (\ref{squid_phase}) into  Eq. (\ref{squid_cur}) and then find the maximum
of $I$ with respect to $\phi_a$. For a symmetric dc SQUID with equal junction 
critical currents $I_a = I_b = I_0$, in the limit of zero loop inductance,
simple algebra gives the following expression for the field dependence
of the maximum supercurrent 
\begin{equation}
I_c(\Phi_{ext}) \; \; = 2 I_0 \left| \cos \left( \pi\frac{\Phi_{ext}}{\Phi_0}
+ \frac{\delta_{ab}}{2} \right) \right| \; .
\label{squid_Imax}
\end{equation}

For finite values of self-inductances the modulation depth of the critical 
current is reduced because the circulating currents generate a flux
contribution in the ring \cite{Wollman93}. From the last equation it is clear
that if YBCO has $s$-wave symmetry then $\delta_{ab}=0$ and the circuit
will behave as an ordinary dc SQUID \cite{Tinkham}: $I_c$ has a maximum
at $\Phi_{ext}=0$. Such a  behavior should be also  exhibited by "edge SQUID"
regardless to the type of symmetry of pairing in the YBCO material.
In contrast, for $d$-wave symmetry the "corner SQUID" should have 
$\delta_{ab} = \pi$ and the critical supercurrent would display a minimum at
zero external flux. This important prediction  was confirmed by the
experiments of Wollman {\em et al} \cite{Wollman93}.

It should be noted that there are several complicating factors in 
interpretation of results obtained by the SQUID experiment of 
Wollman {\em et al} \cite{Harlingen95,Tsuei00}. Since YBCO has an orthorhombic 
crystal structure, it has a tendency to form a twin boundaries at which $a$
and $b$ lattice constants are interchanged. This would randomlize the 
phase that these experiments depend on. However, there are both experimental
\cite{Tsuei94,Mathai95}
and theoretical \cite{Harlingen95} showed that the order parameter maintains
its orientation across twin boundaries and forms a single domain even in 
twined samples. Moreover, experiments on detwined samples 
\cite{Brawner94,Harlingen95} gave consistent results indicating that 
the $d_{x^2-y^2}$ component of the order parameter has the same phase across
twin boundary.

Klemm concerned \cite{Klemm94}
 about the role of the corners in dc SQUID measurements
of the phase anisotropy of YBCO crystals \cite{Wollman93}.
Since flux trapping, demagnetization and field-focusing effects can be strongly
dependent on the sample geometry he argued that the $\pi$ phase shifts 
seen between the corner and edge SQUIDS could result simply from their
geometry differences, even for $s$-wave superconductors. This point
of view was criticized by Wollman {\em et al} \cite{Wollman94} 
showing that corners play no significant role.

An another concern is the flux trapping near the SQUID. This flux can be 
coupled to the SQUID loop and create a shift in the flux modulation pattern 
that is indistinguishable from the intrinsic phase shift. In order to avoid 
this effect one can cool the SQUID many times to get the lowest-energy
state, which will be one with no flux trap.  Magnetic imaging of the $ac$ and
$bc$ plane faces of cuprate superconductors \cite{Kirtley98,Moler98} shows
that there can be vortices trapped between the planes, even when they are
cooled in a very small field. Such trapped vortices could affect the
measured critical-current vs applied field characteristics of the SQUID's 
but not substantially.

\begin{figure}
\epsfxsize=3.2in
\centerline{\epsffile{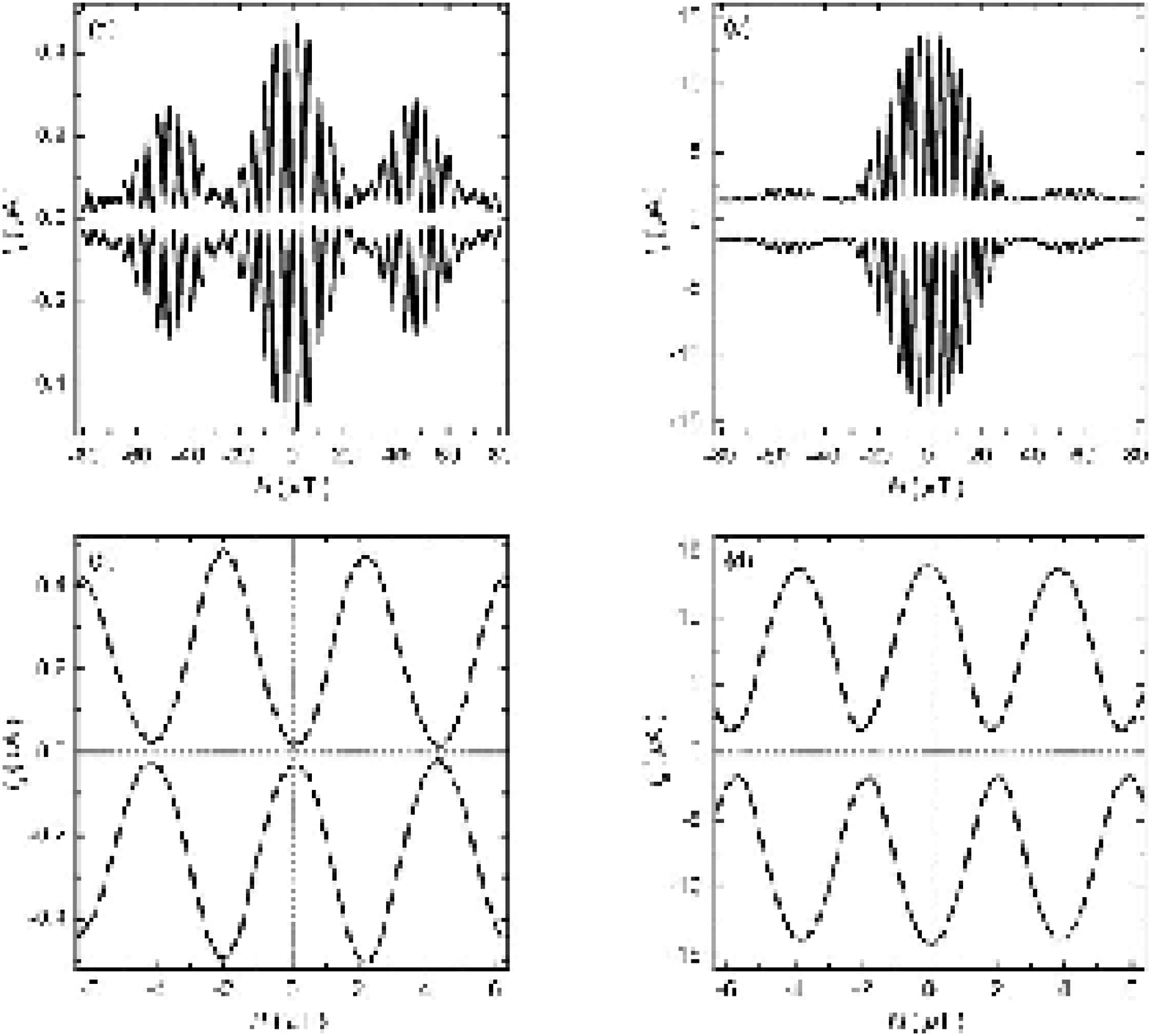}}
\vspace{0.2in}
\caption{Field dependence of the critical current of the $\pi$-SQUID (a) and
(b) and of the standard SQUID (c) and (d) at $T=77$.
The dip at zero field in the corner SQUID is evidence for 
$d_{x^2-y^2}$ pairing symmetry in the YBCO. After Schulz {\em at al.}
\protect\cite{Schulz00}. 
}
\label{squid_Icexp}
\end{figure}

Since Eq. (\ref{squid_Imax}) was obtained for a symmetric ($I_a = I_b$)
SQUID with zero self-inductance. Asymmetries between critical current in the
junctions and the non-zero loop inductance can modify the critical current
pattern significantly. Of particular importance here is a shift in the pattern,
since this could mimic or obscure phase shifts arising from the pairing
symmetry. Since the asymmetry-induced shifts of the $I_c(\Phi_{ext})$
pattern are proportional to the screening, it is vital to design SQUIDs
with negligible inductance. Such a problem was solved by Schulz {\em et al}
\cite{Schulz00} using YBCO thin films epitaxially grown on bicrystal and
tetracrystal substrates. They obtained, as one can see from 
Fig. (\ref{squid_Icexp}), nearly ideal dependences of the
critical current on applied field, with a minimum at $H=0$ for the $\pi$-ring
SQUID, as expected for a $d$-wave superconductor in the tetracrystal
geometry used. A more detailed discussion on the SQUID technique was given
in Ref. \cite{Tsuei00}.

% FIGURE 16

We now discuss single-junction modulation experiments \cite{Wollman95}
the geometry of which
is shown in Fig. \ref{squid}c and \ref{squid}d  for the corner
and the edge case, respectively. These experiments have a 
distinct advantage over the SQUID ones since they are less sensitive to
flux trapping and sample asymmetry. The basic idea is relied on the well 
known effect of an applied magnetic field on the critical current
of Josephson tunnel junctions \cite{Tinkham,Likharev}. The magnetic field
penetrating though the barrier region transverse to the tunneling direction
forces a gradient in the phase of the order parameter across the width 
of the junction, resulting in a variation of the local current density and
a reduction in the total critical current. If the junction is rectangular and
homogeneous with the field applied parallel to one edge of the rectangle, and
if the junction size is much smaller than the Josephson penetration depth, 
the critical current has the standard Fraunhofer 
 form \cite{Tinkham,Likharev}
\begin{equation}
I_c(\Phi) \; \; = \; \; I_0 \left| 
\frac{\sin(\pi\Phi/\Phi_0)}{\pi\Phi/\Phi_0} \right| 
\label{Fraunhofer_s}
\end{equation}
which is familiar from single-lit optical diffraction,
$\Phi$ is the flux threading the junction. The Fraunhofer pattern given by the 
last equation is valid for an edge junction with arbitrary pairing symmetry
and for a corner junction with $xisd$-wave superconductors. Similar
to the edge SQUID, this pattern has a maximum at $\Phi = 0$.

In the case of the corner junction, the order parameter in the $a$ and 
$b$ directions would be of opposite sign, modifying the diffraction
pattern. In a symmetric junction with equal geometries on the $a$ and $b$
faces, the critical current modulates according to \cite{Wollman95,Iguchi94}
\begin{equation}
I_c(\Phi) \; \; = \; \; I_0 \left|
\frac{\sin^2(\pi\Phi/2\Phi_0)}{\pi\Phi/2\Phi_0} \right| \; .  
\label{Fraunhofer_d}
\end{equation}
Contrary to the edge junction, $I_c$ vanishes at zero applied field as the
current through two orthogonal faces cancels exactly.

% FIGURE 17
\begin{figure}
\epsfxsize=3.2in
\centerline{\epsffile{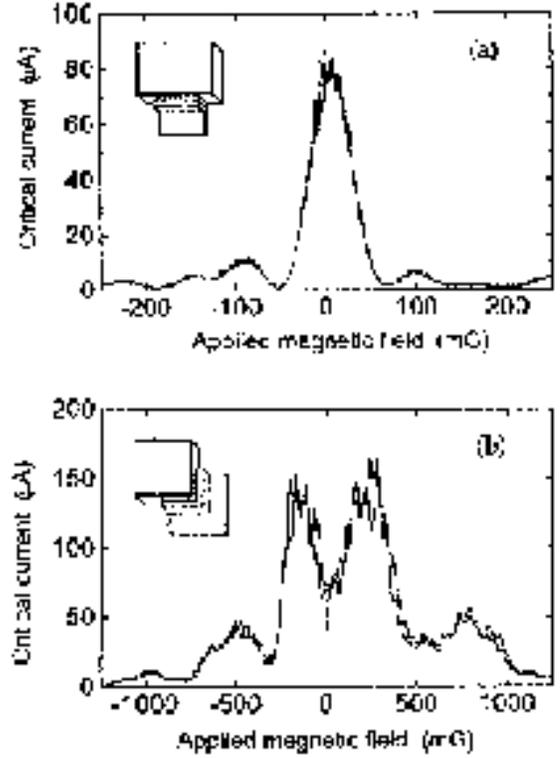}}
\vspace{0.2in}
\caption{Field dependence of the critical current of the edge junction (a) and
of the corner junction (b).
The minimum at zero field in the corner edge is evidence for 
$d_{x^2-y^2}$ pairing symmetry in the YBCO. After Wollman {\em at al.}
\protect\cite{Wollman95}.
}
\label{junction_Icexp}
\end{figure}

The field dependence the critical current of the edge and corner junctions
obtained by Wollman {\em et al} \cite{Wollman95}
is shown in Fig. (\ref{junction_Icexp}). 
Similar patterns were reported by different groups 
\cite{Iguchi94,Miller95,Brawner96}. The pattern shown in
Fig. (\ref{junction_Icexp}b) for $d$-wave corner junction does not agree closely
with the ideal expression (\ref{Fraunhofer_d}). It may be attributed,
as shown by theoretical predictions, 
to the flux trap in the sample \cite{Nappi98} or by the size of the junction
\cite{Kirtley97}. So the experiments on the modulation pattern of the critical
current in the single Josephson junction support the existence of
the $d$-wave pairing symmetry.

To summarize, the phase sensitive, along with phase insensitive experiments
give strong evidence in favour of predominantly $d$-wave pairing symmetry in
a number of cuprates. The identification of this type of symmetry is based
purely on very general principles of group theory and the macroscopic
quantum coherence phenomena of pair tunneling and flux quantization. 
It does not, therefore, necessarily specify a mechanism for high temperature
superconductivity.

\section{Paramagnetic Meissner effect in $d$-wave superconductors}

This section is focused on the $d$-wave mechanism of the PME in the framework
of the single-loop and multi-loop models.

\subsection{One-loop model}

As shown above, the phase across a $\pi$-junction between two $d$-wave
superconductors is shifted by $\pi$ and the corresponding energy
is proportional to $\cos(\phi_1 - \phi_2 + \pi)$. The system tries to minimize
its energy by setting the phase difference ($\phi_1 - \phi_2$) equal to
$\pi$. Thus, there is no way to measure this phase shift directly because
it merely corresponds to the phase change in one of the two superconductors,
say, $\phi _1 \rightarrow \phi _1 + \pi$. This transformation is equivalent,
as one can see from Eq. (\ref{g_orien}), to the exchange of the $x$ and $y$
coordinates in superconductor (1).
In other words, the connection of two $d$-wave superconductors by a single
$\pi$-junction does not, by itself, lead to any special observable effects.
In this sense, whether a junction is 0- or $\pi$-junction is just a matter
of convention.

% FIGURE 18
\begin{figure}
\epsfxsize=2.8in
\centerline{\epsffile{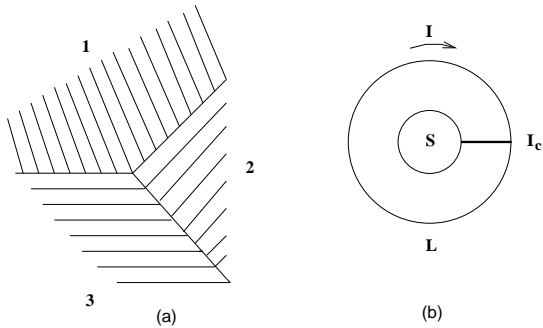}}
\vspace{0.2in}
\caption{(a) Schematic plot of a contact of three grains of superconductors with
unconventional pairing with different orientation of their local crystal
axis. These grains form a
frustrated loop with nonzero spontaneous current in the ground state.
(b) Single $\pi$-junction loop with the self-inductance $L$, area
$S$ and critical current $I_c$.}
\label{loop3_cartoon}
\end{figure}

Physically interesting consequences are, however, expected to arise if
an odd number of $\pi$-junctions are connected in a loop \cite{Sigrist95}.
A cartoon
of a contact from three junctions is shown in Fig. \ref{loop3_cartoon}a.
One can choose the coordinates $x$ and $y$ in two segments to convert
two $\pi$-junctions into two 0-junctions but there is no transformation
to remove the remaining $\pi$-junction without putting one of the two
0-junctions back to a $\pi$-junction. After all redefinitions of the crystal
axis the multi-connected loop has at least one $\pi$-junction. Since there is
no way to minimize the energy of all junctions and at the same time to keep
the phase of the order parameter constant in each segment, such a loop is,
in analogy with spin glass physics \cite{Binder},
frustrated . In
what follows any single loop is supposed to contain effectively
only one $\pi$ junction as shown in Fig. \ref{loop3_cartoon}b
or one 0 junction which is not frustrated.

Assuming that the loop has self-inductance $L$ and the current $I$ flowing
in it is small compared to the critical current $I_c > 0$, the energy is given
by the following form \cite{Sigrist92,Sigrist95}
\begin{equation}
F(\Phi,\Phi_{ext}) \; = \; \frac{1}{2L} (\Phi - \Phi^{ext})^2 -
\frac{I_c\Phi_0}{2\pi} \cos \left(2\pi\frac{\Phi}{\Phi_0} + \delta \right).
\label{E_singleloop}
\end{equation}
Here $\Phi^{ext}$ is the created by the external field
flux threading through the loop, $\Phi = LI$ and the phase shift
$\delta$ is equal to 0 and
$\pi$ for 0- and $\pi$-junction, respectively. The first term in
Eq. (\ref{E_singleloop}) corresponds to the screening whereas the second
term is standard for an Josephson junction.
The important question is if the spontaneous magnetic moment
(flux or supercurrent)
occurs in zero external field. The answer to this question may be obtained
by finding minimum of energy (\ref{E_singleloop}).

% FIGURE 19
\begin{figure}
\epsfxsize=3.2in
\centerline{\epsffile{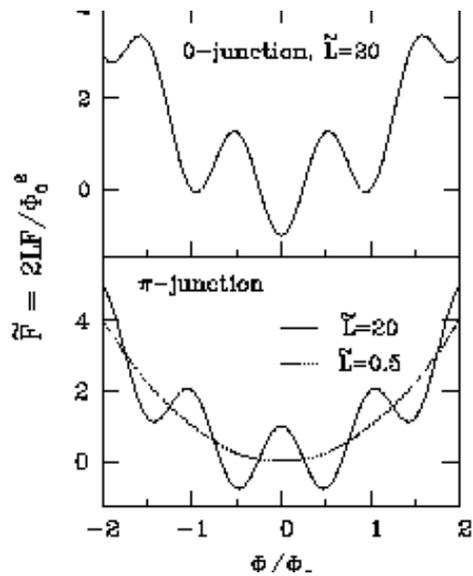}}
\vspace{0.2in}
\caption{Flux dependence of the dimensionless free energy $\tilde{F}$
of the one-loop model. For the 0-junction with $\tilde{L} = 20$ (upper panel)
 and $\pi$-junction
with $\tilde{L} = 0.5 < 1$ (dotted line in lower panel) the function $\tilde{F}$has minimum at
$\Phi/\Phi _0 = 0$. For a $\pi$-junction with $\tilde{L} = 20 > 1$ (solid line
in lower panel) the minimum is located at $\Phi/\Phi _0 \ne 0$ and the
paramagnetic response becomes possible.
}
\label{pme_free_fig}
\end{figure}

In the absence of the external field the dimensionless energy
$\tilde{F}(\Phi/\Phi_0)$ reads as
\begin{eqnarray}
\tilde{F}(\Phi/\Phi_0) \, = \, \frac{2L}{\Phi_0^2} F \, = \, \nonumber\\
\left(\frac{\Phi}{\Phi _0}\right)^2 - \frac{\tilde{L}}{2\pi^2}
\cos \left( 2\pi\frac{\Phi}{\Phi_0} + \delta\right) \, .
\end{eqnarray}
Here the dimensionless self-inductance $\tilde{L}$ is equal to
\begin{equation}
\tilde{L} \; = \; \frac{2\pi I_c L}{\Phi_0} \; .
\label{L_dimless}
\end{equation}
It is easy to show that at $\Phi/\Phi_0 =0$ the energy has extremum and
its second derivative with respect to $\Phi/\Phi_0$ has the simple form
\begin{equation}
\tilde{F}''(0) \; = \; 2(1 \pm \tilde{L}) \; ,
\end{equation}
where the sign plus and minus corresponds to the 0- and $\pi$-junction,
respectively. For an unfrustrated loop $\tilde{F}''(0) > 0$ and the state
without the supercurrent should be stable and the PME is not, as expected,
possible. In the case of the $\pi$-junction the energy
has minimum $\Phi/\Phi_0 = 0$ ($\tilde{F}''(0) > 0$) for $\tilde{L} < 1$
and {\em maximum}($\tilde{F}''(0) < 0$) for $\tilde{L} > 1$.
We come to a very interesting result: the energy of a frustrated loop
has minimum at $\Phi \ne 0$ provided $\tilde{L} > 1$. It is demonstrated
in Fig. \ref{pme_free_fig}. Thus, above the critical value
$\tilde{L}_c = 1$ the spontaneous magnetic moment $M_{sp}$ appears in the single
$\pi$-junction
loop leading to the PME. 

In the $\tilde{L} \rightarrow \infty$
limit $M_{sp}$ may be determined exactly. In this case the second
term in (\ref{E_singleloop}) dominates and the energy has a ladder of
minima at $\Phi = n\Phi_0$ and $\Phi = \left( n+ \frac{1}{2}\right)\Phi_0$
($n$ is integer number) for 0-junction and $\pi$-junction, respectively.
The spontaneous magnetization corresponding to the global minimum
with $n = 0$ is equal to $M_{sp} = \Phi_0/8\pi S$, where $S$ is the loop area.

Since the energy minima of the single $\pi$-junction are located near
$\Phi = \left( n+ \frac{1}{2}\right)\Phi_0$ the PME is, in some sense,
the manifestation of the common half-integer
flux quantum effect \cite{Tsuei00}. The screening plays a key role in the
observation of the paramagnetic response.
Taking into account the temperature dependence of critical current $I_c$
and using the one-loop model (\ref{E_singleloop}), Sigrist and Rice were able
to reproduce the paramagnetic behavior in the FC regime 
\cite{Sigrist92,Sigrist95}. In accord with experiments the paramagnetic signal
is suppressed as the external field is increased.

\subsection{Multi-loop model}

Although the single-loop model considered in the previous chapter predicts the
appearance of the PME, it fails to capture many cooperative phenomena
such as the aging, the compensation effect, a possibility of the existence
of a new chiral glass phase {\em etc} in granular materials.
This shortcoming may be amended in the multi-loop model
\cite{Dominguez,KawLi} where the interaction
between loops is taken into account.

B.1. {\em Hamiltonian}\\

Suppose that weak links connecting the neighboring grains are distributed
sufficiently dense, so that the system can be viewed as an infinite network
of Josephson junctions which are not decomposed into finite clusters.
We model such ceramic superconductors by a hypercubic (two- or
three-dimensional) lattice model of a Josephson junction arrays with finite
self-inductance. Neglecting charging effects of the grain
the Hamiltonian is given by \cite{Dominguez}
\begin{eqnarray}\label{Hamiltonian}
{\cal H} = - \sum _{<ij>} J_{ij}\cos (\phi _i-\phi _j-A_{ij})+ \nonumber\\
\frac {1}{2 L} \sum _p (\Phi_p - \Phi_p^{ext})^2, \nonumber\\
\Phi_p \; \; = \; \; \frac{\Phi_0}{2\pi} \sum_{<ij>}^{p} A_{ij} \; , \;
A_{ij} \; = \; \frac{2\pi}{\Phi_0} \int_{i}^{j} \, \vec{A}(\vec{r})
d\vec{r} \; \; ,
\label{H_multiloop}
\end{eqnarray}
where $\phi _i$ is the phase of the condensate of the grain
at the $i$-th site of a simple hypercubic lattice,
$\vec A$ is the fluctuating gauge potential at each link
of the lattice,
$J_{ij}$ denotes the Josephson coupling
between the $i$-th and $j$-th grains, and
$L$ is the self-inductance of a loop.
The effect of screening currents inside grains is not considered explicitly,
since for large length scales they simply lead to a Hamiltonian ${\cal H}$
with an effective self-inductance $L$ \cite{Sasik97}.
The first sum is taken over all nearest-neighbor pairs and the
second sum is taken over all elementary plaquettes on the lattice.
Fluctuating  variables to be summed over are the phase variables,
$\phi _i$, at each site and the gauge variables, $A_{ij}$, at each
link. $\Phi_p$ is the total magnetic flux threading through the
$p$-th plaquette, whereas $\Phi_p^{ext}$ is the flux due to an
external magnetic field
applied along the $z$-direction,
\begin{equation}
\Phi_p^{ext} = \left\{ \begin{array}{ll}
                   HS \; \;  & \mbox{if $p$ is on the $<xy>$ plane}\\
                   0  & \mbox{otherwise} \; \; ,
                        \end{array}
                  \right.
\label{H_plaq}
\end{equation}
where $S$ denotes the area of an elementary plaquette.
The Hamiltonian (\ref{H_multiloop}) is an extension of the single-loop energy
(\ref{E_singleloop}) to the interacting loops case.

In the $s$-wave ceramics, sign of the Josephson coupling is always positive
($J_{ij} > 0$), while in $d$-wave ceramics it could be either positive
(0-junction) or
negative ($\pi$-junction) depending on the relative direction of the junction
and the crystal grains on both sides. In $d$-wave case, the sign of $J_{ij}$
is expected to appear randomly since the spatial orientation of each crystal
grain would be random. It was also suggested \cite{Sigrist94} that the
Josephson junction between $d$-wave superconductors under certain circumstances
(e.g., near the interface) might have an energy minimum at some
fractional value, neither at 0 or $\pi$, as a result of a spontaneous
time-reversal symmetry breaking at the junction. Such a possibility is not
discussed in this review.

The model with uniform  ferromagnetic couplings ($J_{ij} = J$) studied
by Dasgupta and Halperin \cite{Dasgupta81} exhibits a standard
normal-superconductor transition. Here we deal with two types of bond
distributions. For $d$-wave ceramics $J_{ij}$ is assumed to take the value
-$J$ ($\pi$-junction) with probability $c$ and +$J$ (0-junction) with
probability (1-$c$) \cite{Dominguez}. In the $s$-wave model $J_{ij}$ is always
positive ("ferromagnetic" interaction) but distributed uniformly
between 0 and $2\pi$ \cite{KawLi}.

In the case of $d$-wave superconductors the frustration due to competition
between "ferromagnetic" and "antiferromagnetic" interactions \cite{Binder}
causes the nonzero spontaneous magnetic moment provided $c$ exceeds the
percolation threshold. The global PME becomes, therefore,
possible.  In the absence of an external field, the global minimum of the
energy of the $s$-wave system corresponds
to the configuration with zero supercurrents and the system displays a
standard diamagnetism.

B.2. {\em Critical self-inductance}\\

In the single-loop model with one $\pi$-junction the dimensionless
critical inductance above
which the system shows the PME is equal to $\tilde{L} = 1$. 
The question we as now is
does the interaction between frustrated  loops change this critical value.
For the multi-loop model (\ref{H_multiloop}) $\tilde{L}$ is also given by
Eq. (\ref{L_dimless}) with the critical current \cite{KawLi}
\begin{equation}
I_c \; = \; \frac{2\pi J}{\Phi_0} \; .
\label{crit_curr}
\end{equation}

One can show (see Appendix \ref{sec.lc}) that, contrary to the one-loop model,
the critical inductance above which the PME is observable
is equal to
\begin{equation}
\tilde{L} \, = \, 0.
\label{lc_mloop}
\end{equation}

In the limit of
small inductances the spontaneous flux of the multi-loop model is equal 
to \cite{KawLi}
\begin{equation}
\frac{\Phi}{\Phi_0} \; \approx \; \pm \frac{\tilde{L}}{2\sqrt{2} \pi} \,
\; \; \mbox{for} \; \tilde{L} \ll 1 \; .
\end{equation}
The last expression shows that for small inductance, spontaneous flux remains
nonzero but becomes small. In real ceramic samples one has an additional
diamagnetic contribution from intragranular supercurrents neglected in the
present model, a small paramagnetic contribution from the intergranular
supercurrents may easily be masked. Therefore, in practice, moderately
large inductance would be necessary to observe the paramagnetic behavior.

% FIGURE 20
\begin{figure}
\epsfxsize=3.2in
\centerline{\epsffile{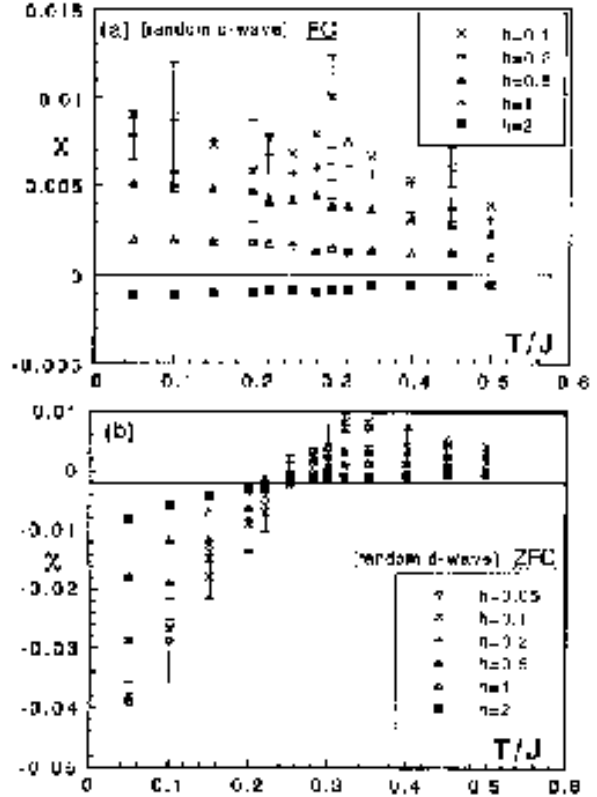}}
\vspace{0.2in}
\caption{The temperature dependence of the FC susceptibility (a) and of the
ZFC susceptibility (b) of $d$-wave ceramics for several values of the
external field. The lattice size is equal to $l = 8$, $\tilde{L} = 1$
and the concentration of $\pi$ junctions $c = 0.5$.
After Kawamura and Li \protect\cite{KawLi}.
}
\label{dw_sus_MC}
\end{figure}

B.3. {\em Simulation results}\\

Total magnetization along the $z$ axis normalized per plaquette $m$ is given by
\cite{Dominguez}
\begin{equation}
m \; = \; \frac{1}{N_p\Phi_0} \sum_{p \in <xy>} ( \Phi _p - \Phi^{ext}_p ) \; ,
\label{magnetization}
\end{equation}
where the sum is taken over all $N_p$ plaquettes on the $<xy>$ plane. The 
linear susceptibility $\chi$ (dimensionless in CGS units)
 may be calculated via fluctuation of the magnetization
as follows
\begin{equation}
\chi \; = \; \frac{\pi JN_p}{k_B T \tilde{L}}
[ < m^2 > - < m >^2 ]_J -
\frac{1}{4\pi} \; ,
\label{suscep_fluc}
\end{equation}
where $<...>$ and $[...]_J$ represent a thermal average and a configurational
average over the bond distribution, respectively. For conventional
superconductors the first term in (\ref{suscep_fluc}) vanishes at low
temperatures and $\chi$ takes on a standard value -$\frac{1}{4\pi}$.

% FIGURE 21
\begin{figure}
\epsfxsize=3.2in
\centerline{\epsffile{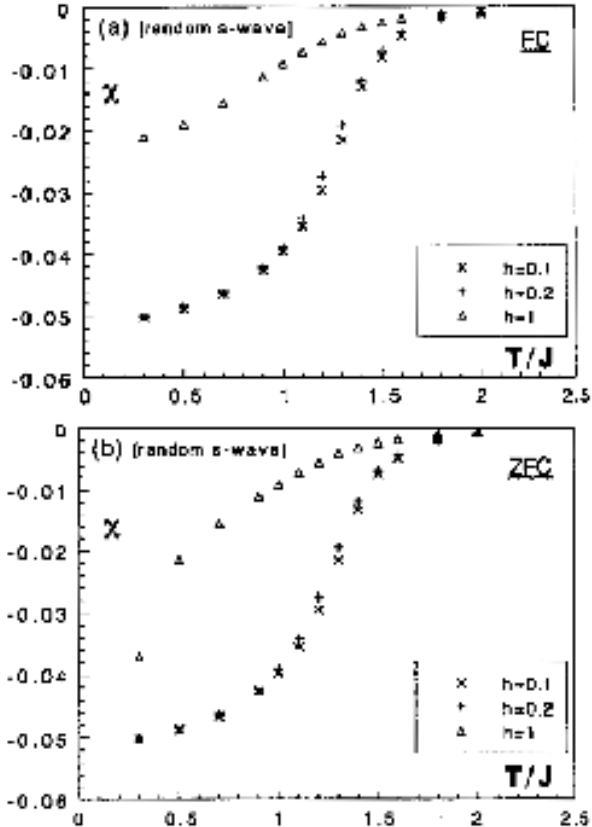}}
\vspace{0.2in}
\caption{The temperature dependence of the ZFC susceptibility (a) and of the
ZFC susceptibility (b) of $s$-wave ceramics for several values of the
external field. The lattice size is equal to $l = 8$ and $\tilde{L} = 1$.
After Kawamura and Li \protect\cite{KawLi}.
}
\label{sw_sus_MC}
\end{figure}

The temperature and field dependence of the FC and ZFC susceptibility was
studied by Langevin dynamics
\cite{Dominguez} and by Monte Carlo simulations \cite{KawLi}. Both these
methods give essentially the same results and we present those obtained
by the latter approach.

As can be seen from
Fig. \ref{dw_sus_MC}, a paramagnetic behavior is shows up in FC regime in the
field range $h \le 1$ (roughly corresponding to $\Phi ^{ext} \sim \Phi_0/4$),
where the dimensionless field $h$ is defined as \cite{KawLi}
\begin{equation}
h \; = \; \frac{2\pi SH}{\Phi_0} \; .
\label{H_dimless}
\end{equation}
By contrast, $\chi_{ZFC}$ remains diamagnetic at low temperatures for any value
of the dc fields studied.

From Fig. \ref{dw_sus_MC} it follows that the remanent magnetization
$m_{rem} \approx m_{FC} - m_{ZFC}$ is positive, consistent with 
experiments on high-$T_c$ ceramics \cite{Malozemoff88,Kawachi94}. This observation
indicates the occurrence of flux trapping in the $d$-wave model.

The Monte Carlo
simulations \cite{KawLi} showed that the PME is clearly visible for a very
small inductance $\tilde{L} = 0.1$. This result compatible with the conclusion in
the previous section that $\tilde{L}_c = 0$ for the multi-loop model.

Fig. \ref{sw_sus_MC} shows the temperature and field dependence of $\chi_{FC}$
and $\chi_{ZFC}$ of the $s$-wave system. We set the system size $l = 8$
and $\tilde{L} = 1$. As expected, the response is always diamagnetic.
Furthermore, there is no appreciable difference between
$\chi_{FC}$ and $\chi_{ZFC}$ for $h < 0.2$. Within the statistical errors
$m_{rm}$ turns out to be zero confirming that the flux does not get  trapped
inside a sample

\subsection{Other models}

Using a Josephson junction model of special geometry \cite{Auletta93}
Auletta {\em et al.} \cite{Auletta95} have shown that
the PME is possible even without $\pi$-junctions.
The similar conclusion was also made by other groups
\cite{Chandran97,Moreira97,Barbara99,Nielsen00,Leo01} for standard
systems of 0-junctions (the effect of mutual inductances
has been considered in \cite{Chandran97,Leo01}).
The paramagnetic behavior can arise either in a
$d$-wave sample or in a conventional sample of confined geometry.
From this point of view, the PME observed in simulations of the 0-junction
networks \cite{Auletta95,Chandran97,Moreira97,Barbara99,Nielsen00,Leo01}
may be due to the finite size effect or it is merely a metastable effect.

Recently, Khalil \cite{Khalil97} assumed that near randomly distributed
defects the localized superconducting states with different orbital moments
could exist. The first-order transitions between these states below
the critical temperature are accompanied by jumps
of the magnetic moments. As a result,
multiple-quanta vortices and
spontaneous moments randomly distributed in the material matrix will be
formed. This provides the explanation of the PME.

Authors of Ref. \cite{He96} proposed that a certain amount
of local moments exist that are becoming partially aligned by an external field.
Based on experimental data, these moments are estimated to be too small
compared to the half flux quanta. They are can not be, therefore, due to
$\pi$- junctions but their nature remains unspecified.

The another interpretation \cite{Chaban00} of the PME is based
on the so called impurity mechanism \cite{Tokmakova90} for high-$T_c$
superconductivity. The validity of this heuristic argument remains
ambiguous.

Finally, the PME was proposed to be related to the vortex 
pinning \cite{Svelindh89} at the Kosterlitz-Thouless transition.
Since this transition is possible in two dimensions \cite{Kosterlitz73} 
it is not clear
if the scenario of Svelindh {\em et al} \cite{Svelindh89} is applied
to three-dimensional systems.

\section{Flux compression and paramagnetic Meissner effect}

In this chapter we discuss the flux compression mechanism for the PME which
is probably applied to conventional superconductors with $s$-wave type of
pairing symmetry. There are two approaches to study the flux trapping in the 
confined geometry. One of them is based on the Bean critical model and was
developed by Koshelev and Larkin \cite{Larkin}. The second approach
\cite{Moshchalkov} relied on
the numerical solution of the Ginzburg-Landay equations, predicts the paramagnetic
response due to spontaneous magnetic moments in giant vortex states which
occur below the third critical field.

\subsection{Bean model}

\begin{figure}
\epsfxsize=3.2in
\centerline{\epsffile{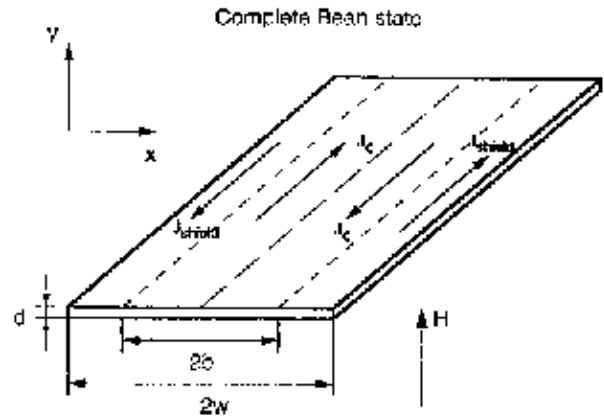}}
\vspace{0.2in}
\caption{Complete Bean critical state in a thin superconducting slab after
field cooling. Reprinted from Koshelev and Larkin \protect\cite{Larkin}.
}
\label{Bean_geom}
\end{figure}

The PME observed in conventional superconductors \cite{Geim} and in 
twined single crystals of YBa$_2$Cu$_3$O$_{7-\delta}$ suggests the possibility
of alternative explanations without exotic $d$-wave pairing.
One of such possibilities is based on the flux compression mechanism
\cite{Larkin}  using the Bean model \cite{Bean62}
for the critical state . A possible origin of the existence of the spontaneous
paramagnetic moment is flux trap caused
by an inhomogeneous superconducting transition.
If the edges of the sample becomes superconducting first due to, e.g., 
inhomogeneity of material
(surface layers have  a higher value of $T_c$ compared to the bulk)
or inhomogeneous cooling, then expulsion of vortices
from this area opens a flux-free region near the edges. This gives rise to the 
flux capture inside the sample. Further cooling leads to a broading of the
flux-free region and further compression of the flux due to, e.g., the vortex
Nernst effect \cite{Huebener91}. As the whole sample becomes superconducting
the critical Bean state develops in the flux region.
For a thin superconducting strip geometry (see Fig. \ref{Bean_geom}) one can 
imagine this scenario as follows. The outer current mimics the 
diamagnetic shielding current, while the inner current models the paramagnetic 
pinning current. If the flux is trapped inside sample, then the flux created by
the outer current should be completely compensated by the flux generated
by the inner current in the flux free region. Since the field  generated by 
currents flowing in the inner region changes sign, these currents give a
smaller contribution to the total flux than do the currents flowing in
the outer region \cite{Larkin}. This should be compensated by a larger magnitude
of the inner paramagnetic currents, which gives rise to the paramagnetic
moment.

We now give a qualitative description of the existence of the PME due 
to flux compression in a
simple case of a thin superconducting slab of width 2$w$ and thickness $d$
situated in a transverse magnetic field \cite{Larkin} as 
shown in Fig. \ref{Bean_geom}.
The Bean state is assumed to be complete with the critical current $j_c$
flowing in the strip of width 2$b$ ($ |x| < b < w$) surrounded by flux
free region of width $w-b$ ($b < |x| <w$) (the incomplete Bean state was
also discussed by Koshelev and Larkin \cite{Larkin}).

% FIGURE 22

Then the tangential ($H_x$) and longitudinal ($H_y$) field components satisfy
the following boundary conditions \cite {Larkin}

\begin{eqnarray}
H_x \; \; &=& \; \; H_J sign(x), \; 0 < |x| < b, \nonumber\\
H_x \; \; &=& \; \; 0, \;  |x| > b, \nonumber\\
H_y \; \; &=& \; \; 0, \;  b < |x| < w, \nonumber\\
H_y \; \; &=&  H_{\infty} \; , \;  r \, \rightarrow \; \infty \; ,
\label{Larkin_boun}
\end{eqnarray}
where 2$H_J = (4\pi/c)J_c, J_c=dj_c$.
The first equation in (\ref{Larkin_boun}) from the Gaussian theorem and
the continuous boundary conditions for the tangential component of the field 
near the slab surface \cite{Landau82,Mikheenko93}. The second and third 
equations are valid for the flux-free region.
The field distribution may be obtained from the Maxwell equations
\begin{equation}
div\vec{B} \; = \; 0 \; , \; rot\vec{B} \; = \; 0 \; 
\label{Maxwell_eq}
\end{equation}
subjected to boundary conditions (\ref{Larkin_boun}.
An elegant way to solve these equations is based on theory 
of complex variable functions \cite{Larkin}. Introducing the complex function
$H(z) = H_x +iH_y$ of complex variable $z=x + iy$, then from Maxwell equations
(\ref{Maxwell_eq}) it follows that $H_x$ and $H_y$ satisfy
the Cauchy conditions for an analytic function. The problem of finding 
$H(z)$ is, therefore,
 reduced to the finding of an analytic function satisfying conditions
(\ref{Larkin_boun}). One can show that \cite{Larkin}
\end{multicols}
\begin{eqnarray} 
H(z) \; = \; -\frac{H_J}{\pi} \left[ \, \left( \frac{z^2 - b^2}{z^2 - w^2}
\right)^{1/2} ln\left(\frac{w + b}{w - b}\right) \, - \,
ln \frac{( 1-z^2/b^2)^{1/2} + (1 - z^2/w^2)^{1/2}}{( 1-z^2/b^2)^{1/2} -
 (1 - z^2/w^2)^{1/2}} \right]
H_{\infty} \left(\frac{z^2 - b^2}{z^2 + b^2} \right)^{1/2} \; .
\end{eqnarray} 
Using this function we obtain the total magnetic flux
\begin{eqnarray}
\Phi \; &=& \; 2 \int_0^b \, H_y dx \; = \; \Phi_J \, + 
\, \Phi_H \, , \nonumber\\
\Phi_J \; &=& \; -\frac{2wH_J}{\pi} \{ [ E(m) - (1-m) K(m) ]
ln \frac{1 + \sqrt{m}}{1 - \sqrt{m}} -2\sqrt{m} K(m) \} \; , \nonumber\\
\Phi_H \; &=& \; 2wH_{\infty} [E(m) - (1-m) K(m) ] \; , \; m = b^2/w^2 \; ,
\label{Larkin_flux}
\end{eqnarray}
\begin{multicols}{2}
\noindent where $E(m)$ and $K(m)$ are complete elliptic 
integrals \cite{Abramowitz}. The
condition $\Phi = 2wH_{\infty}f$, with $f$ being the fraction of trapped flux,
gives the relation between the critical field $H_J$ and the width $b$ of flux
region.

\begin{figure}
\epsfxsize=3.2in
\centerline{\epsffile{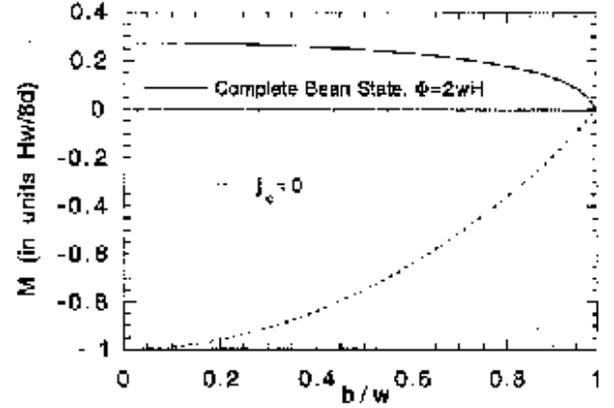}}
\vspace{0.2in}
\caption{Dependence of magnetic moment in the compressed flux state ($f=1$) on 
the relative width of the flux region. For comparison similar dependence for
an ideal superconductor ($j_c=0$) is shown.
After Koshelev and Larkin \protect\cite{Larkin}.
}
\label{Larkin_pme}
\end{figure}

The magnetic moment can be determined from the large-distance asymptotics 
of the field as
\begin{equation}
H(z) \; \rightarrow \; \frac{4wdM}{z^2} \; , \; z \rightarrow \infty \; ,
\end{equation}
which gives
\begin{eqnarray}
M \; \; &=& \; \; M_J \, + \, M_H \; , \nonumber\\
M_J \; \; &=& \; \; \frac{H_Jw}{8\pi d} \left[ (1-m) ln \left(
\frac{1 + \sqrt{m}}{1 - \sqrt{m}} \right) + 2\sqrt{m} \right] \; , \nonumber\\
M_H \; \; &=& \; \; -\frac{H_{\infty}w}{8d} (1-m) \; .
\label{Larkin_mag}
\end{eqnarray}
Parameter $m$ has to be expressed through the critical current $J_c=cH_J/2\pi$
and the fraction of captured flux $f=\Phi/2wH_{\infty}$ using 
Eq. (\ref{Larkin_flux}). In the limit $m << 1$ (strong compression) the magnetic
moment
\begin{equation}
M \; \; = \; \; \frac{H_Jw\sqrt{m}}{8d} \left( \frac{4}{\pi} - 
\frac{1}{f} \right) + O(m) \; 
\end{equation}
which is positive for $f > \pi/4 \approx 0.79$. Remarkably, this value almost
coincides with the critical value $f \approx 0.8$ 
(above which a paramagnetic moment
is possible) obtained numerically \cite{Larkin}.
Fig. (\ref{Larkin_pme}) shows the dependence
of $M$ on the relative width of flux region $b/w$ for $f$=1 obtained from 
Eq. (\ref{Larkin_flux}) and (\ref{Larkin_mag}).

% FIGURE 23

In the weak compression regime $w - b << w$ one can show that the magnetization
is always negative \cite{Larkin}.

\subsection{Ginzburg-Landau equation approach}

In this section we consider the flux compression mechanism by the another
approach \cite{Moshchalkov} which is based on the self-consistent numerical 
solution of the Ginzburg-Landau equations.
One can show show that due to confined geometry
the giant vortex state is formed in the FC regime at the third critical field
$H_{c3}(T)$. The PME appears as a result of the flux trap inside the giant
vortex state with a fixed orbital quantum number $L$.

The basic postulate of the Ginzburg-Landau theory is that if the Cooper
wave function $\Psi$ is small and varies slowly in space, the free energy
${\cal F}$ can be expanded in a series of the form \cite{Ginzburg50}

\end{multicols}
\begin{eqnarray}
{\cal F} \; = \; \int d\vec{r} \, \left\{ \alpha \mid \Psi \mid ^2 +
 \frac{\beta}{2}
|\Psi|^4 + \frac{1}{2m^*} \left| \left( \frac{\hbar}{i}\nabla
- \frac{e^*}{c} \vec{A} \right) \Psi \right|^2 + \frac{H^2}{8\pi} \right\} \, ,
\label{FreeGL}
\end{eqnarray}
\begin{multicols}{2}
where $\alpha (T) \propto (T - T_{c0}), T_{c0} $ is the mean field transition
temperature. Minimizing the Ginzburg-Landau free energy with respect to the
fields $\Psi(\vec{r})$ and vector potential $\vec{A}(\vec{r})$ one can get 
two well-known
equations which will be used to study the PME.

Following Ref. \cite{Moshchalkov} (see also Ref. \cite{Zharkov01})
we consider a infinitely long cylinder
of radius $r_0$ in a magnetic field parallel to its axis.
First we will find $H_{c3}(T)$ above which the nucleation of
the superconducting phase takes place and then study the PME.  
Determination of $H_{c3}(T)$ requires to solve the linearized Ginzburg-Landau
equation for the superconducting order parameter $\Psi$ \cite{Tinkham}

\begin{equation}
\frac{1}{2m} \left(  -i\hbar \vec{\nabla} - \frac{e^*}{c} \vec{A}
 \right)^2 \Psi \; = \; - \alpha \Psi \; ,
\label{GL_linear}
\end{equation}
which is identical to the Schr{\H o}dinger equation for a particle of charge
$e^*=2e$ in a uniform magnetic field given by the vector potential
$\vec{H} = \vec{\nabla}\times \vec{A}$. Parameter $\alpha$ in 
Eq. (\ref{GL_linear}) plays the role of energy $E$ in the corresponding
Schr{\H o}dinger equation and

\begin{equation} 
E \; = \; -\alpha \; = \; \frac{\hbar^2}{2m\xi ^2(T)} \; = \;
\frac{\hbar^2}{2m\xi ^2(0)} \frac{T_{c0}-T}{T_{c}} \; ,
\label{Ener_Mos}
\end{equation} 
where $\xi(T)$ is the bulk coherence length and $T_{c0}$ is the critical
temperature at zero field. 

For finite samples the order parameter $\Psi$ should satisfy the boundary 
condition for a superconductor-insulator interface \cite{deGennes}

\begin{equation}
\left( -i\hbar \vec{\nabla} - \frac{e^* \vec{A}}{c} \right) \Psi |_n \; = \; 0
\; 
\label{BC_Mos}
\end{equation}
which assures that no current passes through the surface.

The common strategy to find $H_{c3}(T)$ is that \cite{Tinkham} one has to solve
the linearized Ginzburg-Landau equation subjected to the boundary condition
(\ref{BC_Mos}). Then $H_{c3}(T)$ is subtracted from the lowest energy
$E(H)$ which gives the highest $T$ in Eq. (\ref{Ener_Mos}) coinciding with
the nucleation phase boundary. In the case of large bulk samples when the
boundary condition (\ref{BC_Mos}) becomes irrelevant, the lowest Landau
level $E=\hbar\omega/2$ (the cyclotron frequency
$\omega=e^*H/mc$) gives the well known upper critical field $H_{c2}$
\begin{equation}
H_{c2} \; = \; \frac{\hbar c}{e^*\xi^2(0)}\frac{T_{c0}-T}{T_{c0}} \;
= \; \frac{\Phi_0}{2\pi \xi^2(T)} \; .
\label{H_c2}
\end{equation}

For samples of confined geometry the boundary condition (\ref{BC_Mos}) must
be taken into account. The crucial question is how small the sample should
be to observe the finite size effect. To answer this question one may refer
to analogy between the Ginburg-Landau equation (\ref{GL_linear}) and the
Schr{\H o}dinger equation for normal electron considered by Dingle
\cite{Dingle52a} for the analysis of quantum oscillations  in small metallic
samples. The finite size 
effect is expected to become visible \cite{Dingle52a}  provided $r_0 \le r_L$,
where $r_L$ is the Larmor radius or
\begin{equation}
H\times r_0 \le 5 (G cm) \; .
\label{size_crit}
\end{equation}
Clearly, the critical size depends on the applied magnetic field. If the size
is equal to, say, $r_0 = 5 \mu$m then the boundary condition (\ref{BC_Mos})
must be taken account for $H < 10^4$ G.

In order to solve the linearized Ginzburg-Landau equation (\ref{GL_linear})
for a long cylinder
we choose the cylindrical coordinates $(r,\phi,z)$ and the following gauge
\begin{equation}
\vec{A} \; = \; (0, Hr/2, 0) \; .
\label{gauge_cyl}
\end{equation}
Then the solution reads as \cite{Dingle52b}
\begin{equation}
\Psi_L(R,\phi) \; = \; e^{\pm iL\phi} R^L \exp (-R^2/2) M(-N,L+1,R^2)
\label{WF_Mos}
\end{equation}
and the energy $E_{\perp}$ of motion in the plane perpendicular to $\vec{H}$ is
given by the orbital quantum number $L$ and parameter $N$:
\begin{equation}
E_{\perp} \; = \; \frac{e^*\hbar H}{2mc} (2N \pm L + L +1) \; .
\label{Ener_LN}
\end{equation}
Here $M(-N,L+1,R^2)$ is the Kummer function \cite{Dingle52b}, the dimensionless
radius $R = \sqrt{\gamma} r, \gamma = e^* H/(2\hbar c)$. From the condition
$\Phi(r,\phi +2\pi) = \Phi(r,\phi )$ it follows that $L$ should be integer 
number. It is important to stress that $N$ is not necessarily integer number
and it must be found from the boundary condition (\ref{BC_Mos}) which with the 
gauge choice (\ref{gauge_cyl}) is reduced to
\begin{equation}
\frac{\partial |\Psi (R)|}{\partial R} |_{R=R_0} \; = \; 0.
\label{BC1_Mos}
\end{equation}

So far as we are looking for the lowest possible energy state for determination
of $H_{c3}(T)$, one should take 
the minus sign in the argument of the exponent $\exp (-iL\phi )$ in the
solution (\ref{WF_Mos}). Then the energy levels become
\begin{equation}
E_{\perp} \; = \; \hbar \omega (N +1/2) \; ,
\label{Ener_LN_lowest}
\end{equation}
where $\omega$ is the cyclotron frequency.
This expression coincides with the well-known Landau levels but now $N$ is 
any real number including negative one.

\begin{figure}
\epsfxsize=3.2in
\centerline{\epsffile{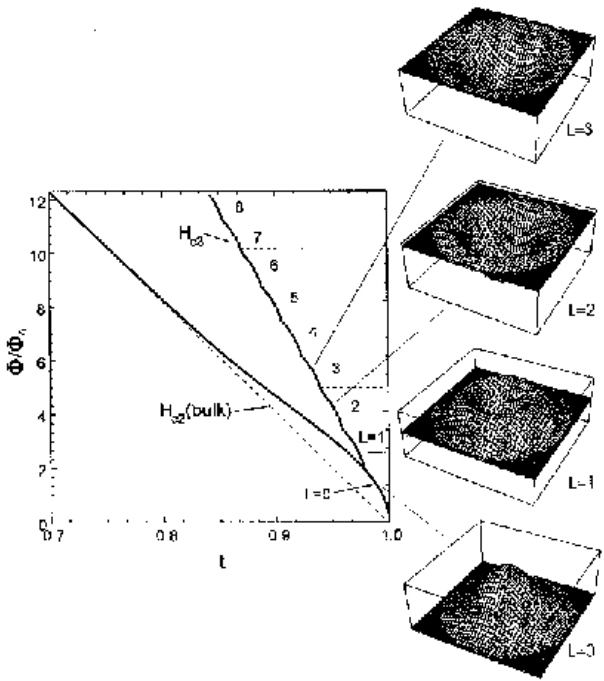}}
\vspace{0.2in}
\caption{The third critical field $H_{c3}$ and the bulk upper critical
field $H_{c2}$ (dashed line) vs normalized temperature $t = T/T_c$. The
cusplike $H_{c3}$ is formed due to the jump of the orbital quantum number
$L$. The three-dimensional plot of $|\Psi |$ is shown for several values 
of $L$. After Moshchalkov {\em et al.} \protect\cite{Moshchalkov}.
}
\label{Hc3_fig}
\end{figure}

From Eqs. (\ref{WF_Mos}) and (\ref{BC1_Mos}) we obtain the equation for
determination of the  parameter $N$ \cite{Moshchalkov}
\begin{eqnarray}
(L^2 - R_0^2) M(-N,L+1,R_0^2) - \nonumber\\
\frac{2NR_0^2}{L+1} M(-N+1,L+2,R_0^2) \; = \; 0 \; .
\end{eqnarray}
Remarkably, the numerical solution of the last equation gives negative values
for $N(L,R_0)$ which immediately lead to the energy $E_{\perp}$ in
(\ref{Ener_LN_lowest}) lower than the bulk value $\hbar \omega/2$. Therefore,
as seen from Fig. \ref{Hc3_fig}, in the presence of
the finite surface the superconductivity can occur
well above $H_{c2}(T)$ line obtained for $N=0$. 
The cusplike phase boundary $H_{c3}$ occurs due to switching between different
orbital momenta $L$. Such a phase boundary was observed experimentally
\cite{Buisson90,Moshchalkov95} for the superconducting disks. The
linear component of the cusplike $H_{c3}$ line is 1.965$H_{c2}$, which is 
in good agreement with the calculations of $H_{c3}$ in the $L \rightarrow
\infty$ limit \cite{deGennes,Saint-James63}.

In order to study the PME one has to answer the
following question: how to mimic the FC and ZFC experiments by the 
Ginzburg-Landau method? In the FC mode when the external field is applied to
a sample the states with orbital quantum numbers $L > 0$ corresponding to
the rotation of the superconducting condensate due to the action of the 
Lorentz force are expected to be formed. Then by cooling down we are crossing
the $H_{c3}(T)$ boundary (see Fig. \ref{Hc3_fig}) at a given $L > 0$ and
reaching a temperature $T < T_{c3}(H)$ at which measurements are carried out. 
It should be noted that \cite{Cruz79}
 $L$ can be retained over a large temperature 
interval, namely, from $T(H_{c3})$ to $T(H_c)$ (near or at $T(H_c)$ the
quantum number $L$ changes abruptly, thereby expelling a large amount of
flux from the sample, as the specimen makes a transition to the Meissner state).
From the physical point of view, the conservation of $L$ is related to
pinning of the giant vortex state by the sample boundary.
Thus, the FC experiments may be analysed by solving the 
Ginzburg-Landau equations below $H_{c3}(T)$ for fixed  $L > 0$. 
The same is valid for the ZFC regime but with $L=0$.

% FIGURE 24

Below $H_{c3}(T)$, one needs to solve the full Ginzgurg-Landau equations
because the superconducting condensate induces the screening due 
to supercurrents (proportional to $\Psi ^2$) which should be taken into account.
Then two coupled dimensionless equations are \cite{Abrikosov88}

\begin{eqnarray}
\left( \frac{\vec{\nabla}}{i\kappa} - \vec{A} \right)^2 \Psi \; &=& \;
\Psi (1 - |\Psi |^2 ) \; , \nonumber\\
\vec{\nabla} \times \vec{\nabla} \times \vec{A} \; &=& \; \frac{1}{2} 
\nonumber\\
\left[ \Psi ^* \left( \frac{\vec{\nabla}}{i\kappa } - \vec{A} \right) \Psi +
\Psi  \left( -\frac{\vec{\nabla}}{i\kappa }  - \vec{A} \right) 
\Psi ^* \right] \; ,
\label{GL_nonli}
\end{eqnarray}
where $\Psi, \vec{A}$ and $r$ are in units of $\Phi _{infty}, 
\sqrt{2}H_c\lambda (0)$ and $\lambda (0)$, respectively.
$\Psi _{\infty}, H_c, \lambda (0)$ and $\kappa$ are the wave function in bulk,
the thermodynamic critical field, the penetration depth at zero field and zero
temperature, and the Ginzburg-Landau parameter, respectively. For the
cylindrical symmetry we choose
\begin{equation}
\Psi (r,\phi ) = e^{iL\phi} F(r), \, \vec{A} = (0, \frac{Hr}{2} + 
\frac{\varphi}{2r}, 0) \; .
\label{gauge_nonli}
\end{equation}
Comparing Eqs. (\ref{gauge_nonli}) and ((\ref{gauge_cyl}) one can see that
the vector potential in the full Ginzburg-Landau equations contains the 
induced field neglected in the linear approximation.
Denoting the local magnetic field by $b$ and using 
$\vec{b} = \vec{\nabla} \times \vec{A}$ we have
\begin{equation}
b \; = \; H + \frac{1}{2r} \frac{d\varphi}{dr} \; .
\label{field_loc}
\end{equation}
Substitution of Eqs. (\ref{gauge_nonli}) and (\ref{field_loc}) into Eq.
(\ref{GL_nonli}) gives two coupled equations for determination of $F$ and 
$\varphi$ \cite{Moshchalkov}
\begin{eqnarray}
\frac{d^2F}{dr^2} \; = \; - \frac{1}{r} \frac{dF}{dr} + \nonumber\\
\left( \frac{1}{2} \kappa Hr + \frac{1}{2} \frac{\kappa \varphi}{r}
- \frac{L}{r} \right)^2 F - \kappa F (1 - F^2) \; , \nonumber\\
\frac{d^2\varphi}{dr^2} \; = \; \frac{1}{r} \frac{d \varphi}{dr} +
\left( Hr^2 + \varphi - \frac{2L}{\kappa} \right) F^2 \; .
\label{GL_final}
\end{eqnarray}
The corresponding boundary conditions are \cite{Fink66}
\begin{equation}
\frac{dF}{dr} \; = \; 0, \; \frac{d\varphi}{dr} \; = \; 0
\label{BC_r=r0}
\end{equation}
at $r = r_0$ and

\begin{eqnarray}
\varphi \; = \; 0 , \; \; F \; &=& \; 0 \; \; \mbox{for $L \neq 0$} \; ,
 \nonumber\\
\frac{dF}{dr} \; &=& \; 0 \; \; \mbox{for $L = 0$}
\label{BC_r=0}
\end{eqnarray}
at $r = 0$. 

The magnetization per unit volume $4\pi M$ can be defined via the flux through
the cylinder and has the following form
\begin{equation}
\frac{4\pi M}{H_c} \; = \; 2 \int_0^{r_0} \, (b(r) - H) r dr \, = \,
\varphi (r_0) \; .
\label{mag_Mos}
\end{equation}

% FIGURE 25
\begin{figure}
\epsfxsize=3.2in
\centerline{\epsffile{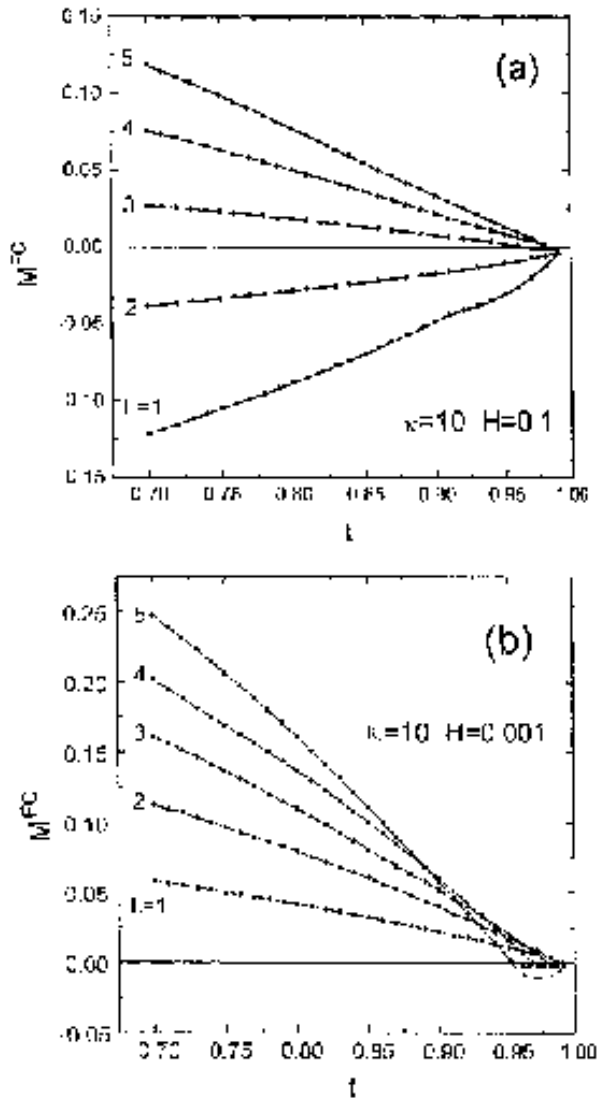}}
\vspace{0.2in}
\caption{Temperature dependence of $M_{FC}$ of a superconductor
with the Ginzburg-Landau parameter $\kappa = 10$ for different values of $L$
in applied magnetic field $H = 0.1$ (a) and $H = 0.01$ (b). The radius of
the cylinder $r_0 = \sqrt{3} \lambda (0)$.
After Moshchalkov {\em et al.} \protect\cite{Moshchalkov}.}
\label{Mfc_Mos}
\end{figure}

As in the linear approximation the orbital quantum number $L$ remains a good
quantum number due to the cylindrical symmetry. For a fixed $L$ one can solve
Eqs. (\ref{GL_final}) - (\ref{BC_r=0}) numerically to obtain the magnetization
in the FC mode. The typical results are shown in Fig. \ref{Mfc_Mos}.
In accord with experiments the PME occurs for the fixed values of $L$
and it gets enhanced as the external field is lowered.
The paramagnetic signal is, however, sensitive to $H$ and $L$. For $L = 1$ and 2
the increase in the field from $H = 0.001$ to $H = 0.001$ changes the sign
of the magnetization from  unusual paramagnetic to conventional diamagnetic.
The reported sensitivity to the surface treatment
\cite{Thompson,Kostic,Lucht95} may be caused by violation of the $L$ 
conservation and recovery of a normal diamagnetic response corresponding to a
transition from large $L$ values trapped at $H_{c3}$ to the state with $L = 0$.

The numerical solution of Eqs. (\ref{GL_final}) - (\ref{BC_r=0}) for the ZFC 
regime with $L = 0$ gives, in agreement with experiments,
 the diamagnetic response at any temperature 
\cite{Moshchalkov}. Thus, the PME and its field dependence can be obtained
from the self-consistent solution of the Ginzburg-Landau equations assuming
that orbital quantum number $L$ is conserved. This very simple and natural
approach does not involve any further assumptions related to the existence
of a $\pi$ junction or $d$- wave superconductivity. The drawback of the
Ginzburg-Landau equation
approach is that it does not allow for studying the PME far from the
transition temperature to the superconducting phase.

The PME in a mesoscopic disk has been also studied
\cite{Schweigert00,Palacios00} by the Ginzburg-Landau method . 
The results are qualitatively the same as for the case of the cylindrical
symmetry.

To summarize this section, according to the Ginzburg-Landau theory, the 
paramagnetism in low-$T_c$ mesoscopic superconductors may be caused 
by moments in giant vortex states.
The vortices formed inside a sample in the FC regime get
pinned to the boundary which is a source of inhomogeneity. The directions
of the external and of the vortex magnetic fields are the same but the
directions of currents screening these fields from the bulk are opposite.
Since the screening currents contribute to the magnetic moment they give
two contributions of opposite sign ($M = M_{dia} + M_{para}$). The
distributions of the fields and currents inside the sample vary with
$H$ and the measured magnetization, therefore, may be either positive
or negative depending on $H$.

\section{Chiral glass}

The existence of spontaneous supercurrents in ceramic superconductors may
lead to, as assumed by Kusmartsev \cite{Kusmartsev92},  a so called
"orbital glass". A natural question arises is what is the nature of the
orbital glass? The picture by Sigrist and Rice \cite{Sigrist92} is essentially
a single-loop picture in which the interactions between loops are irrelevant.
In their scenario, when the temperature is lowered across the superconducting
transition of each grain $T_c^{gr}$, the FC susceptibility changes sign from
negative to positive at a certain temperature $T_0$, sightly below $T_c^{gr}$.
Such a change of sign of $\chi_{FC}$, which is often regarded as a measure
of the orbital glass transition point, is a crossover not related to any
intergranular (interloop) cooperative phenomena. In this picture, the PME arises
as a property of an ensemble of noninteracting loops if there occurs an 
intragranular superconducting transition. By contrast, the possible cooperative
character of the orbital glass was pointed out by Dominguez {\em et al}
\cite{Dominguez} and by Khomskii \cite{Khomskii94}, although these authors did 
not detail the nature of the cooperative phenomena. 

Naturally, one may ask here whether could be a thermodynamically stable orbital
glass state characterized by a spontaneous breaking of certain symmetry over
an entire granular system. An issue to be addressed is whether there could be 
some sort of thermodynamic {\em intergranular} phase transition accompanied
with a divergent length scale over grains, and if it is, what is the order 
parameter of such phase transition. It is reasonable to assume that
the orbital glass is, similar to spin glass, a state of frozen spontaneous
moments or fluxes due to frustration caused by random distribution of
$\pi$ junctions.

An attempt \cite{KawLi97a} to detect such a orbital (or flux) glass by Monte 
Carlo simulations with the use of multi-loop model (\ref{H_multiloop})
gave an ambiguous conclusion. Instead, the simulations
\cite{KawLi97,KawLi97a} showed clearly that a novel thermodynamic phase may 
occur {\em in zero external field} in certain ceramic high-$T_c$ 
superconductors. This phase is characterized by a spontaneously broken
time-reversal symmetry with keeping $U(1)$ gauge symmetry, and is called
"chiral glass phase". The order parameter is then a "chirality", quenched-in
half a vortex, which represents the direction of the local loop-supercurrent
circulating over grains. As one can see below, since  chirality and flux
have the same symmetry we believe that the orbital glass should also exist in
three dimensions. The fact that it was not identified in Monte Carlo
simulations \cite{KawLi97a} may be related to small system sizes. 

In this chapter we introduce the chirality concept in discuss the chiral
glass phase in detail. Experimental search for this new phase will be presented.

\subsection{Chirality concept}

Frustration in vector spin systems often give rise to noncollinear and
noncoplanar spin orderings which may be characterized by the so call
chiral degrees of freedom. 
While the chirality concept has long been a familiar concept in molecular 
chemistry, it was introduced into the field of magnetism first by Villain
\cite{Villain}.

\begin{figure}
\epsfxsize=3.2in
\centerline{\epsffile{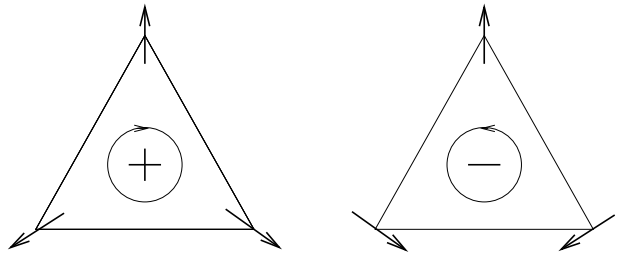}}
\vspace{0.2in}
\caption{The ground state spin configuration of three XY spins
antiferromagnetically coupled on a triangle
lattice. Frustration leads to the canted ordered state. The ground state is
twofold degenerate according to whether the non-collinear spin structure is 
right- or left-handed, each of which is characterized by the opposite chirality
as shown in left and right panel.
}
\label{chiral_villain}
\end{figure}

The simplest way to introduce
the chirality concept \cite{Villain} is to consider the XY model defined by
the following Hamiltonian
\begin{equation}
H \; = \; -\sum_{ij} \, J_{ij} \vec{S}_i. \vec{S}_j \; = \;
-\sum_{ij} \, J_{ij} \cos (\phi_i - \phi_j) \; ,
\label{XY_model}
\end{equation}
where the unit spin vector $\vec{S}_i = (\cos\phi_i, \sin\phi_i)$.
Model (\ref{XY_model}) is the simplest version of (\ref{H_multiloop}) without
the external field and screening. We consider the triangular lattice and 
assume that the spin-spin interaction is
uniform and antiferromagnetic ($J_{ij} = J < 0$), then the ground state of three spins located
at corners of a triangle, as shown in Fig. \ref{chiral_villain} is two-fold
degenerate. Angles between neighboring spins are 120$^o$ and -120$^o$
for the left and right panel, respectively. One may define the chirality
$\kappa$ via a
vector product of two neighboring spins averaged over three spin pairs
\begin{equation}
\kappa \; = \; \frac{2}{3\sqrt{3}} \sum_{<ij>} [\vec{S}_i \times \vec{S}_j ]_z
\; = \; \frac{2}{3\sqrt{3}} \sum_{<ij>} \sin (\phi_i - \phi_j) \; .
\label{chirality_XY}
\end{equation}
Clearly, $\kappa = 1$ for the right-handed (clockwise) configuration
 in left panel
of Fig. \ref{chiral_villain} and $\kappa = -1$ for the left-handed 
(counterclockwise) configuration in right panel. In the case of XY spins
considered here, the chirality $\kappa$ is actually a {\em pseudoscalar}.
It remains invariant under global $SO(2)$ proper spin rotations while
it changes sign under global $Z_2$ spin reflections. In order to transform
the chiral state with $\kappa = 1$ to the state with $\kappa = -1$ one
needs to make a global spin reflection. 

It should be noted that if the interaction is ferromagnetic then the system
becomes geometrically unfrustrated (all spin in the ground state are parallel)
and $\kappa = 0$. So the chirality concept makes sense only for frustrated
systems. The chirality defined by Eq. (\ref{chirality_XY}) is a vector quantity,
while in the literature one discusses also a scalar chirality \cite{Kawa_rev98}
for Heisenberg spins.

% FIGURE 26

The chirality concept plays an important role in physics of classical
\cite{Kawa_rev98,Kawa_Can01} and quantum \cite{Kawa_rev02} spin systems.
Especially since the discovery of high-$T_c$ cuprates
the scalar chirality has been a key theoretical concept in physics of strongly
correlated electrons \cite{Kalmeyer87,Wen89,Lee92}.
In order for the spin chirality to be ordered both the time-reversal
and parity symmetry must be broken. Breaking of these symmetries in two 
dimensions brings about many intriguing physics such as parity anomaly
\cite{Deser82,Nielson83}, anyon superconductivity \cite{Laughlin88}, and
quantized Hall effect in zero external magnetic field \cite{Haldane88}.
In this review we try to attract the attention of readers to a
possibility to observe the  so called chiral glass phase in ceramic materials.

\subsection{Simulation search for the chiral glass phase}

In order to study the nature of ordering of the chiral glass phase in
ceramic superconductors we use the multi-loop Hamiltonian (\ref{H_multiloop})
with the zero external field. The Josephson coupling $J_{ij}$ is assumed to be
an independent variable taking the values $J$ or $-J$ with equal probability
(bimodal distribution). While we present results for this particular 
distribution, one could expect from experience in spin glass studies that 
the results would be rather insensitive to the details of the distribution.

Contrary to the gauge-glass
model \cite{Huse90,Gingras92,Cieplak92}, the Hamiltonian (\ref{H_multiloop})
defined at $H=0$, 
keeps the $Z_2$ time-reversal symmetry in addition to the $U(1)$
gauge symmetry.
Frustration arises  from the random
distribution of both positive and negative Josephson couplings.
This should be contrasted to
the vortex-glass (gauge-glass) problem, where
the  associated Hamiltonian does not possess
the time-reversal symmetry
due to external magnetic fields, while
the frustration arises from  the magnetic field but
not from the $J_{ij}$.

Extending the definition (\ref{chirality_XY}) to the case of the multi-loop 
model the local chirality at each plaquettes is given by the gauge-invariant
quantity \cite{KawLi}
\begin{eqnarray}
\kappa_p=2^{-3/2}\sum_{<ij>}^p
\mbox{sign}(J_{ij}) \sin (\phi_i-\phi_j-A_{ij}),
\end{eqnarray}
where the sum runs over a directed contour
along the sides of the plaquette $p$. Physically, the chirality, $\kappa _p$,
is a half ($\pi $) vortex, being proportional to the
loop-supercurrent circulating round a  plaquette $p$.
If the plaquette $p$ is frustrated, the local chirality
$\kappa _p$ tends to take a finite value, its sign
representing either clockwise or counterclockwise orientation of
circulating
supercurrent. If, on the other hand,
the plaquette is unfrustrated, as mentioned before,
it tends to take a value around zero.
Note that the chirality is a pseudoscalar in the sense that it is
invariant under global $U(1)$ gauge transformation,
$\phi _i\rightarrow \phi _i+\Delta \phi,\ A_{ij}\rightarrow
A_{ij}$,
but changes its sign under global $Z_2$ time-reversal transformation,
$\phi _i\rightarrow -\phi _i,\ A_{ij}\rightarrow
-A_{ij}$.
Due to this symmetry property, chirality can be regarded
as an order parameter of the chiral order.

In analogy with the spin glass theory, the chiral glass phase may be 
characterized by the Edwards-Anderson order parameter \cite{Edwards75}
$q_{CG}^{EA}$:
\begin{equation} 
q_{CG}^{EA} \; = \; [<\kappa_p>^2]_J \, .
\end{equation} 
In the chiral glass phase $q_{CG}^{EA} \ne 0$, while in the chirality-disordered
("paramagnetic") phase $q_{CG}^{EA} = 0$. In the chiral glass state chiralities
are orientated randomly in space but frozen in time. 

The magnetization and linear susceptibility are given by 
Eqs. (\ref{magnetization}) and (\ref{suscep_fluc}) but with $\Phi^{ext} = 0$.
The nonlinear susceptibility, $\chi _2$
is defined as follows \cite{KawLi,KawLi97a}
\begin{eqnarray}
\chi_2 \; = \; \frac{1}{6}\frac{d^3 m}{dH^3} \; = 
\frac{1}{6}\left(\frac{\pi JN_p}{k_BT\tilde{L}}\right)^3
[< m^4> \nonumber\\ - 3<m^2>^2  -4<m><m^3> + \nonumber\\ 12 <m^2><m>^2 - 
 6 <m>^4 ]_J \; .
\end{eqnarray}
Note that $\chi _{2}$,  being
proportional to the minus of the third-harmonic
component of the ac susceptibility, is sometimes denoted as $\chi_3$
in the literature.

Following the standard technique from the spin glass theory \cite{Binder}
we introduce the overlap $q$ between the chiral variables in the two
independent replicas \cite{KawLi97,KawLi97a} 
\begin{eqnarray}
q \; = \; \frac {1}{N_p}\sum_p \kappa _p ^{(1)}
\kappa _p ^{(2)} \; ,
\end{eqnarray}
where upper indices 1 and 2 denote two replicas.
In terms of this chiral overlap
the Binder ratio of the chirality is calculated by
\begin{eqnarray}
g_{{\rm CG}} \; = \; \frac{1}{2}
\left(3-\frac{[ <q^4>]_J}{[<q^2>]_J^2}\right).
\end{eqnarray}
Here $g_{CG}$ is normalized
so that in the thermodynamic limit
it tends to zero above the chiral-glass
transition temperature, $T_{CG}$, and
tends to unity below $T_{CG}$ provided
the ground state is non-degenerate. At the chiral-glass
transition point,
curves of $g_{CG}$ against $T$
for different system sizes $l$ should intersect asymptotically.
It should be noted that one can use not only
the standard Binder function $g$ but also so called A and G function
\cite{Guerra96,Parisi98,Bokil99} to
study the nature of ordering in frustrated systems.

The chiral-glass susceptibility, which is expected to diverge at the
chiral-glass transition point, is given by
\begin{eqnarray}
\chi _{CG}=N_p[<q^2>]_J \; .
\label{sus_CG_defin}
\end{eqnarray}

In the framework of the one-parametric scaling theory the correlation 
length $\xi$ 
diverges as one approaches to the chiral glass transition temperature
by a power law 
\begin{equation}
\xi \; \; \sim \; \; |T - T_{CG}|^{-\nu_{CG}} \; .
\end{equation}
The divergence of the nonlinear and the chiral glass susceptibilities near
$T_{CG}$ is characterized by critical exponent $\gamma_2$ and $\gamma_{CG}$
\begin{eqnarray}
\chi_2 \; \; &\sim& \; \; |T - T_{CG}|^{-\gamma_2} \; , \nonumber\\
\chi_{CG} \; \; &\sim& \; \; |T - T_{CG}|^{-\gamma_{CG}} \; .
\label{suscgeq}
\end{eqnarray}

For the finite size scaling analysis we introduce the  scaling
functions \cite{Binder,KawLi97,KawLi97a}
\begin{eqnarray}
g_{CG} \; \equiv \; g_{CG}
(l^{1/\nu _{CG}}\mid T-T_{ CG}\mid )
\label{scal_BD_CG}
\end{eqnarray}
for the Binder parameter,
\begin{eqnarray}
\chi _2=L^{\gamma_2/\nu _{CG}}
 {\tilde \chi }_2
(l^{1/\nu _{CG}}\mid T-T_{CG}
\mid ),
\label{scal_nlsus}
\end{eqnarray}
for the nonlinear susceptibility and
\begin{eqnarray}
\chi _{CG}=L^{2-\eta _{CG}}
 {\tilde \chi }_{CG}
(l^{1/\nu _{CG}}\mid T-T_{CG}
\mid )
\label{scal_sus_CG}
\end{eqnarray}
for the chiral glass susceptibility. The critical
exponents can be determined from the condition that the
scaling functions, plotted versus argument $l^{1/\nu _{CG}}\mid T-T_{CG}\mid$,
should not depend on system size $l$ \cite{Barber83}. 
In the other words, they must
be collapsed onto a single curve.
Exponent $\eta _{CG}$ determining
the decay of the chiral correlation function at $T = T_{CG}$
($[<\kappa_p(0) \kappa_p(r)>]_J \sim r^{-D+2-\eta_{CG}}$, where $D$ is
spatial dimensionality) and exponent $\gamma_{CG}$ are related by
a simple scaling law
\begin{equation}
\gamma_{CG} \; = \; (2 - \eta_{CG})\nu_{CG} \; .
\label{sclawgamma}
\end{equation}
In the vicinity of the transition point the Edwards-Anderson order parameter
behaves as
\begin{equation}
q_{CG}^{EA} \; \sim \; \mid T - T_{CG} \mid ^{\beta_{CG}} \, .
\end{equation}
Exponent $\beta_{CG}$ is expressed via the other exponents by
\begin{equation} 
2\beta_{CG} \; = \; D\nu_{CG} - \gamma_{CG} \; .
\label{sclawbeta}
\end{equation} 

It should be noted that the chiral glass phase in the multi-loop model
(\ref{H_multiloop}) without screening has been studied 
\cite{Kawamura95a,KawLi2001}
by Monte Carlo simulations. It was shown that thermal fluctuations destroy the
three-dimensional spin glass ordering but leaving the chiral glass ordering 
to be stable at nonzero temperature. Although chiralities and Ising spins
have the same symmetry, the chiral and spin glass phases may belong to different
universality classes. Furthermore, in contrast to spin glasses,
the chiral glass phase is likely to exhibit a 
one-step-like peculiar replica symmetry breaking \cite{KawLi2001}. Similar
behavior was recently observed in the chiral glass state of the 
three-dimensional Heisenberg systems \cite{Hukushima00}.
The one-step replica symmetry breaking was also reported to take place in the
random field model \cite{Korshunov93} and fragile glasses \cite{Parisi97}.

In this review we focus on the chiral glass phase in the 
model (\ref{H_multiloop}) with screening  which captures the PME in ceramic
superconductors. Moreover, the screening effect could be substantial in 
intergranular ordering of these materials since the length unit to be
compared with the penetration depth is the grain size ($\sim 1\mu$m) rather
than the short coherence length of the Cooper pair. As the screening effect
makes the otherwise long-ranged interaction between vortices short ranged
and destabilizes the vortex glass (or gauge glass) phase of type-II
superconductors in a field \cite{Bokil95,Wengel96},
one wonders if it would eventually wash out a sharp phase transition
to the chiral glass phase.

% FIGURE 27
\begin{figure}
\epsfxsize=3.2in
\centerline{\epsffile{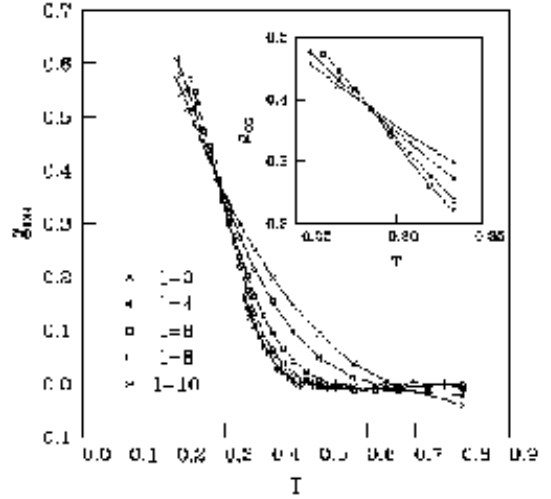}}
\vspace{0.2in}
\caption{
The temperature and size dependence of the Binder
ratio of the chirality, $g_{CG}$, for $\tilde{ L}=1$.
Inset is a magnified view around the
transition temperature $T_{CG}\simeq 0.29$. After Kawamura and Li
\protect\cite{KawLi97}.
}
\label{Binder_CG}
\end{figure}

Because of a rugged energy landscape
the multi-loop model (\ref{H_multiloop}) is very hard to equilibrate. This
difficulty can be partially overcomed by using, e.g., the Monte Carlo 
exchange method 
proposed by Hukushima and Nemoto \cite{Hukushima96}. The work of these authors
was influenced by the simulated tempering method \cite{Marinari92}, which
in turn can be understood as a special case of the method of expanded 
ensembles \cite{Luybartsev92}. In the  
 extended ensemble method of
Ref. \cite{Hukushima96}
one simulates the sample
with a given bond realization
at $N_T$ distinct temperatures at a time
distributed in the range $[T_{min}, T_{max}]$ and
the whole
configurations at two neighboring temperatures
are occasionally exchanged.
Monte Carlo updating
consists of the two parts: The first part is the standard
local Metropolis updating
at each temperature, and the second part is an exchange of the whole
lattices at two neighboring temperatures. Parameters $N_T$, $T_{min}$ and
$T_{max}$ should be chosen in such a way that each replica could  wander
over the whole temperature region and the rate of exchanges between
two neighboring temperatures would be of order of 0.5
and nearly constant \cite{Hukushima96}.

We present mainly the results obtained for the inductance
$\tilde{L} = 1$ by the replica exchange Monte Carlo 
\cite{Hukushima96} simulations. Fig. \ref{Binder_CG} displays the size 
and temperature dependence of the
Binder ratio of the chirality,
$g_{{\rm CG}}$.
The data of $g_{CG}$ for
$l=3,4,6,8$ all cross at almost the same
temperature $T \approx0.29$, strongly
suggesting the occurrence of a  finite-temperature
chiral-glass transition at $T_{CG} \approx 0.29$ (temperature $T$ is
measured in units of $J$). In particular,
the data below $T_{CG}$ show
a rather clear fan out.  The determined  value of $T_{CG}$ is slightly lower 
than the
corresponding chiral-glass transition temperature of the
pure $\pm J$  XY spin glass
determined in Ref. \cite{Kawamura95a}, $T_{CG}=0.32\pm 0.01$ (an estimate in
\cite{KawLi2001} yields a higher value for $T_{CG}$).
Note that
the XY spin-glass model corresponds to the $\tilde{L}\rightarrow 0$
limit of the present model.
The observed suppression of $T_{CG}$ by the screening effect
seems reasonable,
since the latter
makes the long-ranged interaction between
chiralities short-ranged making
the  chiral-glass transition less favorable.

%%%%%%%%%%%%%%%
\end{multicols}

\noindent TABLE 3. Critical exponents of three-dimensional (3D) Ising
spin glass (SG) and
of 3D XY chiral glass (CG) with and without screening. The standard scaling
relations have been used to obtain the full set of exponents from the values
reported in the original references.

\vspace{0.1cm}

\begin{center}
\begin{tabular}{l|l|l|lllr}\hline\hline
type   & distribution  & Ref. & $\; \; \beta \; \; $ & $\; \; \gamma \; \;$ & $\; \; \nu \; \; $ & $\; \; \eta \; \;$ \\ \hline
3D Ising SG   & $\pm J$  & \cite{Kawashima96} & $\; \; \approx 0.55 \; \; $ & $\; \; \approx 4.0 \; \; $ & $ \; \; 1.7(3) \; \; $ & $ \; \; -0.35(5) \; \;$ \\
3D Ising SG   & $\pm J$  & \cite{Palassini99} & $\; \; \approx 0.65 \; \; $ & $\; \; 4.1(5) \; \; $ & $ \; \; 1.8(2) \; \; $ & $ \; \; -0.26(4) \; \;$ \\ 
3D Ising SG   & Gaussian  & \cite{Marinari98} & $\; \; \approx 0.64 \; \; $ & $\; \; \approx 4.7 \; \; $ & $ \; \; 2.0(1.5) \; \; $ & $ \; \; -0.36(6) \; \;$ \\ \hline
3D XY CG, $\tilde{L}=0$   & $\pm J$  & \cite{Kawamura95a} & $\; \; \approx 0.45 \; \; $ & $\; \; \approx 3.6 \; \; $ & $ \; \; 1.5(0.3) \; \; $ & $ \; \; -0.4(2) \; \;$ \\ 
3D XY CG, $\tilde{L}=0$   & $\pm J$  & \cite{KawLi2001} & $\; \; \approx 0.69 \; \; $ & $\; \; \approx 2.2 \; \; $ & $ \; \; 1.2(0.2) \; \; $ & $ \; \; 0.15(20) \; \;$ \\
3D XY CG, $\tilde{L}=1$   & $\pm J$  & \cite{KawLi97} & $\; \; \approx 0.5 \; \; $ & $\; \; \approx 2.9 \; \; $ & $ \; \; 1.3(0.2) \; \; $ & $ \; \; -0.2(2) \; \;$ \\ \hline \hline

\end{tabular}
\end{center}

\vspace{0.2cm}

\begin{multicols}{2}

% FIGURE 28
\begin{figure}
\epsfxsize=3.2in
\centerline{\epsffile{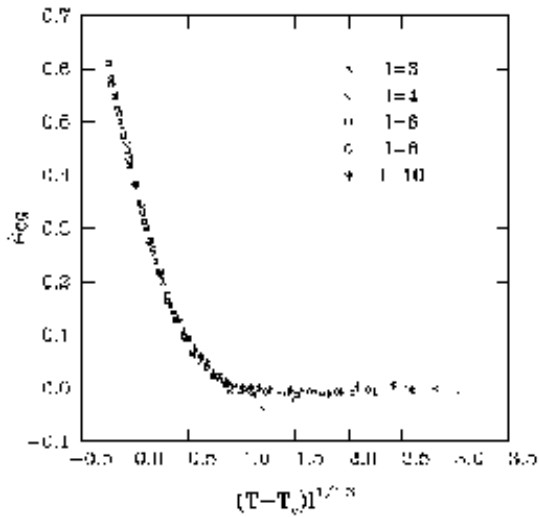}}
\vspace{0.2in}
\caption{Finite-size scaling plot
of $\tilde{g}_{CG}$ with $T_{CG}=0.286$ and $\nu _{CG}=1.3$.
After Kawamura and Li
\protect\cite{KawLi97}
}
\label{Binder_CG_fit}
\end{figure}

Fig. \ref{Binder_CG_fit} shows the scaling function
$\tilde{g}_{CG}(l^{1/\nu _{CG}}\mid T-T_{CG}\mid )$ given by
Eq. (\ref{scal_BD_CG}) with fixing $T_{CG} \approx 0.29$.
The best scaling fit gives the correlation length exponent
$\nu _{CG}=1.3\pm 0.2$ \cite{KawLi97}.

\begin{figure}
\epsfxsize=3.2in
\centerline{\epsffile{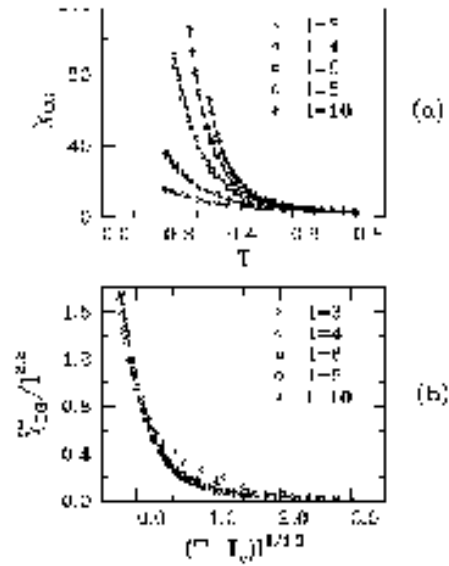}}
\vspace{0.2in}
\caption{(a) The temperature and size dependence of the
chiral-glass susceptibility, $\chi_{CG}$,
for $\tilde{L}=1$.
(b) Finite-size scaling plot
of $\tilde{\chi}_{CG}$ with $T_{CG}=0.286$,
$\nu _{CG}=1.3$ and $\eta _{CG}=-0.2$.
After Ref. 
\protect\cite{KawLi97a}
}
\label{sus_CG_fig}
\end{figure}

The temperature and size dependence of
the chiral-glass
susceptibility, $\chi _{CG}$, defined by Eq. (\ref{sus_CG_defin}),
are shown in Fig. \ref{sus_CG_fig}a.
Finite-size scaling analysis based on the relation (\ref{scal_sus_CG})
is made with fixing $T_{{\rm CG}}=0.286$ and $\nu _{{\rm CG}}=1.3$,
yielding the chiral critical-point-decay
exponent $\eta _{{\rm CG}}=-0.2\pm 0.2$. The resulting
finite-size-scaling plot is displayed in Fig. \ref{sus_CG_fig}b.
Other exponents can be estimated via
the standard scaling relations (\ref{sclawgamma}) and (\ref{sclawbeta})
as
$\gamma _{{\rm CG}}\simeq 2.9$ and $\beta _{{\rm CG}}\simeq 0.5$
\cite{KawLi97a}.

% FIGURE 29

As one can see from Table 3,  for $\tilde{L} = 1$ the exponents
 reported in Ref. \cite{KawLi97} are
comparable to those obtained in Ref. \cite{Kawamura95a}
for $\tilde{L}=0$ but they are clearly different from newer estimates
of Kawamura and Li \cite{KawLi2001}. Therefore, it is not clear whether the 
screening effect is irrelevant or not at the 3D chiral-glass transition.  
Further studies are required to clarify this point.

Within error bars the exponents of chiral glasses obtained in 
Ref. \cite{Kawamura95a}
and \cite{KawLi97} coincide with those for 3D Ising spin glasses,
suggesting that they belong to the same universality class. The results
of \cite{KawLi2001} showed that the chiral glass transition may lie in
a universality class different from the Ising glass.
Regardless to this controversy, the simulations \cite{KawLi97} gave a strong
evidence of the existence of a chiral glass as a new phase in ceramic 
superconductors.

% FIGURE 30
\begin{figure}
\epsfxsize=3.2in
\centerline{\epsffile{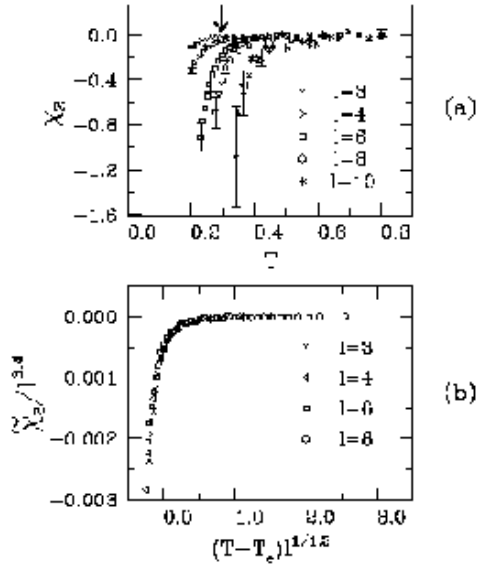}}
\vspace{0.2in}
\caption{(a) The temperature and size dependence of the
nonlinear susceptibility $\chi_2$
for $\tilde{L}=1$. An arrow represents the location of the
chiral-glass transition point.
(b) Finite-size scaling plot
of $\tilde{\chi}_2$ with $T_{CG}=0.286$,
$\nu _{CG}=1.3$ and $\gamma _2=4.4$.
After Kawamura and Li 
\protect\cite{KawLi97a}
}
\label{nlsus_CG_fig}
\end{figure}

From the experimental point of view the nonlinear susceptibility $\chi_2$ plays
a more important role than the chiral glass susceptibility $\chi_{CG}$
(\ref{suscgeq}) because experiments can probe it directly but not $\chi_{CG}$.
On general theoretical grounds,
$\chi _2$
is expected to show a
negative divergence at the transition point where
the time-reversal symmetry is spontaneously broken in a
spatially random manner \cite{Kawamura95}. The results shown in 
Fig.\ref{nlsus_CG_fig}a 
are fully consistent with
this expectation. The exponent associated with the
negative divergence
is estimated via a finite-size scaling analysis 
(see Fig.\ref{nlsus_CG_fig}b) is equal to $\gamma _2\simeq 4.4$ assuming
$T_{{\rm CG}}=0.286$ and $\nu _{{\rm CG}}$=1.3.
This value of $\gamma _2$ is 
larger than
the chiral-glass susceptibility exponent, $\gamma _{{\rm CG}}
\simeq 2.9$.
At present, it is not entirely clear
whether this  deviation reflects a true difference in the
asymptotic critical behavior.

\begin{figure}
\epsfxsize=3.2in
\centerline{\epsffile{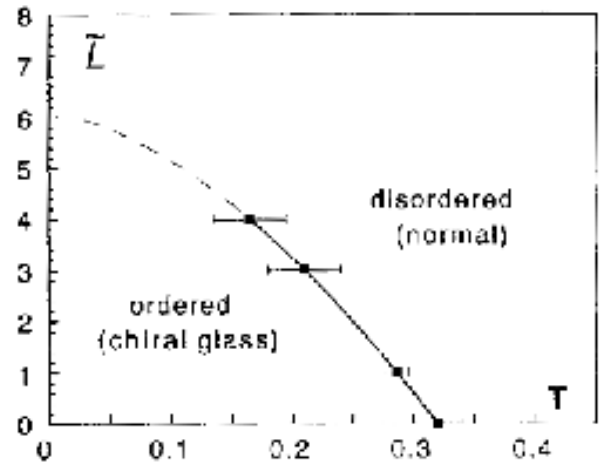}}
\vspace{0.2in}
\caption{A phase diagram in the $T$-$\tilde{L}$ plane.
Renormalized inductance $\tilde{L}$ is defined by Eq.(\ref{L_dimless}).
After Kawamura and Li
\protect\cite{KawLi97}
}
\label{diagram_CG}
\end{figure}

The tendency that the chiral-glass ordering is suppressed at
larger inductances was seen from simulations \cite{KawLi97a}.
The phase diagram in the $T-\tilde {L}$
plane is sketched in Fig.\ref{diagram_CG}.
There appears to be a finite critical value of the inductance,
$\tilde {L}_c$,
above which there is no equilibrium chiral-glass transition.
Although
it is difficult to precisely locate $\tilde {L}_c$
due to the
extremely slow relaxations at low temperatures,
it appears to lie around $5 \leq \tilde{L}_c \leq 7$.
It was also shown \cite{KawLi97a} that the paramagnetic tendency is more 
enhanced
for larger $\tilde{L}$, while the chiral-glass ordering
itself is suppressed
for larger $\tilde{L}$.

An attempt to observe the flux glass (or orbital glass) was made in 
Ref. \cite{KawLi97a} by the Monte Carlo simulations.
Naively, one expects that the flux defined by
Eq. (\ref{H_multiloop}),
should behave in the same way as the chirality, since it is
also a pseudoscalar variable sharing the same symmetry
as the chirality. Indeed,
the flux-glass susceptibility, e.g.,
shows a divergent behavior \cite{KawLi97a}  similar to its chiral glass 
counterpart $\chi_{CG}$.
However, in contrast to the naive expectation,
clear crossing of the Binder ratio
as observed in $g_{CG}$
is not observed in the flux glass case at least
in the range of lattice sizes $l \le 10$.  
Rather, the ordering tendency seems
more enhanced in the sense that
the flux glass susceptibility $g_{{\rm FG}}$ tends to increase 
with increasing $L$
exhibiting a feature of the ordered phase
even above
$T_{{\rm CG}}\approx 0.29 $ \cite{KawLi97a}.
As the flux is an
{\em induced\/} quantity generated by the finite inductance effect,
we believe this behavior to be a finite-size effect.
Presumably, for inductance $\tilde{L} = 1$
the flux hardly reaches its asymptotic critical behavior
in rather small lattices studied there. We tend to believe that the flux
or orbital glass can occur in the multi-loop model (\ref{H_multiloop})
of larger system sizes and it should accompany with the PME.

% FIGURE 31

\subsection{Differences between chiral glass and other glassy phases in 
         disordered type II superconductors}

The point quenched randomness destabilizes the Abrikosov flux line lattice in
a pure type-II superconductors, yielding novel
glassy phases such as Bragg, gauge and
vortex glass (the so called Bose glass \cite{Tauber97}
caused by columnar or planar defects will not be discussed here).
The aim of the present chapter is 
to make a clear distinction 
between these phases and the chiral glass. We consider the weak and strong
disorder cases separately.

% FIGURE 32
\begin{figure}
\epsfxsize=3.2in
\centerline{\epsffile{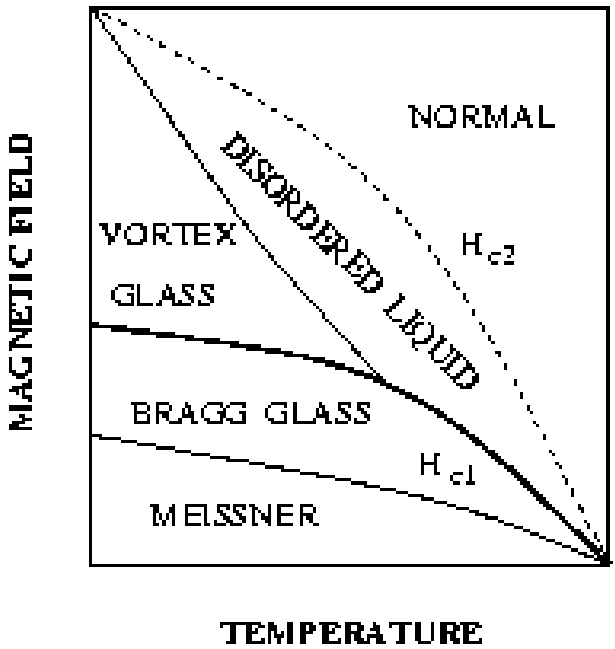}}
\vspace{0.2in}
\caption{A phase diagram of a weakly
 disordered type-II superconductor. In three dimensions the vortex glass
phase should disappear in the presence of the screening.
}
\label{diagram_VG}
\end{figure}

\noindent {\em Bragg and vortex glass}. The starting point to describe the
Bragg \cite{Giamarchi94,Giamarchi95,Giamarchi97} and vortex glass 
\cite{Fisher89} is the 
Ginzburg-Landau free energy (\ref{FreeGL}) with {\em a weak disorder}. 
The later can be introduced via the mean field transition temperature
$T_{c0}$ in the expression of $\alpha (T)$ in Eq. (\ref{FreeGL}), i.e,
$T_{c0} \rightarrow T_{c0} + \delta T_{c0}(\vec{r})$, where 
$\delta T_{c0}(\vec{r})$ is a random quantity. 
In the absence of an external magnetic field and disorder
the model  (\ref{FreeGL})
may be mapped to the uniform XY model with the specific heat exponent
$\alpha < 0$ \cite{Guillou80,Lipa96}.
Then, according to the Harris criterion \cite{Harris74}, the weak randomness
does not change the zero-field critical behavior of the type-II
superconductors.
 
The situation becomes quite different if the external field is applied
to a sample. In this case the question about the $H-T$ phase diagram is 
still under debate
but its currently popular view 
\cite{Giamarchi97,Nattermann00,Kierfeld00,Khaykovich97}
is shown in Fig. \ref{diagram_VG}. As shown first by Larkin \cite{Larkin70},
disorder spoils the translational long-range order in the Abrikosov lattice 
since the randomness in the local
values of the critical temperature
 leads to a random potential acting on the vortices.
Although the disordered averaged order parameter $[<\exp(i\vec{Q}.\vec{u})>]$
($\vec{Q}$ and $\vec{u}$ denotes a reciprocal lattice vector of the Abrikosov
lattice and the vortex displacement, respectively) vanishes in the thermodynamic
limit, the correlation function
\begin{equation}
S(\vec{Q},\vec{r}) \; = \; [<\exp\{i\vec{Q}.[\vec{u}(\vec{r}) - u(0)]\}>]
\label{BG_OP}
\end{equation}
may still obey an algebraic decay \cite{Giamarchi94,Nattermann00}
 provided the external field is weak enough. Since this algebraic decay should
show up in Bragg peaks  the corresponding phase is called
Bragg glass \cite{Giamarchi95} or
elastic vortex glass \cite{Fisher91}). So the Bragg glass
is characterized by translational correlations
(\ref{BG_OP}) decay asymptotically as power laws. It reflects some
residual order of the Abrikosov lattice. After the Bragg glass was first
proposed \cite{Giamarchi95} (one should be noted that the algebraic decay
of $S(\vec{Q},\vec{r})$ of three-dimensional vortex lattices in impure superconductors 
was first predicted by Nattermann \cite{Nattermann90}) its existence was
supported by further analytical \cite{Carpentier96,Kierfeld97,Fisher97}
and numerical calculations \cite{Gingras96,Otterlo98}.
 The strong experimental evidence
for the existence of the Bragg glass phase was provided by the neutron
diffraction data of Klein {\em et al} \cite{Klein01}. 

On increasing external field
the Bragg glass becomes unstable to the vortex glass phase in which 
translational correlations (\ref{BG_OP}) decay exponentially.
In the vortex glass phase, although there is no off-diagonal long-range order
one may expect a nonzero Edwards-Anderson order parameter 
\cite{Fisher89}
\begin{equation}
q_{VG}^{EA} = \frac{1}{V}\int d\vec{r} |\tilde{\Psi} (\vec{r})|^2 \, ,
\label{VG_OP}
\end{equation}
where gauge-invariant
$\tilde{\Psi} (\vec{r}) = \Psi (\vec{r})\exp[\int^x i\vec{A}.d\vec{l}]$
and $V$ is the volume of the system.
This means that the vortices in this state are completely frozen at random
positions dictated  by the disorder as spins or chiralities in the spin
or chiral glass phase. The pinning of vortex lines prevents the system from
the energy dissipation and the linear resistivity should, therefore,
vanish in the vortex glass phase.
Interesting transport properties at and below the critical temperature were
intensively discussed by Fisher {\em et al} \cite{Fisher91} by scaling
arguments. Their predictions were confirmed by measurements of the $I-V$
characteristics \cite{Koch89,Gammel91}. The existence
of the vortex glass phase was also supported by simulations \cite{Otterlo98}.
The evidence of the Bragg glass-vortex glass transition was provided by
experiments \cite{Deligiannis97,Safar95} and simulations \cite{Otterlo98}.
This transition appears to be very sensitive to the disorder strength
\cite{Safar95}.

Recently, it has been proposed that low-$T_c$ materials would be better
captured by a phase diagram different from what proposed in 
Fig. \ref{diagram_VG} \cite{Menon02}. Namely, instead of the vortex glass phase
the so called multi-domain glass would intervene between Bragg glass and
disordered liquid phase. Translational correlations in this new phase are
argued to exhibit a significant degree of short-range order.

Nonomura and Hu have found the
vortex slush phase \cite{Worthington92}
to exist between the vortex glass and disordered liquid
phases by simulations \cite{Nonomura01}. Such a new phase was not detected
by similar simulations of Olsson and Teitel \cite{Olsson01}
but it seems to be supported by the resistivity  
experiments \cite{Nishizaki00,Wen01} and by the analytical argument
\cite{Ikeda01}. Further studies are required to to confirm
or reject the existence of the multi-domain glass and vortex slush phases. \\

\noindent {\em Gauge glass}. So far the disorder was assumed to be weak. 
In the case of {\em strong disorder} gauge glass like models were proposed
to describe granular superconductors \cite{Ebner85,John86}. Assuming that each
superconducting grain $i$ is described by the phase $\phi _i$ of the order
parameter and the Josephson couplings $J_{ij}$ are positive and the 
same for all of grain pairs ($J_{ij}=J > 0$),
the Hamiltonian is given by \cite{Wengel96}
\begin{eqnarray}
{\cal H} \; = \; -J \sum_{<ij>} \cos (\phi _i - \phi _j - A_{ij} -
\lambda_0^{-1} a_{ij} ) + \nonumber\\
 \frac{1}{2} \sum_{P} (\vec{\nabla}\times \vec{a})^2 \, .
\label{H_GG}
\end{eqnarray}
Here $\sum_{<ij>}$ and $\sum_{P}$ denotes the summation over nearest neighbours
and over plaquettes of a cubic lattice, respectively. The influence of the external field and randomness is accounted 
by gauges $A_{ij}$ which are taken to be independent random variables with
a distribution between 0 and $2\pi$. 
$a_{ij} = \int_{\vec{r}_i}^{\vec{r}_j} \vec{a}(\vec{r}) d\vec{r}$, where 
$\vec{a}$ denotes the fluctuations of the vector potential which are limited by
the bare screening length $\lambda_0$.

The gauge glass phase is characterized by a vanishing averaged condensate
wave function $[<\Psi >] = 0$, but a finite Edwards-Anderson order
parameter $[|<\Psi >|^2]$. The transition to this phase is accompanied
with a diverging gauge-glass susceptibility defined in Ref. \cite{Huse90}.

In the absence of the screening ($\lambda _0 \rightarrow \infty$) the gauge
glass was shown to be ordered 
at finite temperatures in three dimensions, while in two dimensions the
ordering takes place only at $T=0$
\cite{Gingras92,Cieplak92,Fisher91a,Reger91,Hyman95,Kosterlitz98,Maucourt98,Olson00}. The analysis based on
the determination of the scaling behavior of
the domain wall energy ${\cal E}_{DW}(l) \propto l^{\theta}$ gave
exponent $\theta (3D) \approx 0$ 
\cite{Reger91,Fisher91a,Maucourt98}, indicating that the 3D gauge glass is
more marginal than the Ising spin glass which have $\theta (3D) \approx 0.2$. 
A detailed discussion of the relation between the vortex glass and the gauge
glasses was given by Blatter {\em et al} \cite{Blatter94}. 
The vortex glass occurs in the model with weak disorder but the disorder
becomes effectively strong as the field is increased \cite{Menon94}.
If one assumes (\ref{VG_OP}) as a definition of the vortex glass order parameter
then the gauge glass would be a vortex glass.

In ceramic superconductors the length unit associated with the intergranular
ordering is the mean grain size which is of order micron, screening effect is
generally non-negligible.
The influence of screening (finite $\lambda _0$)
on the nature of ordering of the gauge glass was
considered by several groups \cite{Bokil95,Wengel96,Kisker98}. It turns out
that screening destroys the gauge glass transition in three dimensions.
This conclusion remains unchanged if one
includes the effect of anisotropy into the gauge glass model 
\cite{Kawamura00,Pfeiffer99} by assuming an extra contribution to 
$A_{ij}$ arising from the external field.

Contrary to the Bragg and vortex glass, the gauge and chiral glass may occur
in systems with strong disorder. Comparing Hamiltonian (\ref{H_multiloop})
for the chiral glass ($\Phi_p^{ext}=0$) and (\ref{H_GG}) for the gauge glass
one can see two main
differences. First, the former is supposed to appear in the zero external 
magnetic field, while the latter cannot exist without the field which
serves as a source of frustration. Second, opposed to
the gauge glass case where the couplings are always "ferromagnetic",
the couplings in the chiral glass model may be positive or negative.
This means that the chiral glass may be observed only in a system with
anisotropic pairing. Since the $U(1)$ gauge symmetry is not broken,
in this phase, unlike in the vortex or gauge glass, the phase of
the condensate is {\em not} ordered, even randomly, on sufficient long length
and time scales: The thermodynamic ordering occurs only in the 
loop-supercurrents, or in the chiralities. Therefore, the chiral glass is
not thermodynamically superconducting state.
The modest screening destroys the gauge glass in three dimensions but leaving
the chiral glass stable against thermal fluctuations.

\subsection{Experimental search for the chiral glass phase}

First we will  discuss  some requirements for the appropriate
ceramic samples where one could expect the chiral-glass phase.
One important parameter characterizing the sample is
the dimensionless inductance, $\tilde{L}$, given by Eq.(\ref{L_dimless}).
From Fig. \ref{diagram_CG} it follows that
an equilibrium chiral-glass state could be realized in the
samples with
not large $\tilde{L}$.
If one models a loop as
a cylinder of radius $r$ and height $h$, its inductance is given by
$L=4\pi ^2r^2/h$.
Putting $r\sim 1\mu $m,
$h/r\sim 0.01$ and $J\sim 20$K
(these values are chosen to mimic the sample used in Ref.\cite{Matsuura1}),
one gets $\tilde{L}\sim 10^{-2}$.
Since this value is considerably smaller than $\tilde{L}_c$,
an equilibrium chiral-glass phase may well occur
in such samples. By contrast, if the sample
has too large a grain size or too strong Josephson coupling,
an equilibrium chiral-glass phase will  not be realized, or at
least largely
suppressed.
Another requirement for the sample is
that the grains must be connected via weak links into an
infinite cluster,
not decomposed into finite clusters. This means that the concentration
of $\pi$ junctions should exceed the percolation threshold.
The finite-cluster
samples cannot exhibit a chiral-glass transition, although
the paramagnetic Meissner effect is still possible 
\cite{Sigrist92,Magnusson95a,Magnusson95b}.

Once appropriate samples could be prepared, the chiral-glass
transition is detectable in principle via the standard
magnetic measurements by looking for a negative divergence of
nonlinear susceptibility $\chi _2$ \cite{KawLi97} as in the case of
spin glasses.
In such magnetic measurements, care has to be taken to keep the
intensity of applied ac and dc fields weak enough, typically much
less than 1G, so that the external flux per loop is sufficiently
smaller than $\Phi _0$.

A sharp negatively-divergent
anomaly of $\chi _2$ was reported in a YB$_2$C$_4$O$_8$ ceramic
sample by the ac method by Matsuura {\em et al} \cite{Matsuura1}
which might be a signal of the chiral-glass
transition. This study was extended by Deguchi {\em et al} \cite{Deguchi02}
to the case when the pressure is applied to a sample
to vary the Josephson couplings.
The transition temperature $T_{CG}$ was found to be shifted towards higher
temperatures as the pressure increases but the character of the transition
to the chiral glass state remains unchanged.
This observation is consistent with the theoretical prediction
\cite{KawLi97,KawLi97a} that $T_{CG}$ is determined entirely by the strength
of the Josephson interaction which should grow with the pressure. 

The opposite conclusion from  the susceptibility data
 was reported by Ishida {\em et al}
\cite{Ishida} who showed that
the nonlinear susceptibility of
a (Sr$_{0.7}$Ca$_{0.3}$)$_{0.95}$CuO$_{2-x}$ sample negatively
diverges but it oscillates as a function of temperature near the critical
point. As argued by these authors, such a behavior can be
explained by the Bean critical state model but not by the chiral glass one.

Recently, Papadopoulou {\em et al} \cite{Evie_thes}
has also observed the negative divergence of $\chi _2$ with exponent
$\gamma _2 \approx 3.4$ for a melt-cast Bi$_2$Sr$_2$CaCu$_2$O$_8$ sample.
This value of $\gamma _2$ is smaller that the theoretical value
$\gamma _2 \approx 4.4$ but close to the chiral glass susceptibility
exponent $\gamma_{CG} \approx 2.9$ \cite{KawLi97,KawLi97a}. The temperature
region to observe the divergence of $\chi _2$ is, however, too narrow that
the result of Papadopoulou {\em et al} \cite{Evie_thes} cannot be
considered as a good evidence for the existence of the chiral glass phase.

As in the case of spin glasses,  measurements
of dynamic susceptibilities such as $\chi' (\omega )$ and
$\chi'' (\omega )$ would also give useful information on the
possible chiral-glass ordering, particularly
when combined with the dynamic scaling analysis. 
Near the  chiral-glass transition point,
the imaginary part of the linear susceptibility,
$\chi'' (\omega )$, is expected to satisfy
the  dynamic scaling relation of the form,
\begin{eqnarray}
\chi'' (\omega ,T,H)\approx \omega ^{ \beta _{{\rm CG}}
/z_{{\rm CG}}\nu_{{\rm CG}} } \bar \chi'' (
\frac { \omega }{t^{z_{{\rm CG}}\nu _{{\rm CG}} } },\
\frac {H^2 }
{t^{ \beta _{{\rm CG}} + \gamma _{{\rm CG}} } } ),
\label{scaling_chi}
\end{eqnarray}
where $t\equiv \mid (T-T_{{\rm CG}})/T_{{\rm CG}}\mid $
and $z_{{\rm CG}}$  is a dynamical chiral-glass exponent. From
Ref. \cite{KawLi97} (see also Table 3), we get the static chiral-glass
exponents to be $\nu _{{\rm CG}}\simeq 1.3$,
$\beta _{{\rm CG}}\simeq 0.5$ and $\beta _{{\rm CG}}+
\gamma _{{\rm CG}}\simeq 3.4$.
Using the dynamical exponent $z_{CG} = 6.3 \pm 1.7$ \cite{Kawamura00a}
we obtain
take a value around $z_{{\rm CG}}\nu _{{\rm CG}} \approx 8.2$.

A dynamic scaling analysis was made by
Leylekian, Ocio and Hammann
for LSCO ceramic samples \cite{Leylekian91,Leylekian94}. These authors performed
both the ac susceptibility and the noise measurements, and
found an
intergranular cooperative transition  even in
zero field at a
temperature about 10\% below the superconducting
transition temperature of the grain.
Note that the noise measurements enable one to
probe truly zero-field phenomena where
one can be free from the extrinsic pinning effects
such as the ones envisaged in the so-called critical-state
model \cite{Bean64}.
It was then found that
the data of $\chi''$ satisfied the
dynamic scaling relation (\ref{scaling_chi}).
Here note that one is {\em not\/} allowed to invoke
the standard vortex-glass scenario to
explain such intergranular cooperative transition
{\em in zero field\/}, since in the standard vortex-glass picture
frustration is possible only under finite external fields.
By contrast, the experiment seems
consistent with the chiral-glass picture.

Meanwhile, when the intragranular superconducting transition
and the
intergranular transition  take
place at mutually close
temperatures as in Ref. \cite{Leylekian91,Leylekian94},
the Josephson coupling, $J$, which has been assumed
to be temperature independent in the present model, is
actually strongly temperature dependent
in the transition region. In such a case, care has to be taken
in analyzing the experimental data, since the temperature
dependence of $J$ might modify the apparent
exponent value from the true asymptotic value
to some {\em effective value\/}.
In fact, the dynamical exponent $z\nu \simeq 30$
determined by Leylekian {\em
et al\/} were  different from the
standard spin-glass value,
which might be due to
the proximity effect of the intragranular
superconducting transition \cite{Leylekian91,Leylekian94}.
If, on the other hand,
the temperature dependence of $J$ was taken into account
in the fit, a
more realistic value $z\nu \simeq 10-15$ was obtained in Ref.
\cite{Leylekian91,Leylekian94}. As claimed by Leylekian {\em et al}
they data are compatible with the transition to the gauge glass state.
Using $\nu_{GG}= 1.3 \pm 0.04$ and $z_{GG} = 4.7 \pm 0.7$ from Ref.
\cite{Reger91}, we have
$z_{GG}\nu_{GG} \approx 6.2$.
The experimental value seems to be  more consistent with the
theoretical estimate $z_{CG}\nu_{CG} \approx 8.2$ for the chiral glass than
with the gauge glass (or vortex glass) counterpart. In view of large error
bars, such a conclusion remains , however, ambiguous.

\begin{figure}
\epsfxsize=2.6in
\centerline{\epsffile{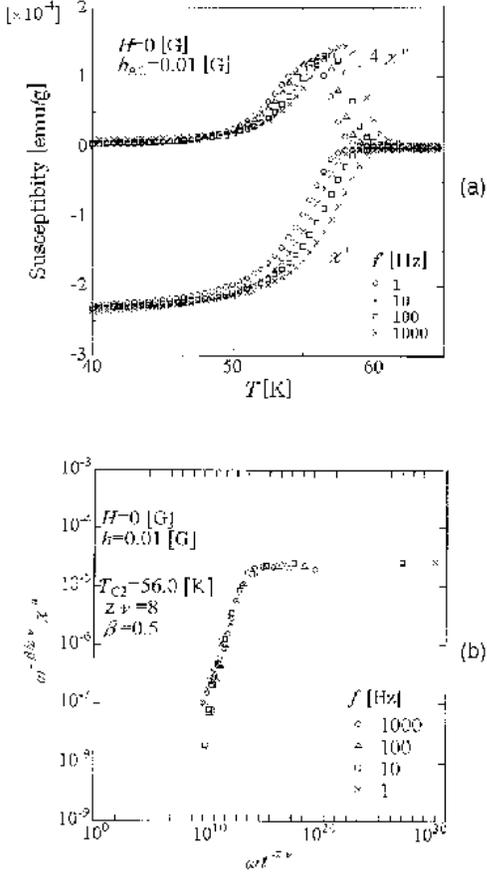}}
\vspace{0.2in}
\caption{(a) Temperature dependence of the linear susceptibility for different
values of frequency indicated next to the curves.
(b) The scaling plot for $\chi ''$ which gives $z\nu = 8$ and $\beta = 0.5$.
 After Deguchi {\em et al} \protect\cite{Deguchi02a}.
}
\label{Deguchi_exp}
\end{figure}

The most convincing evidence for the chiral glass phase from the
dynamical scaling analysis of susceptibility was reported by 
Deguchi {\em et al} for YBCO ceramic \cite{Deguchi02a}. Their results 
shown in Fig. \ref{Deguchi_exp} give $z_{GG}\nu_{GG} = 8.2$ and $\beta _{CG} = 0.5$ which are 
well accounted by the chiral glass model.

Experimental data from the out-of-phase component $\chi''(\omega)$ of the ac 
susceptibility and the scaling analysis may be employed to obtain the 
relaxation time which diverges on approaching $T_{CG}$. In this case, a 
"freezing" temperature $T_f$, can be defined as a frequency-dependent
temperature $T_f(\omega)$ at which the maximum relaxation time is $1/\omega$.
If there is a phase transition at $T_{CG}$, the relaxation time 
$\tau = 1/\omega$ behaves as
\begin{equation}
\frac{\tau}{\tau_0} = 
\left| \frac{T_f(\omega) - T_{CG}}{T_{CG}}\right|^{-z_{CG}\nu_{CG}} =
t^{-z_{CG}\nu_{CG}} 
\label{RT_scaling}
\end{equation}
as $\, T \rightarrow T_{CG}$.
Here $T_f(\omega)$ is chosen as the temperature at which the peak of
$\chi''(\omega)$ is located.
With the help of the scaling relation (\ref{RT_scaling}) Papadopoulou
{\em et al} have obtained $ z_{CG}\nu_{CG} \approx 11$ for the melt-cast Bi2212
sample supporting the transition to the chiral glass phase 
\cite{Papadopoulou}.
Moreover, a similar analysis on the sintered Bi2212 sample, that does not
display magnetic aging, gave unphysical results for $\tau_0$ and 
$z_{CG}\nu_{CG}$ \cite{Papadopoulou}. This result strengthens the evidence
for the existence of the chiral glass phase.

% FIGURE 33

Another way to probe the existence of the chiral glass phase is based on
transport properties \cite{Kawamura00a,Matsuura2}. As mentioned above, 
in this phase a sample should not be true superconductor, with a small but
nonvanishing linear resistivity $\rho_L$ even at and below $T_{CG}$.
Rough estimates of the residual $\rho_L$ were given in \cite{KawLi97a}.
The dynamical scaling theory for transport properties near the chiral glass
transition may be developed with the help of its vortex glass counterpart
\cite{Fisher91}. Assume that under the external current of density $j$,
there occurs a voltage drop, or electric field of intensity $E$. The voltage
drop comes from two nearly independent sources: One from the motion of integer
vortex lines, $E_v$, and the other from the motion of chiral domain walls,
$E_{\kappa}$ \cite{Kawamura00a}. Then we have
\begin{equation}
E(j,T) \, = \, E_v(j,T) + E_{\kappa}(j,T) .
\end{equation}
The first part is essentially a regular part, while the chiral part should obey
the dynamic scaling law associated with the
chiral-glass transition, {\it i.e.\/},
\begin{eqnarray}
E_\kappa \approx \mid t\mid ^{(z_{CG}+1)\nu_{CG} }\bar E_\kappa (j/
\mid t\mid ^{2\nu_{CG} }),
\label{E_scaling}
\end{eqnarray}
where he spatial dimension has been set equal to $D=3$.
Employing the Fisher-Fisher-Huse argument \cite{Fisher91} yields
the following asymptotic behaviors
of the scaling function
\begin{equation}
\bar E_\kappa (x)\approx
\left\{ \begin{array}{ll}
ax, & t>0,  \\ a'\exp[-bx^{-\mu_{CG}}], & t<0, \end{array}
\right. \ \ \ \ \ \ {\rm as}\ x\rightarrow 0,
\end{equation}
where $a$, $a'$ and $b$ are positive constants. $\mu_{CG} $ is an unknown
 exponent
describing the chirality dynamics in the chiral-glass state
but it is expected to be positive (in the vortex glass
case $0 < \mu \leq 1$).

The linear
resistivity $\rho _L$ and nonlinear resistivity $\rho _{NL}$ can be
written as a sum
of the two nearly independent contributions \cite{Kawamura00},
\begin{eqnarray}
\rho _L\equiv \left. \frac{{\rm d}E}{{\rm d}j}
\right | _{j=0}=
\rho _{L,v}+\rho _{L,\kappa} \, , \nonumber\\
\rho _{NL}\equiv \frac{1}{6} \left. \frac{{\rm d}^3E}{{\rm d}j^3}
\right | _{j=0}
=\rho _{NL,v}+\rho _{NL,\kappa}.
\end{eqnarray}
At the chiral-glass transition point, the vortex
terms
$\rho _{L,v}(T)$ and $\rho _{NL,v}(T)$ stay finite without prominent anomaly,
while the chiral parts exhibit a singular behavior associated with the 
transition. Then from Eq. (\ref{E_scaling}) we have
\begin{eqnarray}
\rho _{L,\kappa}(T) \approx
\left\{
\begin{array}{ll} 0, & t<0, \\ c't^{(z_{CG}-1)\nu_{CG} }, & t>0, \end{array}
\right. \nonumber\\
\rho _{NL,\kappa}(T)\approx
\left\{
\begin{array}{ll} 0, & t<0, \\ c''t^{(z_{CG}-5)\nu_{CG} }, & t>0.
\end{array}
\right.
\label{R_scaling}
\end{eqnarray} 
Since $(z_{CG}-1)\nu_{CG} \approx 6.9$ is a large positive number 
$\rho _{L,\kappa}(T)$ given by Eq. (\ref{R_scaling}) vanishes toward $T_{CG}$ 
sharply and the behavior of $\rho _L$ is dominated by the regular vortex term. 

As shown in the next section, Yamao {\em et al} have observed that the nonlinear
resistivity  and the nonlinear magnetic susceptibility of YB$_2$Cu$_4$O$_8$
diverge at the temperature
where the magnetic remanence
sets in \cite{Matsuura2}. Meanwhile, 
the linear resistivity remains finite 
(see Fig. \ref{v1_exp_fig} below) without any appreciable anomaly. This behavior is hard to understand
if one regards the transition to be a transition to the
Meissner or the vortex glass phase, where $\rho _L$ should vanish.
The experimental data obtained for the linear resistivity \cite{Matsuura2}
 seem to be
compatible with the chiral glass picture where
the total linear resistivity remains finite 
at $T=T_{{\rm CG}}$.

As follows from Eq. (\ref{R_scaling}), the nonlinear resistivity 
$\rho _{NL}$ shows, however,
stronger anomaly than $\rho _L$ and it becomes divergent provided
$z_{CG} < 5$.
Using the phase representation Kawamura \cite{Kawamura00a} obtained
$z_{CG}=6.3 \pm 1.7$ for the self-inductance $\tilde{L}=1$ and
it is hard to justify that $z_{CG} < 5$.
Thus, it is not clear if the nonlinear susceptibility diverges at the
chiral glass transition or not. So far as $z_{CG}$ generally depends 
on the kind of dynamics,
it may take a smaller value in the vortex representation than
in the phase representation. Another possibility is that $z_{CG}$
may depend on the self-inductance and other values of $\tilde{L}$
could lead to smaller values of $z_{CG}$. More accurate estimates of
the dynamical exponent are needed to make a full comparison with experiments.

The chiral glass phase is expected to display off-equilibrium phenomena
similar to those in spin glasses, such as aging and memory effects.
As mentioned above, recently Papadopoulou {\em et al} \cite{aging}
 have observed
the aging effect in ceramic BSCCO sample at very weak fields
(see Fig. \ref{aging_exp99} below). It may have some relation to 
the chiral glass order.

Taken together, the problem of the existence of the chiral glass phase
in ceramic materials remains open. Further theoretical and experimental studies
of off-equilibrium dynamical  and transport properties of the chiral
glass ordered state would be of much interest.

\section{Dynamical phenomena: Experiments and Simulations}
\label{sec.dynamics}

In this section we present results of experimental studies of
the  dynamical phenomena 
related to the PME. Monte Carlo and Langevin dynamics simulations show that
all of these phenomena may be captured by 
simple multi-loop model (\ref{H_multiloop}).

\subsection{AC susceptibility} 

\noindent {\em Experiments}. 
It is well known that dc susceptibility provides important information on
static magnetic properties of a investigated material whereas
ac-susceptibility may shed light on both static and dynamic behaviors and
provide the clue to the underlying mechanism for the PME. 

\begin{figure}
\epsfxsize=2.6in
\centerline{\epsffile{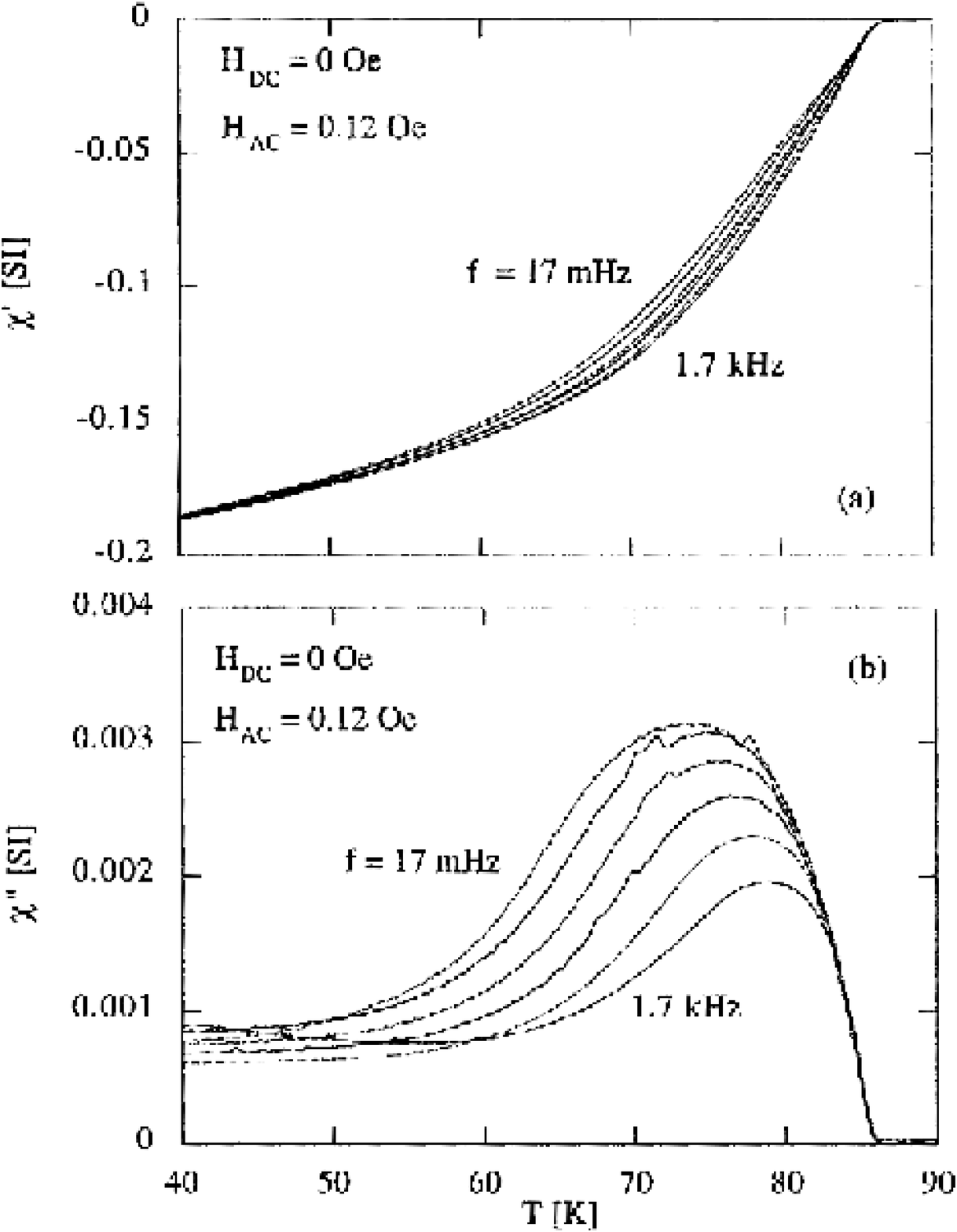}}
\vspace{0.1in}
\caption{ \label{acsus0}
Ac-susceptibility of a sintered Bi-2212 PME sample
($T_c = 87$ K) at different frequencies:
17 mHz, 170 mHz, 1.7 Hz, 170 Hz and 1.7 kHz. From Magnusson
{\em et al} \protect\cite{Magnusson98a}.
}
\end{figure}

Fig. \ref{acsus0} shows
the ac susceptibility at different frequencies of a sintered Bi-2212
PME-sample \cite{Magnusson98a} measured in an ac field $H_{ac} = 0.12$ Oe at 
different frequencies. The in-phase component $\chi '(\omega)$ shows shielding
properties that directly correspond to the ZFC dc susceptibility. The 
out-of-phase component $\chi ''(\omega)$ measures the dissipation in the
sample and its sharp appearance takes place at a temperature very close but
somewhat below $T_c$. The onset temperature for dissipation in the sample
is almost frequency-dependent within the resolution of the
measurement \cite{Magnusson98a}, only at an appreciably lower temperature
an observable frequency dependence occurs. The broad frequency-dependent 
maximum in $\chi ''(\omega)$ 
from Fig. \ref{acsus0} and the corresponding frequency dependence of
$\chi '(\omega)$ presumably
come from spontaneous magnetic moments
with widely distributed relaxation times.

\begin{figure}
\epsfxsize=2.6in
\centerline{\epsffile{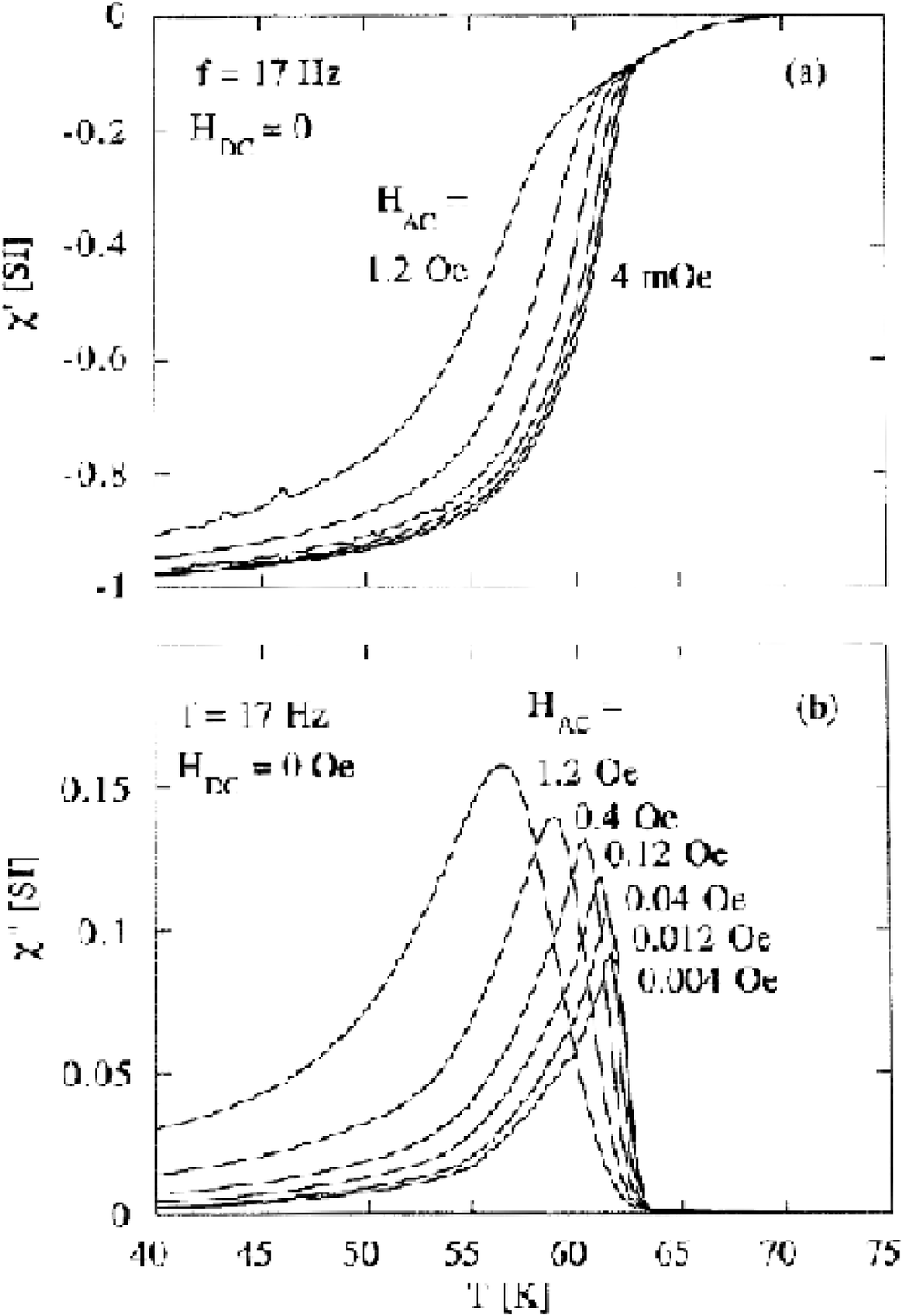}}
\vspace{0.1in}
\caption{ \label{acsus1}
Ac-susceptibility of a sintered Bi-2212 PME sample a frequency 17 Hz in
different superposed dc-fields.
From Magnusson {\em at al.} \protect\cite{Magnusson98a}.
}
\end{figure}

In Fig. \ref{acsus1} $\chi '$ and $\chi ''$ of the above sintered Bi-2212
sample is shown at a frequency 17 Hz but in different superposed dc fields.
One remarkable observation here is that at temperature close to
$T_c \approx 87 K$ the $\chi ''$ curves are strongly suppressed even by
a dc-field as small as 0.1 Oe. This behavior contradicts the picture of
conventional flux penetration where an increasing of $H_{dc}$ should increase
the dissipation and shifts $\chi ''$ to lower temperatures. The present
suppression of the out-phase component may be understood assuming that the low
field dynamics are governed by the response from the orbital magnetic
moments. On one hand, the spontaneous moments with a relaxation time
$\omega$, where $\omega$ is the angular frequency of the ac field, are flipped
back and forth by the ac-field causing the energy loss. On the other hand,
the dc-field polarizes many of these moments along its direction preventing
them from flipping, and thus reducing $\chi ''$.

% FIGURE 34

The increase of $\chi ''$
at high dc field again is due to ordinary intragranular flux penetration
\cite{Magnusson98a}.
An interesting difference between sintered and melt-cast samples was observed
\cite{Evie_thes,Magnusson98a,Papadopoulou02}: the dynamics of the former is
dominated by
spontaneous moments whereas the motion of thermally excited vortices plays
also an important role for the later ones.
Thus, the ac susceptibility measurement points to the existence of
spontaneous moments responsible for the PME but it does not clarify their
origin.

\begin{figure}
\epsfxsize=2.6in
\centerline{\epsffile{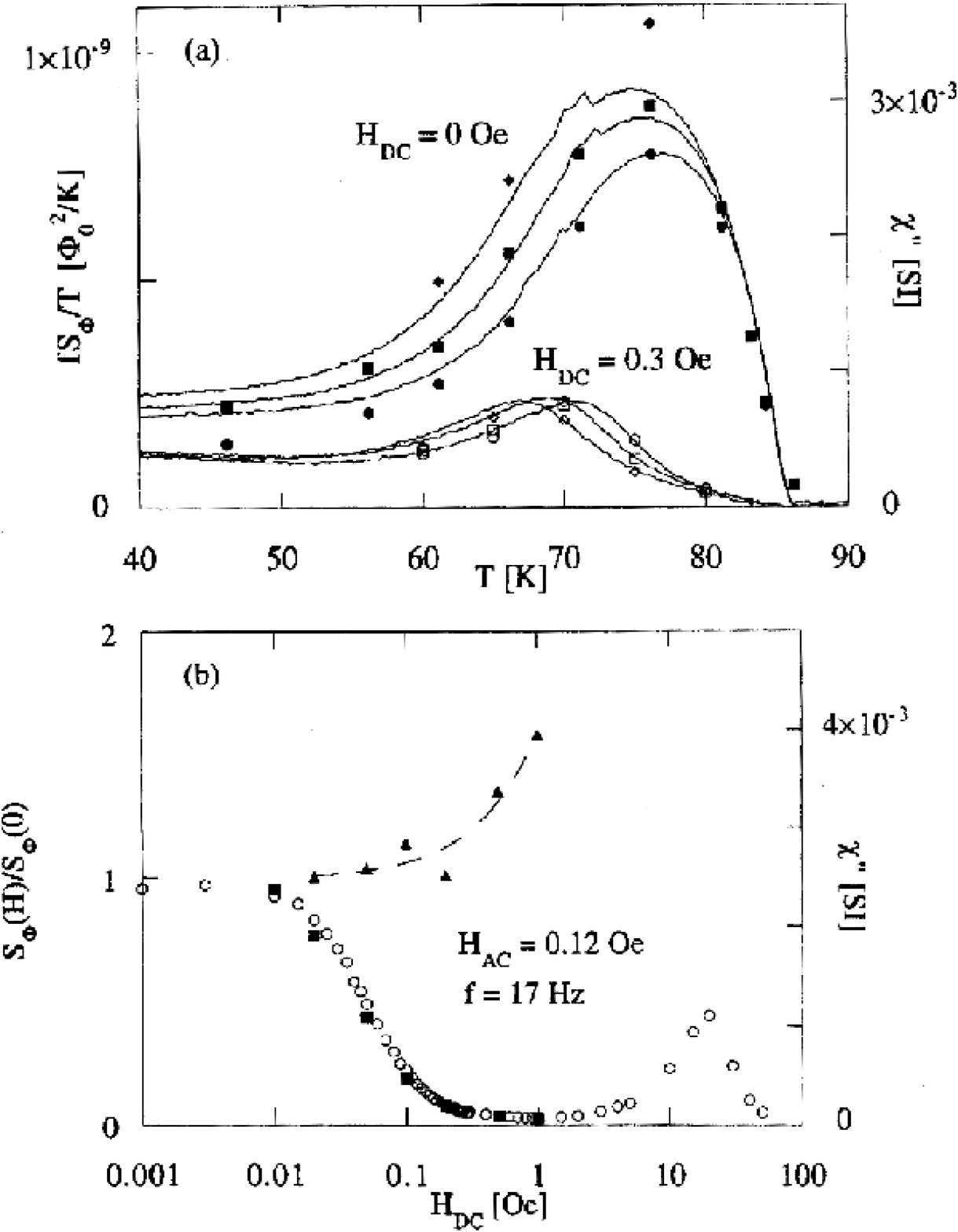}}
\vspace{0.1in}
\caption{ \label{acsus_FDT}
(a) Magnetic flux noise $fS_{\Phi}/T$ (left-hand scale) at 0.2 Hz (diamonds),
2 Hz (squares), and 20 Hz (circles)
and $\chi ''$  at three different
frequencies, from left to right 0.17, 1.7 and 17 Hz
(right-hand scale, solid lines) versus $T$.  Filled symbols correspond
to $H_{dc}$ = 0 and open symbols to $H_{dc} = 0.3$. PME sample.
(b) Magnetic noise $S_{\Phi}(H)/S_{\Phi}(0)$ (left-hand scale, solid squares)
 and $\chi ''$ (right-hand scale) at 17 Hz and $T = 80 $K vs. $H_{dc}$
for the PME sample. The maximum at higher
fields arises from intragranular flux penetration. $S_{\Phi}(H)/S_{\Phi}(0)$
at $f$ = 17 Hz and $T$ = 62 K versus the superimposed dc field for the non-PME
sample (solid triangles) is also shown. The perfect agreement with the
fluctuation dissipation theorem is seen.
From  Magnusson {\em et al} \protect\cite{Magnusson98b}.
}
\end{figure}

Another way to probe the flux dynamics is to study the flux noise spectra
$S_{\phi}(f)$ \cite{Magnusson98b}. The suppression of the flux noise
in PME samples by
the dc field \cite{Magnusson98b,Evie_thes}
(see upper panel of Fig. \ref{acsus_FDT}) also confirms the existence of the
spontaneous supercurrents.
For large enough dc fields the noise is expected to be dominated by
vortices starting to grow with increasing field.
The dc field dependence of the flux noise
resembles that, at the same temperature and field ranges,
of $\chi ''$.

In a region where the response of the ac susceptibility is linear
in field, $\chi ''$ and $S_{\phi}(f)$ are related by the
fluctuation-dissipation theorem

\begin{equation}
S_{\phi}(f,T,H_{dc}) \; \; = \; \; 2k_B T \frac{\chi''(f,T,H_{dc})}{\pi f} \; ,\label{FDT}
\end{equation}
where $k_B$ is Boltzmann constant.
Fig. \ref{acsus_FDT} shows a comparison
between the magnetic noise and the ac-susceptibility
 via the fluctuation dissipation
theorem.
It is remarkable to note the almost perfect mapping between the ac
susceptibility and the zero field noise was achieved.
The validity of the fluctuation dissipation theorem
constitutes that the noise measurements have been done in the equilibrium.

% FIGURE 35

Since $\chi ''(f)$ depends on $f$ weakly, from Eq. (\ref{FDT}) we have 
$S_{\phi}(f) \sim 1/f$. The experiments of Magnusson {\em et al} show
that this quantity scales with $f$
as $1/f^{\alpha}$ where $\alpha$ is close to 1 indicating that the noise
may be a flicker noise and arose from the distribution of activation
energies of the vortex hopping. Since the $1/f$ noise has been also observed
in spin glasses \cite{Nordblad_Young} one may
suggests that the PME is relevant to some glassy phase like a chiral
glass one.

\noindent {\em Simulation}. AC susceptibilities may be monitored by applying 
the external magnetic ac
field. Then 
the external field $H$ containing the dc and ac parts
is given by
\begin{equation}
H \; \; = \; \; H_{dc} + H_{ac} \cos(\omega t) \; \; .
\label{HdcHac}
\end{equation}
In general, the dc field is is necessary to generate even harmonics.
The real and imaginary parts of $n$-th order harmonics
$\chi'_n(\omega)$ and $\chi''_n(\omega)$ are calculated as
\begin{eqnarray}
\chi'_n(\omega) \; \; &=& \; \; \frac{1}{\pi h_{ac}}
\int_{-\pi}^{\pi} \; m(t) \cos(n\omega t)d(\omega t) \; \; , 
\nonumber\\\chi''_n(\omega) \; \; &=& \; \; \frac{1}{\pi h_{ac}}
\int_{-\pi}^{\pi} \; m(t) \sin(n\omega t)d(\omega t) \; .
\label{harmonics}
\end{eqnarray}

For the system described by the multiloop model (\ref{H_multiloop}) the
magnetization $m(t)$ is given by Eq. (\ref{magnetization}), the dimensionless
ac field $h_{ac}$ is related to $H_{ac}$ via (\ref{H_dimless}). 

% FIGURE 36

The linear ac susceptibility and  higher harmonics of ceramic $s$-wave
superconductors have been intensively studied using 
the Bean model \cite{Ji89}, Kim model 
\cite{Ishida90} and  a network of resistively shunted Josephson junctions
\cite{Wolf93,Majhofer91}. Very little theoretical and simulation work
has been done in this direction for $d$-wave superconductors, although
such work would be vital to interpret experimental data 
\cite{Magnusson98a,Magnusson98b,Papadopoulou02}.

The ac susceptibility of a $d$-wave ceramic superconductor was computed
\cite{KawLi} by Monte Carlo simulations with the help of Hamiltonian 
(\ref{H_multiloop}) and Eq. (\ref{harmonics}) with $n=1$. 
While Monte Carlo simulations involve
no real dynamics, one can still expect that they give useful information on
the long-time behavior of the system. In fact,
the characteristic time for the sintered samples,
which are believed to be captured
by our model, is of order $10^{-12} s$\cite{Wolf93}. This time has the same
order of magnitude as a single Monte Carlo step. In general,
the period of oscillations is much longer than
the characteristic time \cite{KawLi}.
For such a slowly changing ac field the system can be regarded
as being in
quasi-equilibrium and the Monte Carlo updating may be applied.
A priori, the validity of this approximation is
not clear but it may be justified \cite{Li99} by comparing
the Monte Carlo results with those obtained by other approaches to the
dynamics such as considered in Ref. \cite{Wolf93}.

A qualitative agreement between simulations
\cite{KawLi} and experiments \cite{Magnusson98a,Papadopoulou02}
was achieved, e.g., for the dependence of $\chi '$ and $\chi ''$ 
on temperature and the ac field. Namely, as $H_{ac}$ is increased, $\chi ''$
grows and its peak shifts towards lower temperatures, while $\chi '$ becomes 
less and less diamagnetic. Simulations \cite{KawLi} showed that at 
high frequencies
there is very little difference between the $\chi$ of the $s$- and $d$-wave
models. At low frequencies, clear differences have been seen: The magnitude
of $\chi ''$ of the $s$-wave model becomes an order of magnitude smaller than
that of the $d$-wave model, indicating that the $d$-wave system exhibits much
stronger dissipation. A close similarity observed between the $\chi ''$ of 
$d$-wave superconductors and that of the spin glass \cite{KawLi} is consistent
with the chiral glass picture as well as with experimental findings
\cite{Magnusson98a}.

It would be highly desirable to check the validity of the fluctuation 
dissipation theorem (\ref{FDT}) and verify whether the $1/f$-noise occurs
in a system showing the PME by simulations. This may be done in
the framework of the multiloop model (\ref{H_multiloop}). 

\subsection{Compensation effect}

\noindent {\em Experiment}.
Heinzel {\em et al.}\cite{Heinzel} have shown that the PME may
be analysed by the compensation technique based on the measurement of the
second harmonics of the magnetic ac susceptibility.
The essence of the so called
compensation effect is that the second harmonics signal vanishes
at some compensation field $H_{com}$ even a nonzero dc field is applied.
This is a bit surprising because even harmonics should be generated in
the presence of the dc field due to $M(H) = -M(-H)$ symmetry.

% FIGURE 37
\begin{figure}
\epsfxsize=2.6in
\centerline{\epsffile{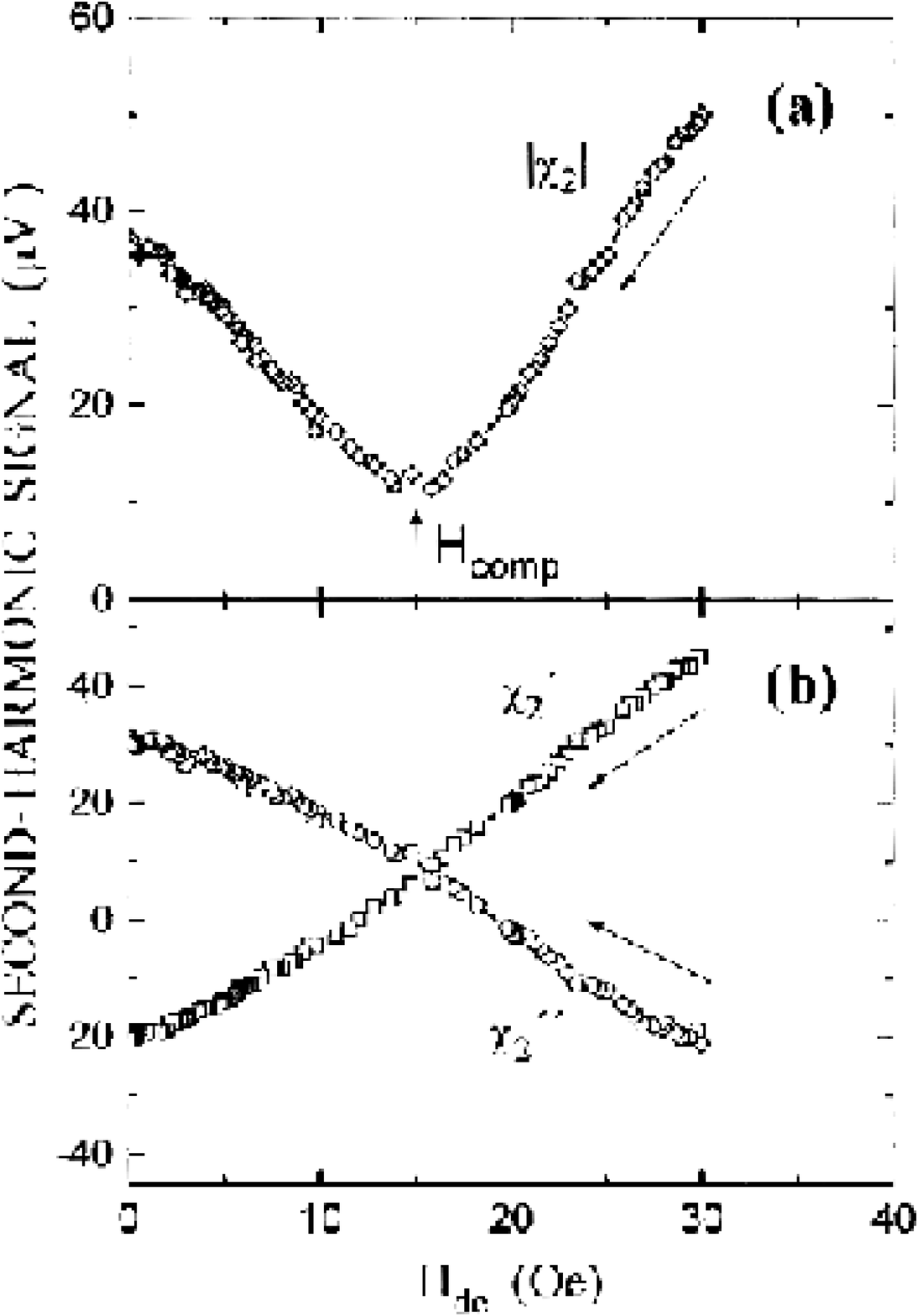}}
\vspace{0.1in}
\caption{ \label{compensation_exp}
Second harmonics of the magnetic ac susceptibility ($H_{ac}$=1 Oe) obtained
for YBa$_2$Cu$_3$O$_7$ after field cooling in a dc field $H_{dc}$=30 Oe to
$T=77 K$ and stepwise reducing $H_{dc}$. (a) Modulus $|\chi _2 |$ of the
complex harmonics. (b) Real ($\chi '_2$) and imaginary
($\chi ''_2$) parts of the second
harmonics. After Heinzel {\em et al.} \protect\cite{Heinzel}.
}
\end{figure}

The compensation effect  may be detected in the following way.
The sample is cooled in
the external dc field down to a low temperature and then the field is
switched off. At the fixed low $T$ the second harmonics are monitored
by applying the dc and ac
fields to the sample. 

The results obtained for PME sample YBa$_2$Cu$_3$O$_7$
are shown in Fig. \ref{compensation_exp} where
$H_{com} \approx 15$ Oe. The observed intersection of $\chi '_2$ and
$\chi ''_2$ at $H_{dc}=H_{com}$ indicates a change of their sign.
The key observation is that the compensation effect
appears only in the
samples which show the PME but not in those which do not. It may be explained
in the following way.
Due to the presence of non-zero spontaneous orbital
moments and the frustration, the remanent magnetization or, equivalently,
the internal field is generated
in the cooling process.
If the direction of the external dc field is
identical to that during the FC procedure, the induced
shielding
currents will reduce the remanence. Consequently, the absolute value
of the second harmonics $|\chi_2|$ decreases until
the signal of the second harmonics is minimized
at a field $H_{dc}=H_{com}$.
Thus the study of the compensation effect
 allows us to establish not only the existence of
the spontaneous orbital moments but also to determine
internal fields generated during the cooling process.
The compensation effect is a collective phenomenon and it may be captured by the multi-loop
model of 0 and $\pi$-junction network.

% FIGURE 38
\begin{figure}
\epsfxsize=3.6in
\centerline{\epsffile{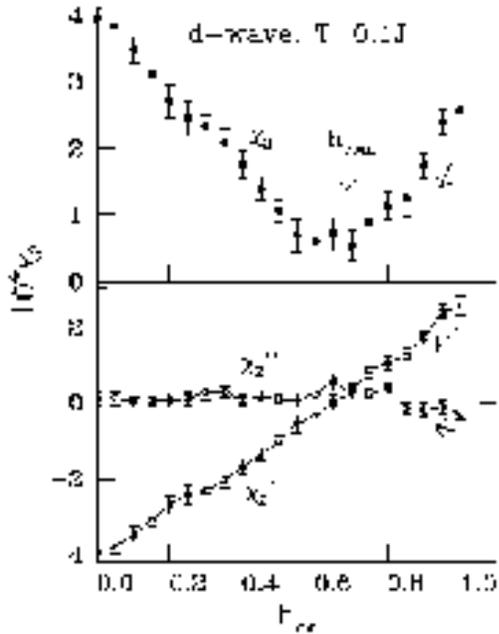}}
\vspace{0.1in}
\caption{ \label{compensation_sim}
The second harmonics of the $d$-wave model (\ref{H_multiloop}) obtained
after field cooling in a dc field $h_{dc}=1$ from $T=0.7$ to $T=0.1$.
At the lowest $T=0.1$ ($< T_{CG} \approx 0.17$ for $\tilde{L}=4$)
the dc field used in cooling is switched off
and the second harmonics are generated by applying the combined dc and ac
fields. The $dc$ field is stepwise reduced from $h_{dc}=1$ to $h_{dc}=0$.
The inductance is chosen to be equal to $\tilde{L}=4$ and system size
$l=8$.
The arrows indicate the sense of the changes in the
dc field. The results are
qualitatively the same as those presented in Fig. \ref{compensation_exp}.
After Ref. \protect\cite{Li99}.}
\end{figure}

\noindent {\em Simulation}.
We now present the Monte Carlo simulation results \cite{Li99} explaining 
the compensation effect \cite{Heinzel} in ceramic superconductors.
Again we employ model (\ref{H_multiloop}) and Eq. (\ref{harmonics}) ($n=2$)
 to compute
the second harmonics. The
calculations follow exactly
the experimental procedure of Heinzel {\em et al}\cite{Heinzel}.
In the FC regime,
first the system is cooled in some dc field $h_{dc} \neq 0$
from a high $T$ down to a low $T$ which is below
the paramagnet-chiral glass transition temperature \cite{Li99}.
When the lowest
temperature is reached the dc field used in cooling is switched off and
we apply the combined field given by Eq. (\ref{HdcHac}).
 We monitor the 
second harmonics
reducing the dc field from a high dc field to zero stepwise.
Fig. \ref{compensation_sim} shows
the typical behavior of the second harmonics for the $d$-wave
ceramic superconductors.  $|\chi_2|$ reaches minimum at
the compensation field $h_{com}=0.7\pm 0.05$. At this point,
similar to the experimental findings presented in Fig. \ref{compensation_exp}
\cite{Heinzel},
the intersection of $\chi'_2$ and $\chi''_2$ is observed. This fact
indicates that at $H_{com}$ the system is really in the compensated state.
Furthermore, in accord with the experiments, at the compensation point the
real and imaginary parts should change their sign\cite{Heinzel}.
Our results show that $\chi'_2$ changes its sign roughly at $h_{dc}=h_{com}$.
A similar behavior is also displayed by $\chi''_2$
but it is harder to observe due to a smaller amplitude of $\chi''_2$.

Performing simulations in the same way as for the $d$-wave case, one cannot
observe the compensation effect
 for the $s$-wave model \cite{Li99} because the cooling
of a unfrustrated "ferromagnetic" system could not produce any 
remanent dc field which would compensate the external one. This result 
is again in accord with the experimental data\cite{Heinzel}.

Thus, the compensation effect may be explained, at least
qualitatively,
by using the multiloop model of the ceramic superconductors with anisotropic
pairing symmetry. The existence of the compensation phenomenon only in those
samples showing the PME indirectly support the chiral glass phase in
these materials.

\subsection{Aging phenomenon}

\noindent {\em Experiment}. The aging phenomenon
observed first in spin glasses \cite{Lundgren83} have
been studied both theoretically
\cite{Theory_aging} and experimentally \cite{Nordblad_Young} in detail.
In this phenomenon the physical quantities depend not only on
the observation time but also on the waiting time (or age), 
$t_w$, i.e. how long one
waits at constant field or temperature before measurements.
The origin of such memory phenomena  relates to the rugged energy landscape
which appears due to disorder and frustration \cite{Binder}.

\begin{figure}
\epsfxsize=3.2in
\centerline{\epsffile{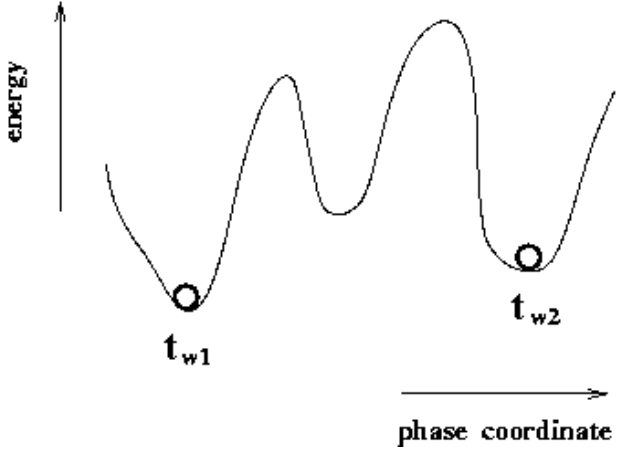}}
\vspace{0.1in}
\caption{ \label{aging_LM_fig}
Schematic energy landscape of a frustrated system in the phase space.
After waiting times $t_{w1}$ and $t_{w2}$ the system gets trapped in
different local minima.
}
\end{figure}

Overall, one can explain the aging effect qualitatively in the following
way.
After the waiting time $t_{w1}$ and $t_{w2}$, e.g., the system gets trapped , 
in principle, in different local minima as shown in Fig. \ref{aging_LM_fig}. 
Since the corresponding energy barriers are different the physics should depend
on in what local minimum the system was before one starts to do measurements.
It should be stressed that both the waiting and measurement times are
typically much
shorter than the relaxation time which may be astronomically large for
a frustrated system \cite{Binder}. The aging is, therefore, a nonequilibrium 
phenomenon.

\begin{figure}
\epsfxsize=2.8in
\centerline{\epsffile{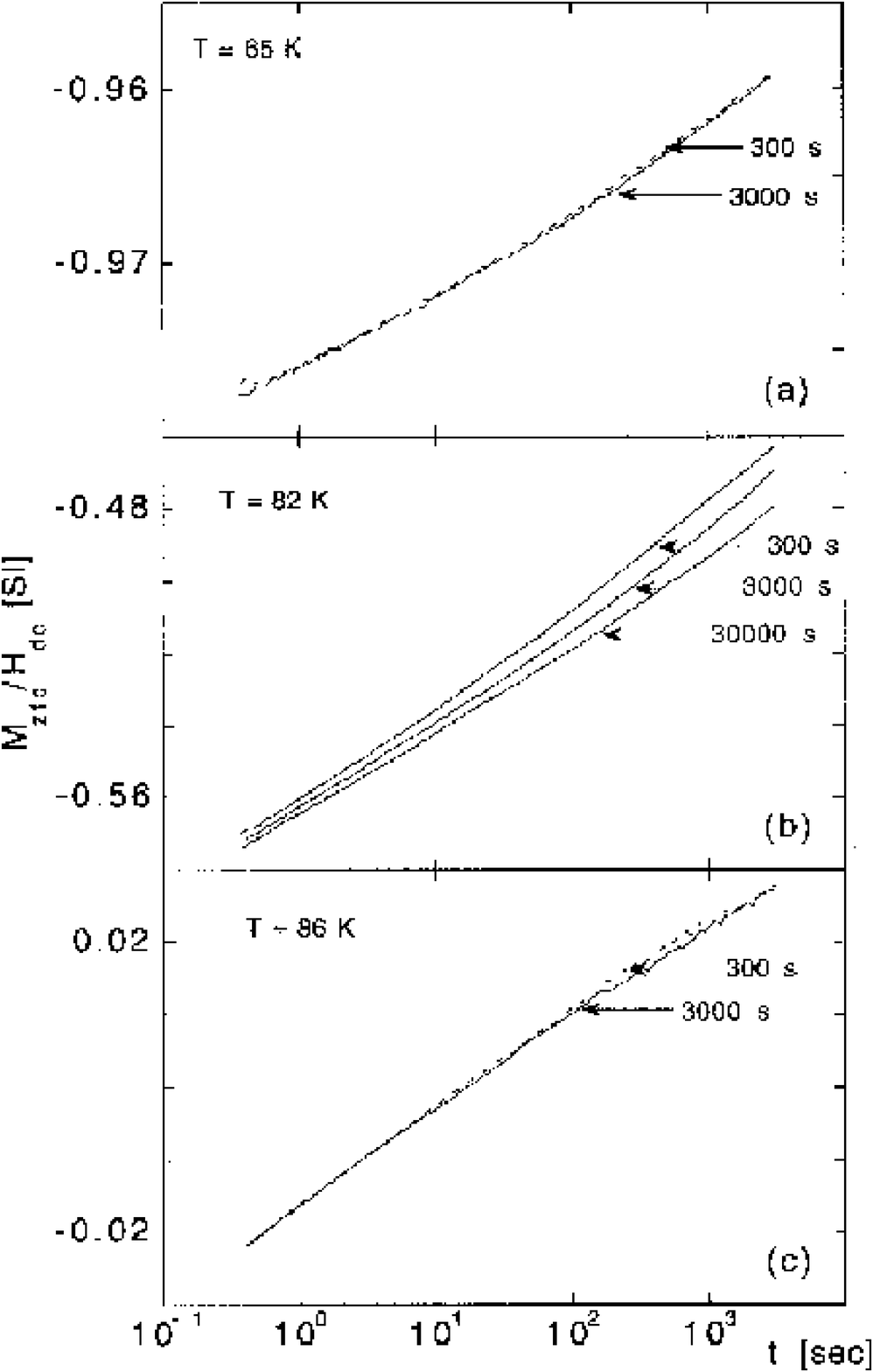}}
\vspace{0.1in}
\caption{ \label{aging_exp99}
$M_{zfc}/H_{dc}$ of the melt-cast Bi$_2$Sr$_2$CaCu$_2$O$_8$ sample
is plotted vs observation time for $H_{dc} = 0.02 G$ at
temperatures (a) 65 K, (b) 82 K, and (c) 86 K. The waiting time used were
300 and 3000 s; in (b) the result for $t_{wait}$ = 30000 s is also included.
From Papadopoulou {\em et al} \protect\cite{aging}.
}
\end{figure}

Recently, Papadopoulou {\em et al} \cite{aging,Papadopoulou01}
has observed  the aging effect in the PME sample
 Bi$_2$Sr$_2$CaCu$_2$O$_8$
monitoring the ZFC magnetization. The relaxation of
the ZFC magnetization has been measured by cooling the sample in zero field
to the measuring temperature, allowing the sample to stay
at that temperature for a certain time $t_w$ and then applying the probing
field and recording the change of the magnetization with observation time
at constant temperature.
Papadopoulou {\em et al} have made two key observations. First,
the aging effect is not observed at high fields and at high temperatures.
Second, this effect occurs at some intermediate temperature interval
(see Fig. \ref{aging_exp99}) and it disappears again at
low temperatures.
Thus the aging phenomenon exists only for weak enough fields and
in an intermediate temperature region.

% FIGURE 39

 The first
observation is trivial because at high fields or high temperatures the
roughness of the energy landscape does not play any crucial role and
the system looses its memory. At low temperatures the role of
the energy landscape becomes important,
the second result of Ref. \cite{aging} is, therefore, not trivial
from the point of view of the standard spin glass theory.
Papadopoulou {\em et al} suggested that at low temperatures the external
field is screened from the bulk of the sample and it cannot probe the
collective behavior of the Josephson junction network.
As one can see below, the correctness of this idea may be confirmed using
the multi-loop model (\ref{H_multiloop}).

The results from aging measurements are in immediate
contrast to models explaining the PME as a flux compression phenomenon
\cite{Larkin,Moshchalkov} which are only applicable to the positive FC
magnetization observed in some conventional superconductors.
The aging phenomenon in the melt-cast samples \cite{aging}, with
similarities to the aging in spin glasses, implies the interaction between 
spontaneous magnetic moments causing them to behave collectively. 
Inside each
of the large grains of a sample there can exist many thousands
of domains, and the intragrain regions can be, therefore, modeled as Josephson
networks containing a random distribution of ordinary Josephson junctions
and $\pi$ junctions.

\noindent{\em Simulation.}
The study of the aging phenomenon in spin glasses by computer simulations lasts
for may years \cite{Rieger95}. Our aim is not
to present those results  but to focus on the explanation \cite{Li01} of
specific features of the aging in granular superconductors observed by
Papadopoulou {\em et al} \cite{aging}.

To understand the experiments \cite{aging} we use the
model (\ref{H_multiloop}) for a $d$-wave superconductor.
The dependence of the magnetization $m(t)$ given by 
Eq. (\ref{magnetization}) on the
waiting time was computed by the Monte Carlo simulations \cite{Li01}.
In order to mimic the aging effect in the ZFC
regime we quench the system from a high temperature to
the working temperature. The system is there evolved in zero field during
a waiting time, $t_w$. Then the external field $h$ is turned on and
the subsequent growth of the magnetization $M(t,t_w)$ is monitored.

% FIGURE 40

One can demonstrate \cite{Li01} that
there are three screening regimes for the aging phenomenon. In the
strong screening limit when $ \tilde{L} > \tilde{L}_2^* = 9\pm 0.5$,
 the aging is suppressed at any temperature.
In the weak screening regime, $ \tilde{L} < \tilde{L}_1^* = 3.5\pm 0.5$, 
it is observable even at low
temperatures. The intermediate screening regime, $\tilde{L}_1^* < \tilde{L}
< \tilde{L}_2^*$, is  found
to be the most interesting: the aging is present only in an
intermediate temperature interval and it does not appear at
low temperatures.

\begin{figure}
\epsfxsize=3.6in
\centerline{\epsffile{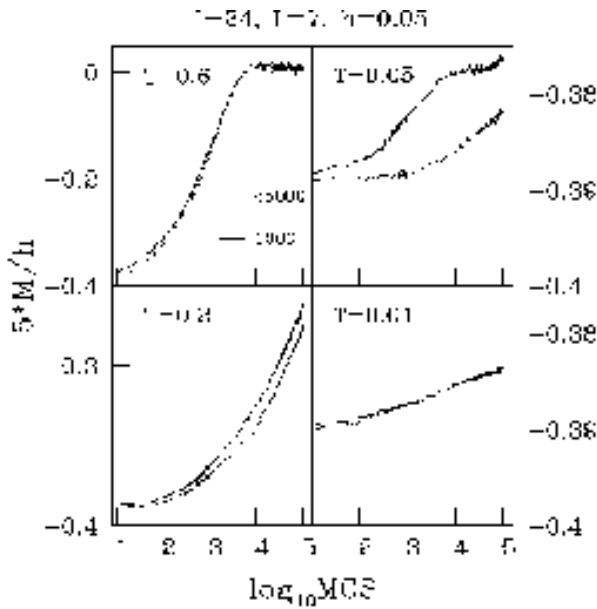}}
%\vspace{0.1in}
\caption{ \label{aging_sim_fig}
The temperature and time dependence of $M$ for
$t_w=1000$ and $t_w=25000$.
$l=24, \tilde L=7$ (intermediate screening regime) and $h=0.05$.
The aging disappears at low temperatures.
The results are averaged over
60 -- 120 samples. From Li {\em et al} \protect\cite{Li01}.
}
\end{figure}

Fig. \ref{aging_sim_fig} shows the results for $\tilde L$=7 belonging to the 
intermediate screening regime.
In agreement with the experiments \cite{aging},
the aging effect appears only for the intermediate temperature interval.
At low $T$'s ($T \le T^* = 0.02 \pm 0.01$)
 the effect is suppressed due to the screening of the magnetic
field from the bulk.  The results have been obtained for the observation 
times comparable with the
waiting ones but we believe that they should be valid for longer observation
time scales \cite{Li01}.

It should be stressed that the experimental finding of
Papadopoulou {\em et al} \cite{aging} cannot be explained by the
standard XY model where the screening effect is not taken into account.
The mechanism of aging may be understood
from studies the spatial distribution of flux inside a sample \cite{Li01}.
In fact, in the strong screening limit
the aging effect does not occur
at any temperature because the external field is screened entirely.
For a fixed screening strength the magnetic field is expelled more
and more from the bulk as $T$ is lowered. Therefore, for the intermediate values
of screening one could not observe the aging at low $T$'s.

The fact that the increase of the self-inductance would wash out the aging
effect may be explained
by  exploring the effect of screening on the
energy landscape \cite{Li01a}.
Studies of local minima at $T=0$ showed that the energy landscape
gets smoother and smoother as the screening is enhanced  and
the glassy effects would become less pronounced. 

% FIGURE 41

So, the non-trivial aging phenomenon in ceramic superconductors 
may be understood
in the framework of the model (\ref{H_multiloop}) where the anisotropy
of the pairing function plays a key role.
The observation of aging and the PME in the same material does
yield support, therefore,
 for the existence of $d$-wave superconductivity. Furthermore
the PME dynamics owns the characteristics features of the non-equilibrium
dynamics of spin glasses. The results on aging
 are thus indicative of the formation of
a low temperature glassy phase in a system showing a pronounced PME, a phase
that resembles the chiral glass phase
\cite{Kawamura95,KawLi97}.

\subsection{Anomalous microwave absorption}

\noindent {\em Experiment}. 
MWA measurements are a sensitive tool for detection of the
transport properties of high-$T_c$ superconductors below the critical
temperature, when superconducting paths are risen and dc measurements cease
to give information. Soon after the discovery of high-$T_c$ superconductivity
it has been established \cite{Stankowski87,Durny87,Blazey87,Peric88,Marcon89},
that in ceramic samples there exists a large
non-resonant MWA in zero and low applied magnetic field
that can be associated with the transition to the superconducting phase.
In fields below $H_{c1}$ the MWA of high-$T_c$ materials
is due to dissipative flux motion in the link network or, equivalently, due
to the junction's resistance.

\begin{figure}
\epsfxsize=3.6in
\centerline{\epsffile{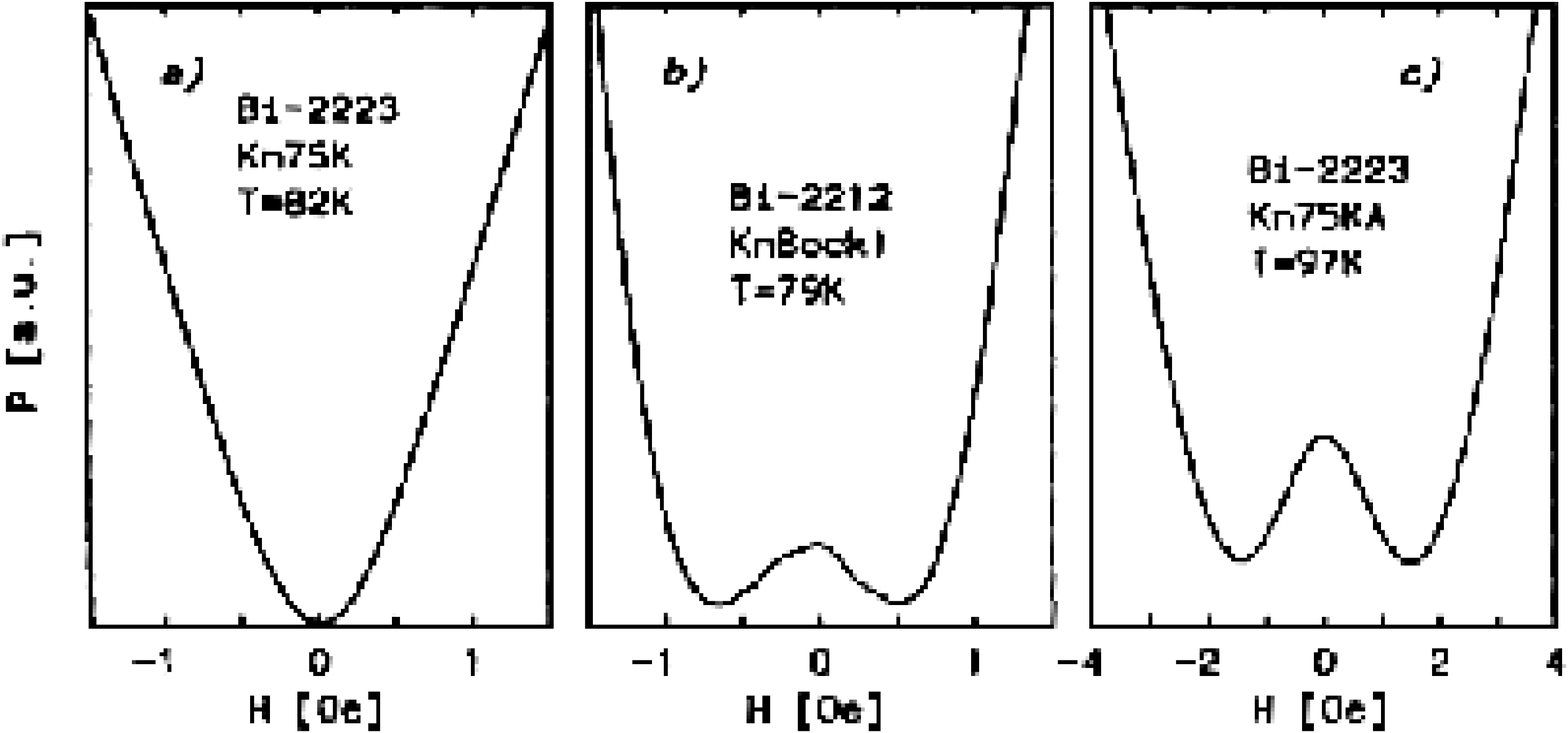}}
%\vspace{0.1in}
\caption{\label{MWA_exp}
(a) The field dependence of the MWA for a powdered HTSC sample which
does not show the PME. The changeover from parabolic to linear occurs around
0,6 Oe. (b) The same as in (a) but for the PME sample. The minima are
located at $\pm 0.6$ Oe, near the field where the FC signal changes from
diamagnetic to paramagnetic. (c) The MWA signal of a Bi-2223 sample showing
a weak PME.
From Ref. \protect\cite{Braunisch93}.
}
\end{figure}

 Braunish {\em et al.} \cite{Braunisch92,Braunisch93,Knauf98} found
a striking
correlation between the field dependence of the MWA and
the PME. Their typical results are shown in Fig. \ref{MWA_exp}.
Clearly, in the absence
of the PME the MWA has a conventional minimum at $H=0$ and in low fields
$P(H) \sim H^2$. The change over from the quadratic to linear behavior
appears at $H_{c1}^*$ associated with the critical field, at which the magnetic
flux breaks into the weak-link network \cite{Braunisch93}.
Contrary to non-PME samples the MWA of PME samples show a maximum at $H=0$
and it reaches minimum at a field near which
the FC signal changes from paramagnetic to diamagnetic behavior \cite{Kataev91}.

% FIGURE 42

\noindent {\em Simulation}.
The unusual behavior of the MWA may be explained in the
framework of a single-loop model \cite{Sigrist92}. If the loop contains a
$\pi$ junction, then the spontaneous internal field is generated. 
In the absence of
the external field, the microwave field is coupled to
this field and dissipates energy. As a static field is turned on, the  
oscillatory field loses gradually its access to the spontaneous moments
and the absorption power decreases. If the dc field is strong enough the
$\pi$ junction loop would behave like a 0 junction one, leading to the
increase of the power. The absorption minimum should be located at
$H \approx \Phi_0/4S$ \cite{Sigrist92}, i.e. between the fields making the centers
between the first maximum and minimum of the free energy shown in Fig.
\ref{pme_free_fig}.

The general scheme to study the MWA
in an Josephson junction networks may be found, for example, in Ref.
\cite{Rycers01}. Here we follow a simpler approach \cite{Dominguez94}.
Namely, in order to reproduce the anomalous behavior of the absorption power
we use the resistively shunted junction (RSJ) model
\cite{Dominguez96} combined with
the linear response theory \cite{Kubo}.
In this formalism the linear response to the external electromagnetic field 
is proportional to a voltage -- voltage correlation
function. To get insight on the MWA we calculate
quantity $P$ which is proportional to the frequency integral of this
correlation function. Then, we have \cite{Dominguez94}
\begin{equation}
P \; \; = \; \; \frac{1}{R} \sum_i \; < V_i^2 > \; \; ,
\label{MWA_def}
\end{equation}
where  $ < V_i^2 >$ is a mean value of the square of the voltage induced by the
thermal noise on each junction
and $R$ is the normal resistance of the links.

% FIGURE 43
\begin{figure}
\epsfxsize=3.6in
\centerline{\epsffile{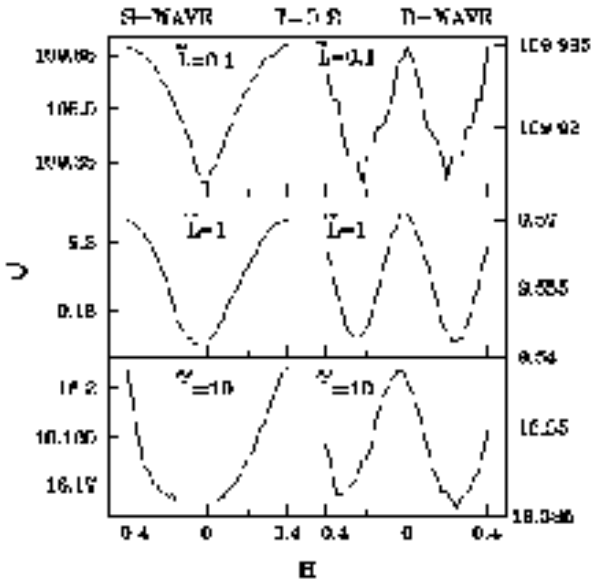}}
%\vspace{0.1in}
\caption{\label{MWA_sim}
The field dependence of $P$ for $s$- (left panel) and $d$-wave
(right panel) ceramic superconductors. We choose $T=0.2$ and
$\tilde{L}=0.1, 1$ and 10.
The results are averaged over
20  samples. After Li \protect\cite{Li01a}.
}
\end{figure}

To calculate $V_i$ we use the RSJ model\cite{Dominguez}
for the current flowing between two grains.
The derivation of dynamical equations
for this model is given in Appendix \ref{sec.rsj}.  
The full set of equations for the gauge invariant phases $\theta$
(see Eq. (\ref{ginvphase}) is as follows
\begin{eqnarray}
\frac{\hbar}{2eR}\frac{d\theta_\mu({\bf n})}{dt} \; = \;
-\frac{2e}{\hbar}J_\mu({\bf n})\sin\theta_\mu({\bf n})
\nonumber\\
-\frac{\hbar}{2eL}\Delta_\nu^{-}\left[
\Delta_\nu^{+}\theta_{\mu}({\bf n})-\Delta_\mu^{+}\theta_\nu({\bf n})\right]
 -\zeta_\mu({\bf n},t) \, \, .
\label{RSJ_eq}
\end{eqnarray}
Here we use following notation:
the site of each grain is at position ${\bf n}=(n_x,n_y,n_z)$
(i.e. $i\equiv{\bf n}$); the lattice directions are
$\mu={\hat{\bf x}}, {\hat{\bf y}}, {\hat{\bf z}}$;
the link variables are between sites ${\bf n}$ and ${\bf n}+\mu$
(i.e. link $ij$  $\equiv$ link ${\bf n},\mu$);
and the plaquettes $p$ are defined by the site ${\bf n}$ and
the normal direction $\mu$ (i.e plaquette $p$ $\equiv$ plaquette
${\bf n},\mu$, for example the plaquette ${\bf n}, {\hat{\bf z}}$ is
centered at position ${\bf n}+({\hat{\bf x}}+{\hat{\bf y}})/2$).
The forward difference operator $\Delta_{\mu}^{+}\theta_\nu({\bf n})=
\theta_\nu({\bf n}+\mu)-\theta_\nu({\bf n})$
and the backward operator  $\Delta_{\mu}^{-}\theta_\nu({\bf n})=\theta_\nu({\bf n})-
\theta_\nu({\bf n}-\mu)$ (see Appendix \ref{sec.rsj}
for more details). Critical currents $J_\mu({\bf n})$ are the same
as $J_{ij}$ in Hamiltonian (\ref{H_multiloop}).
The left part of Eq. (\ref{RSJ_eq}) describes the
normal current, while the first and second terms in the right part
correspond to the supercurrent and screened current, respectively.
The Langevin noise
current $\zeta_{\mu}({\bf n},t)$ has Gaussian correlations
\begin{equation}
\langle\zeta_{\mu}({\bf n},t)\zeta_{\mu '}({\bf n}',t')\rangle \; = \;
\frac{2k_BT}{R}\delta_{\mu ,\mu'}
\delta_{{\bf n} , {\bf n'}}\delta(t-t') \; .
\end{equation}
The local voltage $V_i$ is then given by
\begin{equation}
V_i \; \; = \; \; \frac{d\theta_i}{dt} \; \; .
\end{equation}

Eq. (\ref{RSJ_eq}) describes the overdamped dynamics. 
One can show that the inertia
(capacitive) terms do not change results qualitatively and they are neglected.

The system of differential equations (\ref{RSJ_eq}) is integrated numerically 
by a second order Runge-Kutta-Helfand-Greenside algorithm for stochastic
differential equations \cite{Dominguez91}. The time step is chosen to depend 
on $\tilde{L}$ and
is equal to $\Delta t = 0.1\tau _J$ and
$\Delta t = 0.1\tau _J \times \tilde{L}$ for  
$\tilde{L} > 1$ and $\tilde{L} < 1$,
respectively.

The field dependence of the MWA in two-dimensional $s$-
and $d$-wave disordered superconductors was studied by Dominguez {\em et al}
\cite{Dominguez94}. Similar results have been obtained for three dimensions
\cite{Li01a}.
Fig. \ref{MWA_sim} shows the field dependence of the MWA for $T$=0.2 
and for various values
of $\tilde{L}$. In the case of $s$-wave superconductors we have the
standard minimum
at $H=0$ for any value of inductance and $T$. As expected,
$P \sim H^2$ at weak fields. 

For the $d$-wave samples $P$ has
the unconventional peak at $H=0$. Contrary to the one-loop model
\cite{Sigrist92} such peak is seen not only for $\tilde{L} > 1$
but also for $\tilde{L} \le 1$.

In our model (\ref{RSJ_eq}) the temperature dependence of the critical
current is neglected.
However, one can show that the dimensionless temperature $T$ chosen in
Fig. \ref{MWA_sim}
corresponds to the relevant to experiments real
temperature, $T_R$. In fact, the critical current depends not only on
temperature but also on
conditions under which samples were prepared.
The typical value of the critical current density for
ceramic superconductors
is $ \sim 10^6 A/m^2$ ( see, for example,
Ref. \onlinecite{Marcon89,Larbalestier88,Peuckert89,Murayama89}).
Since the typical size of grains is about $1\mu m$
we have the critical current $I_c \sim 10^{-6}A$.
Using $T_R=JT/k_B=\hbar I_cT/2ek_B$ one obtains $T_R/T \sim 100K$.
Clearly, the dimensionless $T$ chosen in Fig. \ref{MWA_sim}
 correctly describes the experimental values of temperature
\cite{Braunisch92}.

Comparing Fig. \ref{MWA_sim} with Fig. \ref{MWA_exp} one can see that the
multi-loop model (\ref{H_multiloop}) correctly captures the experiments
of Braunish {\em et al} \cite{Braunisch92,Braunisch93}. Furthermore,
the height of the peak shown in Fig. \ref{MWA_sim} is very small:
($P(H=0) - P_{min})/P_{min}$
is of order of $10^{-3}$.
This is also  in qualitative agreement with experimental 
findings \cite{Braunisch92}
that the peak should be low.

\subsection{AC resistivity}

\noindent {\em Experiment}. In recent experiments
Yamao {\em et al.}\cite{Matsuura2} have measured
the ac linear resistivity, $\rho_0$, and the nonlinear resistivity, $\rho_2$,
of ceramic superconductor YBa$_2$Cu$_4$O$_8$. $\rho_0$ and $\rho_2$ are
defined  as the first and third coefficient of the expansion
of the voltage $V(t)$ in terms of the external current $I_{ext}(t)$:
\begin{equation}
V \; \; = \; \rho_0 I_{ext} + \rho_2 I_{ext}^3 + ... \; \; .
\end{equation}
When the sample is driven by an ac current
 $I_{ext}(t) = I_0 \sin(\omega t)$, one can relate $\rho_0$
and $\rho_2$ to the first harmonics $V'_{\omega}$ and third harmonics
$V'_{3\omega}$ in the following way
\begin{eqnarray}
\rho_0 \; \; = \; \; V'_{\omega}/I_0, \; \;
\rho_2 \; \; = \; \; -4 V'_{3\omega}/I_0^3 \; , \nonumber\\
V'_{n\omega} \; \; = \; \; \frac{1}{2\pi}
\int_{-\pi}^{\pi} \; V(t) \sin (n\omega t)
d(\omega t) \; , \; n \; = \; 1, 3 \; \; .
\end{eqnarray}

\begin{figure}
\epsfxsize=3.6in
\centerline{\epsffile{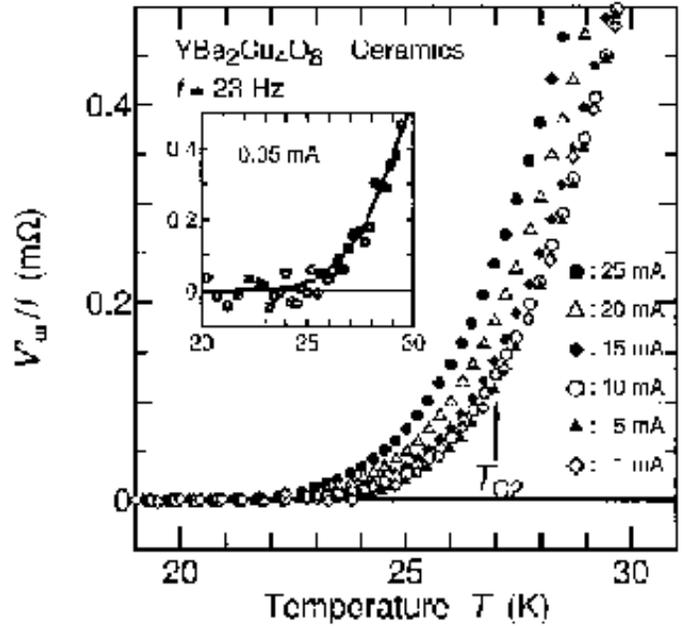}}
\vspace{0.1in}
\caption{ \label{v1_exp_fig}
The temperature dependence of $V'_{\omega}/I$ around $T_{c2}$. Frequency
$f = 23$ Hz and values of $I$ are shown in the figure. The inset shows the
data measured at $I$ = 0.05 mA. The solid line indicates the limiting
$\rho_0 - T$ curve. After Yamao {\em et al} \protect\cite{Matsuura2}.
}
\end{figure}

\begin{figure}
\epsfxsize=2.6in
\centerline{\epsffile{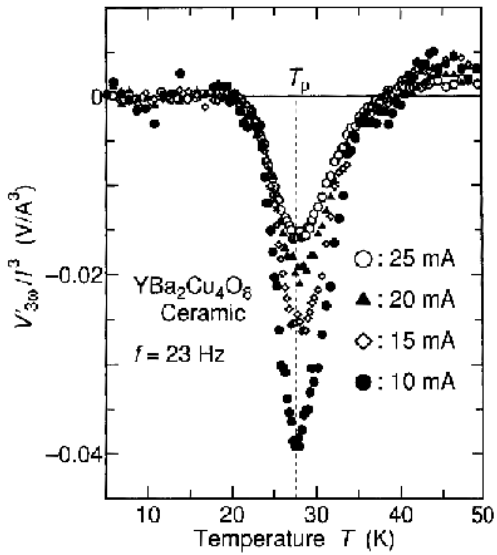}}
\vspace{0.1in}
\caption{ \label{v3_exp_fig}
The temperature dependence of $V'_{3\omega}/I^3$. Frequency
$f = 23$ Hz and values of $I$ are shown in the figure.
The peak is located at $T_p \approx $ 27.5 K. After Ref.
\protect\cite{Matsuura2}.
}
\end{figure}

Fig. \ref{v1_exp_fig} shows the temperature dependence of $V'_{\omega}/I$
for YBa$_2$Cu$_4$O$_8$ sample around $T_{c2}$ below which the remanent
magnetization becomes nonzero \cite{Matsuura2}.
The key observation from  this figure
is that $\rho_0$ does not vanish
even at and below the intergrain ordering temperature $T_{c2}$.
On the other hand, $\rho_2$ has a peak near this temperature, which
was found to be negative \cite{Matsuura2} (see Fig. \ref{v3_exp_fig}).

In the chiral glass phase the $U(1)$ gauge symmetry
is not broken and the phase of the condensate
remains disordered.\cite{KawLi97}
The chiral glass phase, therefore, should not be superconducting but
exhibit an Ohmic behavior with a finite small resistance.
Based on these theoretical predictions Yamao {\em et al.}\cite{Matsuura2}
speculated that their results give further
support to the existence of the chiral glass phase, in addition to previous
results from magnetic susceptibility measurements \cite{Matsuura1}.

Another interesting result of Yamao {\em et al.}\cite{Matsuura2}
is the power law dependence
of $|V'_{3\omega}(T_p)/I_0)^3|$ (or of  $\rho_2$)
at its maximum position $T_p$
on $I_0$: 
\begin{equation}
|V'_{3\omega}(T_p)/I_0^3| \sim I_0^{-\alpha} \, .
\label{exp_alpha}
\end{equation}
The experimental value of the power law exponent is
$\alpha\approx1.1$.

% FIGURE 44

\noindent {\em Simulation}. In order to reproduce
the experimental results of Yamao {\em et al.} \cite{Matsuura2} we
again use the multi-loop model (\ref{H_multiloop}) combined with
the Langevin dynamics equations of the RSJ model which are given
by Eq. (\ref{RSJ_eq}). We should add to the right part of these equations
the external oscillatory electric field 
$E_{ext}(t) \propto I_{ext}$ \cite{Li2000}.
The details of simulations are given in 
\cite{Li2000}.

\begin{figure}
\epsfxsize=2.6in
\centerline{\epsffile{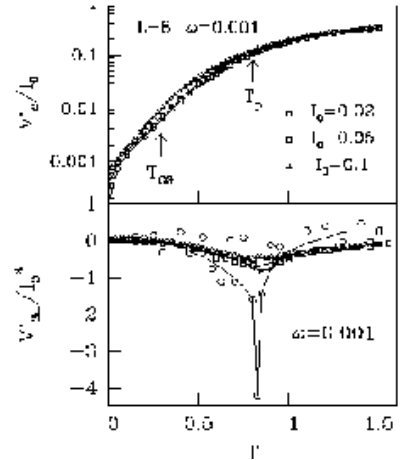}}
\vspace{0.1in}
\caption{ \label{v1v3_sim_fig}
The temperature dependence of $V'_{\omega}/I_0$ and $V'_{3\omega}/I_0^3$
of a $d$-wave superconductor
for $l=8, \tilde{L}=1$ and $\omega =0.001$.
The open triangles, squares and hexagons correspond to
$I_0 = 0.1, 0.05, 0.02$. The arrows correspond to $T_p=0.8$
and $T_{CG}=0.286$, respectively. The results are averaged over
15 - 40 disorder realizations.
From  Li and Dominguez \protect\cite{Li2000}.
}
\end{figure}

Fig. \ref{v1v3_sim_fig} shows the temperature dependence of the linear resistivity
$\rho_0=V'_{\omega}/I_0$ and the non-linear resistivity
$\rho_2=-4V'_{3\omega}(T)/I_0^3$
for different values of $I_0$. The linear resistivity is nonzero
at the chiral glass transition temperature $T_{CG}$ as well as at
the point $T_p$ where the nonlinear 
resistivity has maximum.
The results for
 $\rho_0$ presented in Fig. \ref{v1v3_sim_fig}
 were obtained for the system size $l=8$  but one can show
\cite{Li2000} that they are valid in the thermodynamic limit.
Thus, the fact that $\rho_0$ does not vanish at $T_p$ is in full agreement
with the experiments \cite{Matsuura2}. The other observation from
simulations \cite{Li2000} is that $T_p$ 
coincides with a temperature where the FC magnetization starts to
departure from the ZFC one. This is also consistent with the the experimental
findings of Yamao {\em et al.} \cite{Matsuura2}.   

% FIGURE 45

As said above Yamao {\em et al.} have made an interesting
conjecture that the peak of
the nonlinear resistivity should correspond to the phase transition to the
chiral glass phase.  $T_{CG}$ and
$T_p$ are, however,  well separated (see Fig. \ref{v1v3_sim_fig})
 suggesting that the peak in the nonlinear
resistivity may not be indicative for the existence of the chiral glass
state. One of possible interpretations of simulation results is that
$T_p$ just separates the normal state phase from a ``chiral paramagnet"
where there exist local chiral magnetic moments. These moments can
be polarized under an external magnetic field, an therefore
one can observe the PME under
a low external field below $T_p$.  At a lower temperature, collective
phenomena due to the interactions among the chiral moments will start
to be important, leading to the transition to the chiral glass state.
This last transition should show up in the nonlinear chiral glass
susceptibility which diverges at $T_{cg}$ \cite{KawLi}.

At present, it is not entirely clear whether $T_p$ corresponds 
to the chiral glass
transition. Its departure from $T_{CG}$ may be  
just a shortcoming of the RSJ dynamics. 

From lower panel of Fig. \ref{v1v3_sim_fig} one
can see that the height of the peak in $max|V'_{3\omega}/I_0^3|$ grows
as the current amplitude decreases supporting the experimental
result presented in  Fig. \ref{v3_exp_fig}.
A detailed study \cite{Li02} shows
that this dependence can be
described by the power law (\ref{exp_alpha}) which is valid not only for
both $d$-wave  and  $s$-wave ceramic superconductors.
The exponent $\alpha$ was found to be not universal but depend on the
self-inductance and current regimes \cite{Li02}. 

In the weak current 
regime $\alpha$ is independent of the self-inductance
and $\alpha = 0.5 \pm 0.1$
for both of $s$- and $d$-wave materials. In the s-wave case, since $T_p$ 
corresponds to a true continuous metal -- superconductor
 phase transition, one can use the scaling
argument \cite{Fisher91} to obtain the dynamical exponent $z$ via $\alpha$
\cite{Li02}. In three dimensions $z = 5-2\alpha \approx 4$ which is higher than
$z \approx 3$ for the corresponding  uniform model \cite{Wengel96}. 
For the $d$-wave system, $T_p$
is the temperature where there is an
onset of positive magnetization, i.e. the paramagnetic  Meissner effect
starts to be observed, but it does not seem to correspond to a
phase transition \cite{Li2000}. The scaling analysis is not, therefore,
applied. 

% FIGURE 46

In the strong current regime \cite{Li02} the exponent $\alpha$ depends
on the screening strength. For the $d$-wave superconductors and
$1 < \tilde{L} < 5$ we have $\alpha \approx 1$ which is close to the
experimental value \cite{Matsuura2}.
In order to make the comparison with experiments to be meaningful
we show that $\alpha$ was measured in the strong current regime.
In fact, the real current is $I = \frac{2eJ}{\hbar}I_0$, $J \sim 10^2$ K and
$I_0 \sim 10^{-1}$ we, therefore, have $I \sim 10^{-2}$ mA which is much
weaker than the current $I \sim 10$ mA  used in experiments of
Yamao {\em et al}.
The another issue is that the
interval of inductance where $\alpha \approx 1$ is
realistic for ceramics \cite{Marcon89} because typical
values of
$\tilde{L}$ are bigger than 3 \cite{Marcon89}. An accurate comparison between theory and
experiments requires, however, the knowledge of
$\tilde{L}$ which is not known for the compound of YBa$_2$Cu$_4$O$_8$ studied
in Ref. \onlinecite{Matsuura2}.

\subsection{Enhancement of critical current}

\noindent {\em Experiment}. The question about enhancement 
of the critical current of superconductors
is important 
from both the theoretical and application  points of view.
Several experiments \cite{Mannhart91,Xi92} in ultrathin (5 - 10 nm thick)
YBa$_2$Cu$_3$O$_{7-x}$ films
showed that an electric field $E$ may affect $T_c$ as well as the critical
current $I_c$. The later may increase or decrease depending on the polarity
of $E$ and this is usually
 attributed to the changes in the critical current due to
variations in the charge density or due to a redistribution of carriers which
appear at the surface layer with depths of the order of the electrostatic
screening length $d_E$ (in high-$T_c$ superconductors, $d_E \approx 5 \AA$).
 For any experimentally
measurable field-induced effects, $d_E$ should be larger than the
superconducting coherence length and this happens to the the case in ceramics
which have a reduced carrier density
\cite{Mannhart91,Xi92,Mannhart96}.

\begin{figure}
\epsfxsize=2.5in
\centerline{\epsffile{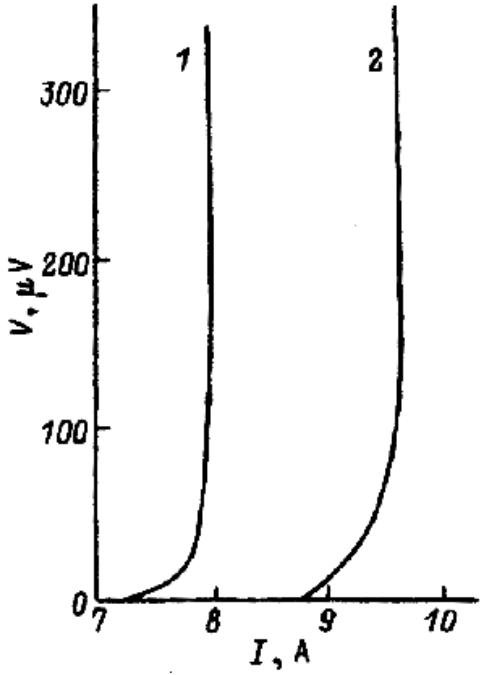}}
\vspace{0.1in}
\caption{\label{enhan_Jc_exp}
I-V characteristics for (BiPb)$_2$$_2$Sr$_2$Ca$_2$Cu$_3$O$_x$
sample for $E=0$ (1) and $E=120$ mV/m (2). $J_{c0} = 320$ A/cm$^2$ and
$T=77$ K. After applying the electric field the IV curve is shifted
towards higher currents, i.e. the apparent critical current increases.
After Smirnov {\em et al} \protect\cite{Smirnov96}.}
\end{figure}

Recent experiments
\cite{Orlova99,Smirnov92,Smirnov93,Orlova94,Smirnov94,Smirnov96,Orlova2001,Smirnov2000} revealed, however, that the enhancement of $J_c$ is also possible
in bulk (1.5 mm thick) ceramic high-$T_c$ superconductors when $E$ is
applied trough an
insulating layer (see Fig. (\ref{enhan_Jc_exp}).
This unusual behavior cannot be explained by a surface effect
\cite{Mannhart91,Xi92}. Moreover, for high enough electric fields, the critical
current always increases regardless of the polarity of the field.

% FIGURE 47

Using the SNS contact model one can show that an electric field can induce
a change in the critical currents of the Josephson junctions in granular
samples \cite{Rakhmanov96}. However, in this model there is either
an enhancement or a deleption of $J_c$ depending on
the sign of $E$. Authors of Ref. \cite{Sergeenkov98} have proposed that an
electric field applied to a granular superconductor can produce a
magnetoelectric-like effect, which could be
indirectly related to the behavior of the critical current observed in
\cite{Smirnov92,Smirnov93,Orlova94,Smirnov94,Orlova95,Smirnov97,Orlova98},
but no comparison with the experiments was given.

\noindent {\em Simulation}.
Following Dominguez {\em et al} \cite{Dominguez99} one can explain the 
experimental findings on enhancement of the critical current of bulk 
granular superconductors using the $d$-wave multi-loop model.
In fact, we could employ the full Langevin equations (\ref{RSJ_eq}) but
the unusual electric field effects may be captured even neglecting the
screening effect ($L=0$) and thermal fluctuations ($\eta_\mu({\bf n},t)=0$). 
Then the current $J_{\vec{\mu}}({\bf n})$ between to grains {\bf n} and
{\bf n} +$\vec{\mu}$ is a sum of the Josephson supercurrent plus a 
dissipative Ohmic current \cite{Dominguez99}
\begin{equation}
J_{\mu}({\bf n}) \, = \, J^0_{{\bf n},\mu} \sin\theta_{\mu}({\bf n})
+ \frac{\Phi_0}{2\pi R} \frac{d \theta_{\mu}({\bf n})}{dt}, 
\label{RSJ_elec_eq}
\end{equation}
where $\theta_{\mu}({\bf n}) = \phi ({\bf n} + \vec{\mu}) - \phi ({\bf n})
- A_{\mu}({\bf n},t)$ is the gauge invariant phase difference, and
$A_{\mu}({\bf n},t) = \frac{2\pi}{\Phi _0} \int_{{\bf n}}^{{\bf n}+\mu} 
{\bf A}. d{\bf l}$. Together with the conditions of current conservation,
$\sum_{\mu} [J_{\mu}({\bf n}) - J_{\mu}({\bf n}-\vec{\mu}) = 0$, 
equations (\ref{RSJ_elec_eq}) defines dynamics of the system. 

We assume that an electric field {\bf E} is applied in the $z$ direction and
the network is driven by an external current density $I_{ext}$ along the 
$y$ direction. Then, in the absence of the external magnetic field, the
vector potential $A_{\mu}({\bf n},t) = - \delta_{\mu,z}\omega _Et -
\delta_{\mu,y}\alpha_y(t)$, where the electric field frequency 
$\omega_E = 2\pi Ed/\Phi_0$ with $d$ being the intergrain distance or junction
thickness \cite{Dominguez99}. As follows from Eq. (\ref{RSJ_elec_eq}), the 
external current density determines the evolution of $\alpha_y(t)$ as
\begin{equation}
I_{ext} = \frac{1}{l^3} \sum_{{\bf n}} I^0_{{\bf n}y} \sin \theta_y({\bf n})
+ \frac{\Phi_0}{2\pi R} \frac{d\alpha_y}{dt} .
\label{RSJ_elec_eqy}
\end{equation}
The average voltage per junction induced by the driving current is then
obtained as $V = (\Phi _0/2\pi)<d\alpha_y/dt>$.

It should be noted that one does not take into account the effects of intergrain and intragrain capacitances
in model (\ref{RSJ_elec_eqy}) and the screening of the electric field is,
therefore, neglected \cite{Dominguez99}. 
The electric field scale is $E_0 = RI_0(T=0)/d = \pi \Delta(0)/2ed$, where
$\Delta (0)$ is the superconducting energy gap; for the YBaCuO ceramics one
has $\Delta (0) \approx 20$ meV, $d \sim 10 - 20 \AA$, which gives 
$E_0 \sim 30$ MV/m, i.e. in the same range as the fields used in 
\cite{Smirnov92,Smirnov93,Orlova94,Smirnov94,Orlova95,Smirnov97,Orlova98}.
Dominguez {\em et al} \cite{Dominguez99} studied two models of disorder.
(i) Granular $s$-wave superconductor: $I^0_{{\bf n}\mu}$ is assumed to
be a random variable uniformly distributed in the interval
$[I_0(1-\Delta_c), I_0(1+\Delta_c)]$ with $<I^0_{{\bf n}\mu}> = I_0$ 
and $\Delta_c < 1$. 
(ii) Granular $d$-wave superconductor involving 0 and  $\pi$ junctions
has the same bimodal distribution
for $I^0_{{\bf n}\mu}$ as discussed in previous sections.
 
\begin{figure}
\epsfxsize=3.2in
\centerline{\epsffile{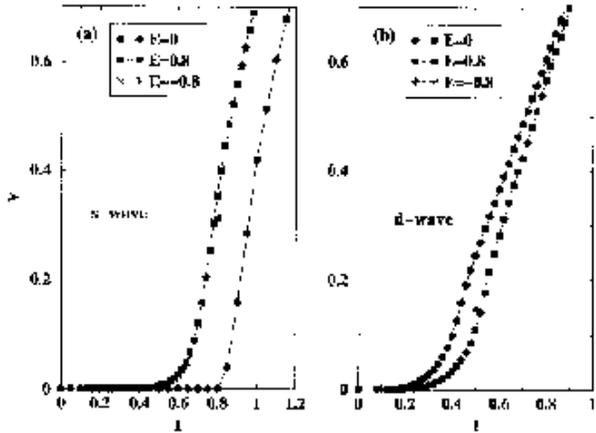}}
\vspace{0.2in}
\caption{\label{enhan_Jc_sim}
Current-voltage characteristics before and after applying an electric field 
{\bf E}. (a) For a granular $s$-wave superconductor with $\Delta_c = 0.6$.
(b) For a $d$-wave superconductor with a bimodal distribution of positive
and negative Josephson couplings coming from 0 and $\pi$ junctions.
Voltages are normalized by $lRI_0$ and currents by $l^2I_0$ and the lattice size
$l=16$.
After Dominguez {\em et al.} \protect\cite{Dominguez99}.}
\end{figure}

Fig. \ref{enhan_Jc_sim} shows typical simulation results 
\cite{Dominguez99} for current-voltage characteristics before
and after applying an electric field {\bf E} for a $s-$ (left panel) and
$d$-wave (right panel) three-dimensional ceramic superconductor. In the former 
case after the field is switched on
the whole $IV$ curve shifts to lower values of the current indicating the 
decrease of the apparent critical current and the voltage change 
$\Delta V = V(E) - V(0)$ is positive for a given current $I > I_c$. This 
contradicts, however, the experimental finding shown in Fig. \ref{enhan_Jc_exp},
where an increase in $I_c$ was seen.

% FIGURE 48

In the $d$-wave case, after application of the electric field
the $IV$ curve moves towards higher currents, i.e. the apparent critical 
current {\em increases}. It is precisely what was observed on experiments 
(see Fig. \ref{enhan_Jc_exp}). Furthermore, if one changes $E \rightarrow -E$
the $IV$ curves overlap showing that the effect is independent of the
polarity of the electric field. The similar effect was also seen in the 
experiments \cite{Orlova94,Smirnov94}.

There is other interesting aspect of experimental findings that if an applied
current is near the critical current then $\Delta V/V_0$ increases as a 
function of $E$  and, after reaching a maximum, it falls and then becomes
negative for large enough fields \cite{Smirnov93}. This behavior can be also 
explained in the framework of the $d$-wave model \cite{Dominguez}. 

Two unexpected results from the experiments on the electric field effects are:
(i) a pronounced electric field effect in  a bulk ceramic sample and
(ii) an increase of the apparent critical current as a function of $E$ 
independent of the field polarity. All these features are qualitatively
captured by the simple $d$-wave model in which the frustration due to
$\pi$ junction and the ac Josephson effect induced by the electric field
are important.
It would be interesting to explore the effect of the 
electric field on the paramagnetic signal, on the chiral glass phase and
glassy dynamics in ceramic superconductors \cite{Orlova94,Smirnov94}.

\section{Summary}

We have reviewed the experimental facts, simulations and theories
of the PME observed in conventional and ceramic high-$T_c$ superconductors.
There exist two main mechanisms for this very interesting phenomenon:
the flux compression and the $\pi$ junction. The later probably occurs due to
anisotropic pairing of electrons but other possibilities like the scattering
on magnetic impurities in Josephson junctions are not excluded. 

From the technical point of view, the paramagnetic signal due to the flux
trapping inside a sample of confined geometry may be obtained either by 
considering the Bean critical model or by solving the nonlinear Ginzburg-Landau
equations. In the case of ceramic superconductors with the $d$-wave symmetry
of the order parameter, the spontaneous magnetic moment leading to the PME
occurs in the loop of odd number of $\pi$ junctions. The screening
plays a crucial role in both the single loop and multi-loop models. It
should be noted that the later can capture not only the PME but also 
many related dynamical phenomena in granular materials.

The flux compression
scenario should work for mesoscopic
conventional materials but it may be also applied
to granular superconductors where the grain surfaces can play an
important role. Therefore, the existence of the PME itself could not
serve as an unambiguous indicator for the $d$-wave symmetry of the 
order parameter of cuprates. It would be very important to prepare
ceramic samples with negligible surface effects to check if the paramagnetic 
signal can occur or not.

In addition to the phase-sensitive and phase-insensitive experiments,
the experiments on the aging, 
anomalous MWA, $1/f$ flux noise,
compensation effect and enhancement of the critical current
which may be described by
simple XY model with screening (\ref{H_multiloop}), tend to
support the $d$-wave pairing symmetry
of the cuprate superconductors. 
From both experimental and
theoretical points of view it would be very exciting to find out whether
these effects appear in conventional superconductors showing the PME. 

The chiral glass phase which may be observed by negative divergence of the
nonlinear susceptibility is an interesting example of the time-reversal
symmetry breaking. In contrast to the gauge and vortex glass, it appears
only in a zero external magnetic field. Furthermore, the moderate screening 
could not destabilize the chiral glass phase at finite
temperatures in three dimensions. Since the experiments 
\cite{Ishida,Matsuura1,Matsuura2,Papadopoulou,Deguchi02,Deguchi02a} gave conflicting
results on 
the existence of this phase in high-$T_c$
superconductors, much more work in this direction is left for 
future studies.

\vspace{0.3cm}

\begin{center}
{\bf Acknowledgement}
\end{center}

\vspace{0.3cm}

We have benefited greatly from discussions with numerous colleagues. We should
particularly like to thank D. Dominguez, H. Kawamura, T. Nattermann,
P. Nordblad and H. Zung. We are very indebted to P. Janiszewski and M. V. Ba 
for their
kind help in preparing many figures. Financial support from the Polish
agency KBN (Grant number 2P03B-146-18) is acknowledged.

\appendix
\section{Critical self-inductance of the multi-loop model}
\label{sec.lc}

\renewcommand{\theequation}{\Alph{section}.\arabic{equation}}

\begin{figure}
\epsfxsize=2.2in
\centerline{\epsffile{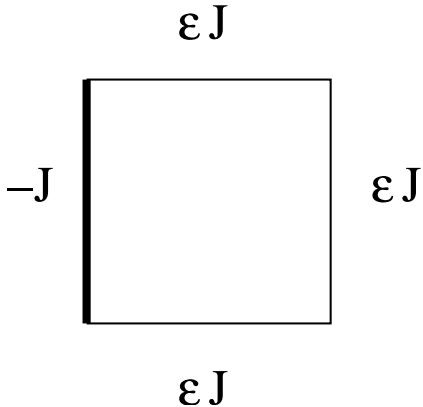}}
\vspace{0.2in}
\caption{An isolated frustrated loop consisting of four junctions, one of which
 is a $\pi$ junction of magnitude -J (thick line) and the other three are
0-junction of
magnitude $\epsilon J$ with $0 < \epsilon < \infty$. The case $\epsilon = 1$
corresponds to an elementary frustrated plaquette of the multi-loop model,
while $\epsilon = \infty$ corresponds to the frustrated single-loop model
analysed in the previous section.
}
\label{plaq_fig}
\end{figure}

In order to demonstrate that the critical inductance of the multi-loop model
is smaller than that of the single-loop model,
we analyse the ground state of an isolated
frustrated plaquette, an elementary unit of our lattice model as shown in Fig.
\ref{plaq_fig}. The plaquette consists of four bonds (junctions) among which
one is a $\pi$-junction with negative coupling $-J < 0$, and other three
are 0-junctions with positive coupling constant $\epsilon J > 0$
($ 0 < \epsilon < \infty$). The multi-loop model considered here corresponds
to $\epsilon = 1$, while  $\epsilon = \infty$ describes a single
$\pi$ junction analysed by Sigrist and Rice \cite{Sigrist92}. The
external field is set equal to 0.

% FIGURE 49

Introducing gauge-invariant phase difference on each bond
$\theta _{ij} = \phi_i - \phi_j - A_{ij}$ ("temporal gauge"
\cite{Dominguez,KawLi}), the dimensionless energy of the plaquette
shown in Fig. \ref{plaq_fig} is
\begin{eqnarray}
F \, = \, {\cal H}/J \, = \, \epsilon \cos \theta_1 - \cos \theta_2
- \cos \theta_3 - \cos \theta_4 + \nonumber\\
\frac{1}{\tilde{L}} (\theta_1 + \theta_2 + \theta_3 + \theta_4 )^2 \; .
\end{eqnarray}
Here, for simplicity, we drop one sub index of the $\theta_{ij}$ and denote
the phases as $\theta_1$ at $\pi$ junction and $\theta_2 - \theta_4$ 
at 0 junctions.
Minimizing $F$ with respect to $\theta_1 - \theta_4$, we get four coupled 
equations. After a little algebra, one can see that the ground-state 
configuration
should satisfy the relations
\begin{equation}
\theta_2 = \theta_3 = \theta_4 = \theta, \; \theta_1 = \mbox{arcsin}(-\epsilon \sin \theta),\end{equation}
with $\theta$ being determined from the following equation
\begin{equation}
\epsilon \sin \theta + \frac{1}{\tilde{L}}(3\theta +
\mbox{arcsin}(-\epsilon \sin \theta )) \, = \, 0 \, .
\end{equation}
One can show that if $\epsilon < 3$, this equation always has doubly degenerate
solutions $\theta = \pm \theta ^*$ which corresponds to supercurrents 
with opposite
directions. A spontaneous moment, therefore, arises for an arbitrary value
of inductance. If $\epsilon$ is larger than three, there appears a finite
critical value of the inductance $\tilde{L}_c(\epsilon) = 1 - 3/\epsilon$
\cite{KawLi}: for a inductance $\tilde{L} > \tilde{L}_c(\epsilon)$, there
again exist doubly degenerate solutions $\theta = \pm \theta ^*$, while for
$\tilde{L} < \tilde{L}_c(\epsilon)$, only trivial solution exists at $\theta = 0$
corresponding to the state without spontaneous supercurrent. In the limit
$\epsilon \rightarrow \infty$, one recovers the critical
value $\tilde{L}_c = 1$ obtained by Sigrist and Rice \cite{Sigrist92}.
Thus, we have shown that, contrary to the one-loop model,
the critical inductance of the interacting loop models with weak links
of nearly the same magnitude is equal to $\tilde{L}_c = 0$.

\section{RSJ model}
\label{sec.rsj}

\begin{figure}
\epsfxsize=3.2in
\centerline{\epsffile{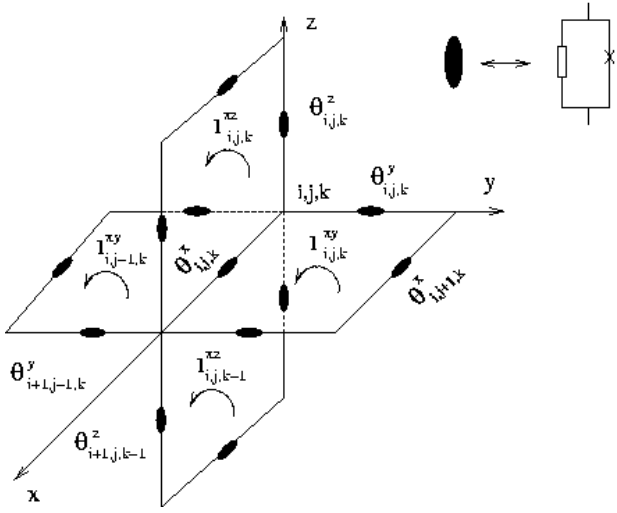}}
\vspace{0.2in}
\caption{Schematic description of the three-dimensional RSJ model of the
Josephson junction network. The gauge invariant phases are
marked on some representative junctions. The $\theta^{\mu}$
values grow in the positive directions the the corresponding axis $\mu = x,y,z$.
The positive direction of currents circulating
in loops is chosen to be counterclockwise. The inset shows a single 
junction containing
the resistive and superconducting parts.}
\label{RSJ_fig}
\end{figure}

In this appendix we derive the equations of the RSJ model which
have been used to study dynamical phenomena such as the MWA, the ac
resistivity and the enhancement of the critical current in chapter 
\ref{sec.dynamics}. As shown in the inset of Fig. \ref{RSJ_fig}
the RSJ model of a single junction contains the superconducting and
resistive parts which are
denoted by a cross and a rectangular, respectively. Then the total current
$I$ through the junction is \cite{Tinkham}
\begin{equation}
I \; = \; I_R + I_S = \; \frac{1}{R}\frac{\Phi_0}{2\pi}\frac{d\theta}{dt} 
+ I_c \sin \theta,
\label{RSJ_cur1}
\end{equation}
where $R$ and $I_c$ are the resistivity and the critical current of the
junction, respectively. The gauge invariant phase $\theta$ between points
1 and 2 is
\begin{equation}
\theta _{12} \; = \; \int_1^2 \, d{\bf l} .\left( \nabla\phi - 
\frac{2\pi}{\Phi_0}{\bf A} \right),
\label{ginvphase}
\end{equation}
where $\nabla\phi$ is the gradient of the phase of macroscopic wave function,
{\bf A} is the vector potential, and the line integration is taken
between two sides of the junction with sign convention in accordance with
the coordinate directions shown in Fig. \ref{RSJ_fig}.

% FIGURE 50

We now consider the three-dimensional network on a cubic lattice shown in 
Fig. \ref{RSJ_fig}. Representing the current flowing from point $(i,j,k)$ 
in direction
$\mu$ to the nearest point
as $I_{i,j,k}^{\mu}$, from Eq. (\ref{RSJ_cur1})
we have \cite{Dominguez96}
\begin{eqnarray}
I^{\mu}_{i,j,k} = \frac{1}{R}\frac{\Phi_0}{2\pi}
\frac{d\theta ^{\mu}_{i,j,k}}{dt}
+ I^{\mu}_{i,j,k c} \sin \theta ^{\mu}_{i,j,k}, \; \mu = x,y,z,
\label{RSJ_curijk}
\end{eqnarray}
where $\theta ^{\mu}_{i,j,k}$ is the gauge invariant phase between points
$(i,j,k)$ and its nearest neighbor in direction $\mu$.
$I^x_{i,j,k}$ is, e.g., the current from $(i,j,k)$ to
$(i+1,j,k)$ (see Fig. \ref{RSJ_fig}) and $I^{\mu}_{i,j,k c}$ is
the local critical current.

The crucial point here is that $I^{\mu}_{i,j,k}$ consists of contributions
from four loops sharing the corresponding junction and we, therefore, have
\begin{eqnarray}
I^x_{i,j,k} \; = \; I^{xy}_{i,j,k} - I^{xy}_{i,j-1,k} + I^{xz}_{i,j,k-1}
-I^{xz}_{i,j,k} \, , \nonumber\\
I^y_{i,j,k} \; = \; I^{yz}_{i,j,k} - I^{yz}_{i,j,k-1} + I^{xy}_{i-1,j,k}
-I^{xy}_{i,j,k} \, , \nonumber\\
I^x_{i,j,k} \; = \; I^{xz}_{i,j,k} - I^{xz}_{i-1,j,k} + I^{yz}_{i,j-1,k}
-I^{yz}_{i,j,k} \, .
\label{loopcurrent}
\end{eqnarray}

Labeling the flux $\Phi^{\mu\nu}_{i,j,k}$ threading through loop
$(i,j,k)$ in $\mu\nu$-plane in the same manner as $I^{\mu\nu}_{i,j,k}$
we have

\begin{eqnarray}
\Phi^{xy}_{i,j,k} &=& \frac{\Phi_0}{2\pi}\left( \theta^x_{i,j,k} + \theta^y_{i+1,j,k} -
\theta^x_{i,j+1,k} - \theta^y_{i,j,k} \right), \nonumber \\
\Phi^{xz}_{i,j,k}  &=& \frac{\Phi_0}{2\pi}\left( \theta^z_{i,j,k} + \theta^x_{i,j,k+1} -
\theta^z_{i+1,j,k} - \theta^x_{i,j,k} \right), \nonumber\\
\Phi^{yz}_{i,j,k} &=& \frac{\Phi_0}{2\pi}\left( \theta^y_{i,j,k} + \theta^z_{i,j+1,k} -
\theta^y_{i,j,k+1} - \theta^z_{i,j,k} \right) .
\label{loopflux}
\end{eqnarray}
For the cubic array containing $l\times l\times l$ junctions fluxes
(\ref{loopflux})
are defined for following values of $(i,j,k)$:
\begin{eqnarray}
\Phi^{xy}_{i,j,k} : 0 \le i \le l-1; 0 \le j \le l-1; 0 \le k \le l;
 \nonumber\\
\Phi^{xz}_{i,j,k} : 0 \le i \le l-1; 0 \le j \le l; 0 \le k \le l-1; 
 \nonumber\\
\Phi^{yz}_{i,j,k} : 0 \le i \le l; 0 \le j \le l-1; 0 \le k \le l-1.
\label{fluxindex}
\end{eqnarray}
The corresponding restrictions on $(i,j,k)$ for phases are
\begin{eqnarray}
\theta^{x}_{i,j,k} : 0 \le i \le l-1; 0 \le j \le l; 0 \le k \le l;
 \nonumber\\
\theta^{y}_{i,j,k} : 0 \le i \le l; 0 \le j \le l-1; 0 \le k \le l;
 \nonumber\\
\theta^{z}_{i,j,k} : 0 \le i \le l; 0 \le j \le l; 0 \le k \le l-1.
\label{phaseindex}
\end{eqnarray}

Assuming the self-inductances of all loops to be identical and equal to $L$
one has

  \begin{equation}
  \label{currentflux}
  I^{\mu\nu}_{i,j,k}  =
    \left\{\begin{array}{cl}
        (\tilde{\Phi}^{\mu\nu}_{i,j,k}-\tilde{\Phi}^{ext}_{\mu\nu})/L &
       \mbox{for  (i,j,k) $\in$ (\ref{fluxindex})}\\
      0 & \mbox{otherwise.}
    \end{array}\right.
  \end{equation}
Here $\Phi_{\mu\nu}^{ext}$ is a flux threading through a loop 
in $\mu\nu$-plane which is equal to $\Phi_{\mu\nu}^{ext} = HS$ if
the ${\mu\nu}$-plane is perpendicular to $\vec{H}$ and $\Phi_{\mu\nu}^{ext} = 0$
otherwise (see also Eq. (\ref{H_plaq})).
Substituting Eq. (\ref{loopcurrent}), (\ref{loopflux}) and (\ref{currentflux})
into Eq. (\ref{RSJ_curijk}) we obtain the full set of equations for 
gauge invariant phases which is often used to study dynamics of the
Josephson junction network at zero temperature. For nonzero temperatures
one has to add the Langevin noise term to mimic the effect of the heat bath.

We now show that in the case when the external magnetic field $H=0$
the Langevin equations for
phases may be written in a compact form. 
Using Eq. (\ref{crit_curr}) and $\Phi_0=hc/2e$ one has
\begin{equation}
I^{\mu}_{i,j,k c} \; = \; \frac{2e}{\hbar c} J^{\mu}_{i,j,k},
\label{loccoupl}
\end{equation}
where $J^{\mu}_{i,j,k}$ is the Josephson couplings. From Eqs.(\ref{RSJ_curijk}),
(\ref{loopcurrent}), (\ref{loopflux}), (\ref{currentflux}) and
(\ref{loccoupl}) we obtain the following equation for $\theta^x_{i,j,k}$: 
\begin{eqnarray}
\frac{\hbar}{2eR}\frac{d\theta^x_{i,j,k}}{dt} = 
-\frac{2e}{\hbar c^2} J^x_{ijk} \sin \theta^x_{i,j,k} \nonumber\\
-\frac{\hbar}{2eL} \big[ (\theta^x_{i,j,k} + \theta^y_{i+1,j,k} -
\theta^x_{i,j+1,k} - \theta^y_{i,j,k}) \nonumber\\
-(\theta^x_{i,j-1,k} + \theta^y_{i+1,j-1,k} -
\theta^x_{i,j,k} - \theta^y_{i,j-1,k}) \nonumber\\
+( \theta^z_{i,j,k-1} + \theta^x_{i,j,k} -
\theta^z_{i+1,j,k-1} - \theta^x_{i,j,k-1}) \nonumber\\
 - ( \theta^z_{i,j,k} + \theta^x_{i,j,k+1} -
\theta^z_{i+1,j,k} - \theta^x_{i,j,k}). \big]
\label{RSJ_mun}
\end{eqnarray}
Equations for $\theta^y_{i,j,k}$ and $\theta^z_{i,j,k}$ can be written 
in a similar form. For the sake of convenience we represent $(i,j,k)$ by
vector {\bf n} and redefine 
$\theta^{\mu}_{i,j,k} \rightarrow \theta_{\mu}({\bf n})$ and 
$J^{\mu}_{i,j,k} \rightarrow J_{\mu}({\bf n})$. We introduce also the
forward operator $\Delta_{\mu}^{+}$ and backward operator $\Delta_{\mu}^-$
which act on $\theta_\nu({\bf n})$ in the following way
\begin{eqnarray} 
\Delta_{\mu}^{+}\theta_\nu({\bf n}) &=&
\theta_\nu({\bf n}+\mu)-\theta_\nu({\bf n}), \nonumber\\
\Delta_{\mu}^{-}\theta_\nu({\bf n}) &=& \theta_\nu({\bf n})-
\theta_\nu({\bf n}-\mu).
\label{operatorfb}
\end{eqnarray}
In the $(i,j,k)$-representation the last equation takes the form
\begin{eqnarray}
\Delta_{x}^{+}\theta^{\nu}_{i,j,k} &=&
\theta^{\nu}_{i+1,j,k} - \theta^{\nu}_{i,j,k}, \nonumber\\
\Delta_{x}^{-}\theta^{\nu}_{i,j,k} &=& \theta^{\nu}_{i,j,k} -
\theta^{\nu}_{i-1,j,k}
\label{operfbijk}
\end{eqnarray}
for component $\mu = x$. Similar equations for $\mu = y$ and $z$ can
be easily written down.
With the help of Eq. (\ref{operfbijk}) one can show that Eq. (\ref{RSJ_mun})
is reduced to Eq. (\ref{RSJ_eq}) in the main text.

\par

\end{multicols}
\end{document}